\documentclass[acmsmall, screen, nonacm]{acmart}

\usepackage{paralist} %
\usepackage[T1]{fontenc} %
\usepackage{glossaries} %
\newacronym{CPU}{CPU}{central processing unit}
\newacronym{GPU}{GPU}{graphics processing unit}
\newacronym{QPU}{QPU}{quantum processing unit}
\newacronym{HPC}{HPC}{high performance computing}
\newacronym{LRZ}{LRZ}{Leibniz Supercomputing Centre}
\newacronym{DSL}{DSL}{domain-specific language}
\newacronym{eDSL}{eDSL}{embedded domain-specific language}
\newacronym{NISQ}{NISQ}{noisy intermediate-scale quantum}
\newacronym{MPI}{MPI}{Message Passing Interface}
\newacronym{QRAM}{QRAM}{quantum random access machine}
\newacronym{HQCC}{HQCC}{heterogeneous quantum-classical computation}
\newacronym{API}{API}{application programming interface}
\newacronym{QPT}{QPT}{quantum programming tool}
\newacronym{QC}{QC}{quantum computing}
\newacronym{ISA}{ISA}{instruction set architecture}
\newacronym{RISC}{RISC}{reduced instruction set computer}
\newacronym{RAM}{RAM}{random access memory}
\newacronym{IR}{IR}{intermediate representation}
\newacronym{SQRAM}{SQRAM}{sequential QRAM}
\newacronym{HPCQC}{HPCQC}{high performance computing - quantum computing}
\newacronym{QIR}{QIR}{Quantum Intermediate Representation}
\newacronym{SQIR}{SQIR}{standardized quantum intermediate representation}
\setacronymstyle{long-sp-short} %
\glsdisablehyper %

\usepackage{braket}
\usepackage{verbatim}
\usepackage{enumitem}
\usepackage{subcaption}
\usepackage{array} %
\usepackage{listings}
\usepackage{units}
\begin{document}

\newcommand{\umbrellaterm}{\gls{QPT}}
\newcommand{\umbrellatermpl}{\glspl{QPT}}
\newcommand{\Umbrellaterm}{\Gls{QPT}}
\newcommand{\Umbrellatermpl}{\Glspl{QPT}}

\title{Integration of Quantum Accelerators with High Performance Computing --- A Review of Quantum Programming Tools}

\author{Amr Elsharkawy}
\email{amr.elsharkawy@in.tum.de}
\orcid{0000-0002-5797-0820}
\affiliation{%
  \institution{Technical University of Munich}
  \country{Germany}}
  \state{Bayern}
  \postcode{85748}
  \city{Garching}

\author{Xiao-Ting Michelle To}
\email{michelle.to@nm.ifi.lmu.de}
\orcid{0000-0002-2407-7892}
\affiliation{%
  \institution{Ludwig-Maximilians-Universität in Munich}
  \country{Germany}}

\author{Philipp Seitz}
\email{philipp.seitz@tum.de}
\orcid{0000-0003-3856-4090}
\affiliation{%
  \institution{Technical University of Munich}
  \country{Germany}}

\author{Yanbin Chen}
\email{yanbin.chen@tum.de}
\orcid{0000-0002-1123-1432}
\affiliation{%
  \institution{Technical University of Munich}
  \country{Germany}}

\author{Yannick Stade}
\email{stya@cit.tum.de}
\orcid{0000-0001-5785-2528}
\affiliation{%
  \institution{Technical University of Munich}
  \country{Germany}}

\author{Manuel Geiger}
\email{manuel.geiger@tum.de}
\orcid{0000-0003-3514-8657}
\affiliation{%
  \institution{Technical University of Munich}
  \country{Germany}}

\author{Qunsheng Huang}
\email{huangq@in.tum.de}
\orcid{0000-0002-1289-6559}
\affiliation{%
  \institution{Technical University of Munich}
  \country{Germany}}

\author{Xiaorang Guo}
\email{xiaorang.guo@tum.de}
\orcid{0000-0003-1697-817X}
\affiliation{%
  \institution{Technical University of Munich}
  \country{Germany}}
  
\author{Muhammad Arslan Ansari}
\email{arslan.ansari@tum.de}
\orcid{0000-0003-1833-6170}
\affiliation{%
  \institution{Technical University of Munich}
  \country{Germany}}
  
\author{Christian B.~Mendl}
\email{christian.mendl@tum.de}
\orcid{0000-0002-6386-0230}
\affiliation{%
  \institution{Technical University of Munich}
  \country{Germany}}

\author{Dieter Kranzlmüller}
\email{kranzlmueller@ifi.lmu.de}
\orcid{0000-0002-8319-0123}
\affiliation{%
    \institution{Ludwig-Maximilians-Universität in Munich and Leibniz Supercomputing Centre}
    \country{Germany}}  

\author{Martin Schulz}
\email{schulzm@in.tum.de}
\orcid{0000-0001-9013-435X}
\affiliation{%
  \institution{Technical University of Munich and Leibniz Supercomputing Centre}
  \country{Germany}}  

\renewcommand{\shortauthors}{Elsharkawy et al.}
\renewcommand{\shorttitle}{Integration of Quantum Accelerators with HPC --- A Review of Quantum Programming Tools}

\renewcommand{\sectionautorefname}{Section}
\renewcommand{\subsectionautorefname}{Section}
\renewcommand{\subsubsectionautorefname}{Section}

\begin{abstract}

\Gls{QC} introduces a novel mode of computation with the possibility of greater computational power that remains to be exploited---presenting exciting opportunities for \gls{HPC} applications.
However, recent advancements in the field have made clear that \gls{QC} does not supplant conventional \gls{HPC}, but can rather be incorporated into current heterogeneous \gls{HPC} infrastructures as an additional accelerator, thereby enabling the optimal utilization of both paradigms.
The desire for such integration significantly affects the development of software for quantum computers, which in turn influences the necessary software infrastructure. %
To date, previous review papers have investigated various \glspl{QPT} (such as languages, libraries, frameworks) in their ability to program, compile, and execute quantum circuits. 
However, the integration effort with classical \gls{HPC} frameworks or systems has not been addressed.
This study aims to characterize existing \glspl{QPT} from an \gls{HPC} perspective, investigating if existing \gls{QPT}s have the potential to be efficiently integrated with classical computing models and determining where work is still required.
This work structures a set of criteria into an analysis blueprint that enables \gls{HPC} scientists to assess whether a \gls{QPT} is suitable for the quantum-accelerated classical application at hand. 
\glsresetall
\end{abstract}

\maketitle

\section{Introduction}\label{sec:introduction}

The rapid development of \gls{QC}, from Shor's algorithm~\cite{shor1994algorithms} in 1994 to recent claims of quantum advantage~\cite{arute2019quantum}, signifies a substantial shift in the paradigm of computation.
Novel quantum technologies may provide a significant advantage for specific computationally expensive applications, such as cryptography~\cite{securecom2020} or quantum chemistry~\cite{chemical2020}. 
Hence, development of quantum systems and \umbrellatermpl{} is a fast-growing area of research, where best practices and optimal strategies have yet to be standardized. 
The term \Umbrellaterm{} embraces quantum programming languages, libraries, and frameworks; this distinction is crucial since it has an impact on the integration process of \gls{QC} with \gls{HPC}, as detailed in \autoref{subsubsec:Type}.

Since quantum computers are sensitive machines with highly complex control hardware, specific expertise and infrastructure are required to build and run them. 
Besides, most implementations of quantum computers offer remote access and sharing of compute resources, which mirrors services provided by existing data centers or \gls{HPC} clusters.
Furthermore, it makes sense to integrate quantum accelerators into such facilities, which already allow for bespoke technical implementations and are experienced in supporting remote access.
This integration offers the possibility of enhancing quantum algorithms with high performance capabilities.
This paper overviews the feasibility and viability of existing \umbrellatermpl{} with respect to \gls{HPC} integration.
We introduce four integration scenarios, with important criteria that have to be taken into account when rating whether a \umbrellaterm{} is suitable for \gls{HPCQC} integration.
For a more straightforward way to rate a \umbrellaterm{}, we propose an analysis blueprint which can be followed by ``answering questions''.
We then use this blueprint to rate six \umbrellatermpl{}: QWIRE~\cite{paykin_qwire_2017}, Quil~\cite{smith_practical_2017}, OpenQASM~3~\cite{cross_openqasm_2022}, Qiskit~\cite{qiskit}, XACC~\cite{mccaskey_xacc_2020}, and OpenQL~\cite{khammassi_openql_2022}.
These were chosen because of their relevance for \gls{QC} in general, or because they have interesting features regarding \gls{HPCQC} integration.

The paper is structured as follows.
In \autoref{sec:background}, we introduce \gls{HPC} and \gls{QC} and discuss what is needed to combine them. 
In \autoref{sec:related_work}, we give an overview of previous survey papers, none of which cover \gls{HPC} integration. 
\autoref{sec:qc_hpc_integration} addresses how \gls{QC} can be integrated into \gls{HPC} on a hardware or software level.
We then introduce important properties that \umbrellatermpl{} need for \gls{HPC} integration in \autoref{sec:taxonomy}, followed by our proposal of an analytical blueprint to evaluate existing \umbrellatermpl{} in \autoref{sec:comparative_analysis}. 
Additionally, we introduce six noteworthy candidates and apply our blueprint to them, rating their suitability to \gls{HPC} integration respectively. 
In the subsequent discussion in \autoref{sec:discussion}, we highlight existing gaps in \gls{HPCQC} integration.
Finally, in \autoref{sec:conclusion} we conclude with a summary of our findings and an outlook for our future work.

\section{Background}\label{sec:background}

This section describes the basics for understanding the integration of quantum computers as quantum accelerators into \gls{HPC} systems.
We provide a brief overview of the typical \gls{HPC} system, toolchain, and workflow. %
Furthermore, we give a short introduction into the basics of \gls{QC}, focusing on the toolchain and workflow of a quantum program.
We thereby highlight the main advantages and shortcomings of utilizing a quantum system before detailing the various challenges in integrating quantum and \gls{HPC} systems.

\subsection{High Performance Computing} \label{subsec:HPC_World}

\gls{HPC} is the term used to describe the aggregation of compute resources to run compute programs, typically large-scale simulations, that would otherwise be intractable on a single machine.
When compared to classical computing on a single machine, \gls{HPC} systems have the additional overhead of \emph{scheduling}, \emph{communication}, and \emph{synchronization}.
Scheduling refers to ordering incoming jobs and allocating resources for maximum throughput. %
As a consequence of using multiple processes, information often must be shared or synchronized at specific time points to satisfy data dependencies or ensure correctness.
When the volume of data shared and/or the regularity of communication increases, the communication and synchronization overhead becomes a significant consideration when reducing time to solution~\cite{hager2010introduction}.

Given that \gls{HPC} systems have been in use since the early 1970s in a wide range of fields,
comprehensive work has been put into designing workload agnostic systems as well as standard user interfaces for requesting and using \gls{HPC} resources \cite{Hey_HPC}.
The following sections detail the specifics of programming paradigms, common parallel programming approaches, toolchains, and workflows in \gls{HPC} systems.

\subsubsection{Programming Paradigms} \label{subsubsec:Programming_Paradigms}
Imperative programming is a programming paradigm in which a program is composed of step-by-step instructions that change its state to achieve the desired outcome. 
This is in contrast to the functional paradigm, in which functions which map one value to another are applied and composed to form programs~\cite{Abelson1996-qw}.
Reviews and studies indicate that \gls{HPC} is strongly dominated by imperative, object-oriented implementations~\cite{FARHOODI2013, Johanson2018}.
However, this trend may not reflect a preference for a particular programming paradigm, as systemic studies of \gls{HPC} publications indicate that functional paradigms are examined in current research as well~\cite{AMARAL2020102584}.

Recent surveys~\cite{Arvanitou2021} and the most commonly used \gls{HPC} benchmarks show a strong focus on reducing time to solution, with additional emphasis on scalability, memory efficiency, and compute efficiency~\cite{Ihde2022}.
Then, considering that a vast majority of libraries or implementations that support massively parallel distributed computing are only available in imperative languages, such as C, C++, and Fortran~\cite{Ciccozzi2022, Laguna2019}, the observed behavior is not surprising.
Hence, future integration of quantum systems with \gls{HPC} applications will likely need to take this into consideration when designing the interface to a typical \gls{HPC} application.

\subsubsection{Parallel Programming Approaches} \label{subsubsec:Programming_Models}
\gls{HPC} relies heavily on effectively exploiting the given hardware when distributing tasks, performing communication, and ensuring data correctness.
While there is a multitude of parallel programming models, they largely fall into two main categories: \emph{shared} memory models and \emph{distributed} memory models.
Shared memory models assume that multiple processing units can access, read, and write from the same location in memory; such models include %
OpenMP~\cite{openmp08}. 
Distributed memory models assume that explicit commands must be used to pass data between processing units; such models include the \gls{MPI}~\cite{mpi40}. %

In summary, the seamless integration of quantum computers into HPC systems demands a meticulous balance between established parallel programming approaches and innovative quantum-specific programming models and frameworks. With the steady evolution of \gls{QC} technology, we can anticipate the emergence of novel parallel programming approaches that are tailored to the unique characteristics of quantum hardware.

\subsubsection{Classical HPC Workflow \& Toolchain} \label{subsubsec:classical_toolchain}
\begin{figure}
    \includegraphics[width=0.60\textwidth]{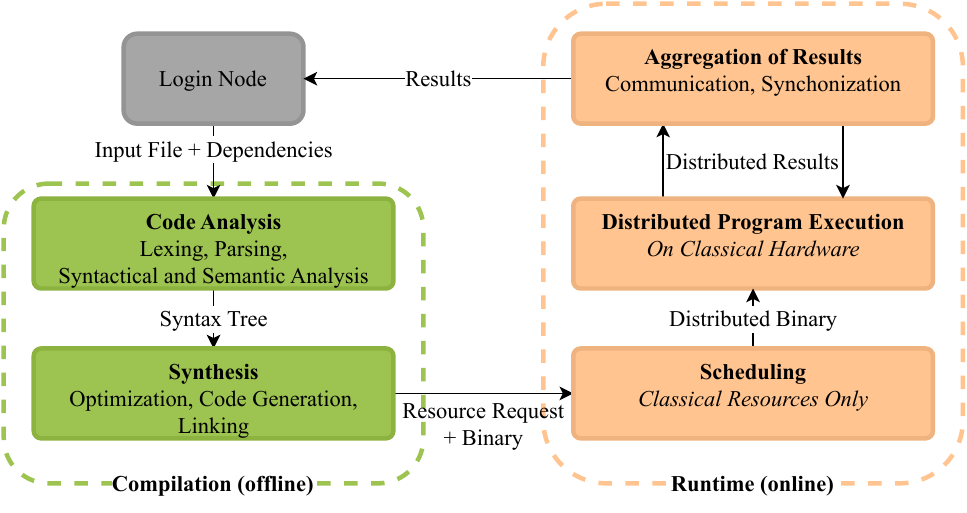}
    \caption{Abstracted \gls{HPC} workflow. Users are responsible for specifying their build dependencies and compilation toolchain, while service providers are responsible for scheduling and hardware execution. The \gls{HPC} workflow starts from the Login Node followed by the Code Analysis and Synthesis included in the Compilation.
    The binaries are distributed among \gls{HPC} computing resources, i.\,e., online, and executed on those.}
    \Description{The figure shows the abstract structure of a general HPC workflow. On the left-hand side three tasks are depicted that are performed at compile time, i.\,e. the login node, code analysis (including lexing, parsing, syntactical, and semantic analysis), and synthesis (including optimization, code generation, and linking). The figure shows on the right-hand side also three boxes that show the different stages of the execution of a program. The binaries that are the output of the synthesis, are scheduled (first stage), then they are executed on classical hardware in a distributed manner (second stage). Finally, the results are aggregated (stage three) and new jobs might be executed, hence, stage two and three work in cooperation.}
    \label{fig:Toolchain_Workflow_HPC}
    \vspace{-4mm}
\end{figure}

The classical \gls{HPC} \textit{workflow} describes a sequence of steps to perform a computation on an \gls{HPC} system.
A general abstracted workflow is shown in \autoref{fig:Toolchain_Workflow_HPC}, using the \gls{HPC} systems of the \gls{LRZ} as a baseline.
\par \gls{HPC} systems typically provide an external access point, commonly referred to as a \textit{login node}. 
They provide a flexible build environment, supported by a package manager, so that users can define their required dependencies and specify their compiler or interpreter; this is also where code optimization and generation take place.
The location of the compiled binaries and requested resources are wrapped in a job request, which is then sent to a scheduler, such as SLURM~\cite{Yoo2003SLURMSL}.
Tasks are distributed onto different compute nodes; communication and synchronization steps occur intermittently in the program per-user specification.
After termination, the results of the job are made available to the user.
Users may utilize the programming models mentioned in \autoref{subsubsec:Programming_Models} or technologies for automatic parallelization to improve execution efficiency~\cite{Blume1994, POLYCHRONOPOULOS1989}.
On the other hand, the service provider is responsible for scheduling and resource allocation, and there is very little, if any, user interaction in this step.
\par The general workflow described above relies on a series of software tools or components used to compile and execute programs on the \gls{HPC} system, which we then refer to as the classical \gls{HPC} \emph{toolchain}.
The compiler or interpreter, package manager, scheduler, and the required libraries for distributed execution are all examples of components that make up the toolchain.
This provides a basis for comparison with quantum systems; while the classical toolchain is mature and well understood, it must be extended to allow for integration with quantum systems.

\subsection{Quantum Computing} \label{subsec:Quantum_Computing} 
In classical computing, the base unit of computation is the bit, while in \gls{QC}, it is the \emph{quantum bit} or \emph{qubit}. 
Similar to a bit, a qubit has two basis states labeled 0 and 1. 
However, unlike a bit, a qubit can exist in a \emph{superposition} of these two states. 
This means that qubits are more expressive compared to classical bits and allow for computations beyond the classical realm.
Therein lies part of the computational power of a quantum computer.

\subsubsection{Basics of Quantum Computing} \label{subsubsec:Basics_of_Quantum_Computing}
The \emph{state} of a single qubit system is represented as a unit vector $\ket{\psi}$ in the two-dimensional complex Hilbert space $\mathcal{H} = \mathbb{C}^2$. The standard basis vectors are conventionally denoted as $\ket{0} = (1, 0)^T$ and $\ket{1} = (0, 1)^T$, such that we can write
\begin{equation}
    \label{eq:qubit}
    \ket{\psi}=\alpha\ket{0}+\beta\ket{1} = \begin{pmatrix} \alpha \\ \beta \end{pmatrix}, \; \alpha, \beta \in \mathbb{C}
\end{equation}
with the normalization condition
\begin{equation}
    \label{eq:qubit_amplitudes}
    |\alpha|^2+|\beta|^2 = 1.
\end{equation}
In the context of quantum physics, the coefficients $\alpha$ and $\beta$ are called \emph{amplitudes}. 
If both $\alpha$ and $\beta$ are non-zero, then the qubit is in a \emph{superposition} of the basis $\{\ket{0}, \ket{1}\}$.

For a quantum system of $n \ge 1$ qubits, the construction generalizes as follows: a quantum state is a unit vector in the Hilbert space $\mathcal{H} = \mathbb{C}^2 \otimes \cdots \otimes \mathbb{C}^2 = (\mathbb{C}^2)^{\otimes n}$. %
$\mathcal{H}$ is isometric to $\mathbb{C}^{(2^n)}$ and, in particular, has dimension $2^n$. 
This definition becomes more concrete when introducing associated computational basis states $\ket{b_{n-1} \cdots b_0}, b_i \in \lbrace0, 1\rbrace$, or simply $\ket{k}$; $b_{n-1}\cdots b_0$ is the binary representation of the integer $k$. 
An $n$-qubit state is thus a linear combination of these basis states. 
For example, for $n = 2$ qubits there are four basis states $\{ \ket{00}, \ket{01}, \ket{10}, \ket{11} \}$ and the state vector is a linear combination of these: $\ket{\psi} = \alpha \ket{00} + \beta \ket{01} + \gamma \ket{10} + \delta \ket{11}$ with amplitudes $\alpha, \beta, \gamma, \delta \in \mathbb{C}$ normalized as $|\alpha|^2 + |\beta|^2 + |\gamma|^2 + |\delta|^2 = 1$. 
The exponentially growing dimension $2^n$ is often referred to as ``curse of dimensionality'' and underlines the complexity-theoretic non-feasibility of efficiently simulating arbitrary quantum systems on classical computers \cite{Aaronson2017, Haferkamp2020}.

An important concept relevant for multi-qubit systems is \emph{entanglement}: 
a multi-qubit state is entangled if it cannot be represented as a tensor product of single-qubit states.
Besides its lack of a classical analog, entanglement is regarded as a ``resource'' \cite{ Chitambar2019}: for example, so-called cluster states enable an alternative approach for implementing quantum computations based on quantum measurements, starting from an initial state with large entanglement.

On a physical level, a quantum system changes as time progresses.
The time evolution can be formulated as follows: let $\ket{\psi}$ denote the initial quantum statevector at time $t_0$, then the state $\ket{\psi'}$ at some later time $t_1 \ge t_0$ is obtained by applying a certain unitary matrix $U$, i.e., $\ket{\psi'} = U \ket{\psi}$. %
The operation $U$ is also referred to as a \emph{quantum gate}. 
The unitary property in particular implies that the normalization \eqref{eq:qubit_amplitudes} is preserved.
For the purpose of \gls{QC}, one strategically constructs a quantum arrangement to generate a specific time evolution operation, serving the purpose of achieving desired computational outcomes.
Additionally, there are multi-qubit gates, which affect more than a single qubit.
A subset contains controlled gates, in which the application of a quantum operation depends on the state of another qubit. 

There are non-unitary operations that can also have an impact on qubits, causing irreversible changes. %
An example is the measurement operation:
it is used to retrieve information from a quantum state; the measurement outcome is normally either $\ket{0}$ or $\ket{1}$.
The squared absolute values of the amplitudes \(\alpha\) and \(\beta\) in (\ref{eq:qubit}) denote the probability of measuring \(\ket{0}\) or \(\ket{1}\) respectively.
The laws of quantum physics prohibit measuring a quantum state without disturbing it; the quantum state collapses to one of the basis states once it is measured.
Hence, this operation ``destroys'' the previous superposition and thus information about the original quantum state is lost.

\begin{figure}
    \centering
    \includegraphics[scale=0.8]{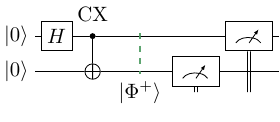}
    \caption{Bell state $\Phi^+$ preparation and subsequent measurement in quantum circuit notation.}
    \Description{Quantum circuit showing two qubit wires, both with initial value $\ket{0}$. On the upper qubit, a Hadamard gate is applied. Then a CNOT-gate is applied on both qubits, the upper one being the control qubit, the lower one the target qubit. After this gate, there is a note saying that the current state is the Bell state $\ket{Phi^+}$. In the end, both qubits are measured.}
    \label{fig:bell_circuit}
    \vspace{-4mm}
\end{figure}

A common representation of quantum programs are \emph{quantum circuits}, the quantum-equivalent to classical circuits.
An example is depicted in \autoref{fig:bell_circuit}, showing the preparation of an entangled quantum state. 
In quantum circuits, qubits are represented by \emph{wires}.
Operations are symbolized by gates on the corresponding wires and are applied on the qubits from left to right.
The example circuit starts with two qubits in the initial state $\ket{00}$. 
A Hadamard gate is applied on one qubit, followed by a controlled X gate on both qubits. %
At the end of the example circuit, the qubits are measured. 
Since measurement results are classical, they are shown as classical bits, i.e., as double wires in a quantum circuit.

For a more thorough introduction to quantum computing from the perspective of a computer scientist, the interested reader is referred to \cite{hidaryQuantumComputingApplied2021}, \cite{merminQuantumComputerScience2007}, and \cite{nielsenQuantumComputationQuantum2012}.

\subsubsection{Challenges within Quantum Computing}\label{sssec:challenges_qc}

The greatest challenge with current quantum hardware is \emph{fidelity}, a measure of how close the output of a quantum system is to the ideal output~\cite{Jozsa1994}.
Due to the sensitivity of quantum systems, they interact with fluctuations in the surrounding environment, also known as \emph{quantum decoherence,}
which introduces error over time. 
Several approaches have been proposed to tackle this issue: 
one approach is \emph{error mitigation}, which passively reduces the impact of errors on the final output, not modifying the circuit during execution~\cite{temme2017error,koczor2021exponential}.
Another approach is \emph{error correction}, which actively modifies the circuit during execution to correct errors~\cite{shor1995scheme, gottesman1997stabilizer, andersen_repeated_2020}.

There are no implementations that completely compensate errors yet, %
and thus, only \emph{\gls{NISQ}} devices are currently available.
Much research is being conducted on how to make use of the already available \gls{NISQ} devices. 
The currently most-known algorithms for this are noise-resistant variational quantum algorithms~\cite{lloyd2014quantum, Romero_2017, romero2021variational}. %
However, it is not certain whether this kind of algorithms is suitable for \gls{NISQ} devices because the errors accumulate in each iteration~\cite{gonzales-garcia2022error}.
Especially variational algorithms run in tight interaction with a classical program.
In \autoref{sec:qc_hpc_integration}, we go into how classical and quantum resources can work together on a structural level.

An additional crucial difference to classical computing is the fact that, according to quantum physics, arbitrary states of qubit cannot be copied. 
This is also known as \emph{no-cloning theorem}~\cite{Wootters1982}.
\Umbrellatermpl{} propose various solutions to enforce adherence to the no-cloning theorem: several functional \umbrellatermpl{}~\cite{green_quipper_2013,paykin_qwire_2017,sivarajah_tket_2020} provide a linear type system, which only allows overwriting the value of a variable, but not to manipulate it on a copy. 
Others prevent cloning by a specific syntax, for instance by only allowing directly assigning the computation result to the variable~\cite{selinger_qfc_2004}.

\subsubsection{Quantum Toolchain \& Workflow} \label{subsubsec:Quantum_Toolchain_Workflow}

Since quantum computers are highly complex and require a specialized infrastructure, specific expertise is necessary. 
As supercomputing centers or IT companies provide the right prerequisites for integration, most \gls{QC} systems reside there.
To make the quantum device accessible to the public, cloud access is often offered.
Through an interface, the user can submit a circuit and queue it for execution.
After execution, the measurement results are returned.

Before the execution, several hurdles must be overcome to properly run a circuit on a quantum system.
First, not every quantum gate can be run natively on a given platform. 
A quantum computer may only provide a certain set of operations, the so-called \emph{native gate set}. 
Thus, a given program must be expressed in the native gate set of the chosen quantum system. 
This conversion process is referred to as \textit{transpilation}.
Second, qubits are often set in a fixed two-dimensional or three-dimensional layout where 
multi-qubit gates can only be applied to adjacent qubits.
Logical qubits are first \emph{mapped} to their physical counterparts and then aligned by SWAP operations (exchanging the states of two qubits) to ensure that multi-qubit operations are applied on adjacent qubits in a process known as \emph{routing}~\cite{cowtan2019}.
Requiring programmers to manually perform these processes is infeasible, as both optimal transpilation and mapping are known to be NP-hard problems. %
Instead, there are automatic tools that transpile given code to match the constraints given by the target machine during compilation time.
Designing implementations or solutions improving these processes is an ongoing area of research~\cite{chhangteMappingQuantumCircuits2022}.

\section{Related Work}\label{sec:related_work}

Review papers of \umbrellatermpl{} go back almost as far as their development, with \cite{goos_brief_2004} (2004) being the earliest. 
The authors provide an overview of three virtual hardware models for \gls{QC}, namely the quantum circuit model, the QRAM model, and the quantum turing machine. 
The operational properties of these models were analyzed and compared to better understand their respective capabilities. 
Furthermore, the two emerging approaches to formulating quantum programs were compared, including imperative and functional languages. 
A similar comparison of these two paradigms has also been repeated in multiple reviews~\cite{simon_j_gay_quantum_2006, unruh_quantum_2006, jorrand_programmers_2007}. 
These studies have discussed the distinctive features of each approach and listed different programming languages that belong to them.
More recently, other programming paradigms were also included, such as process algebras used, for example by QPAlg~\cite{jorrand2003, lalire_qpalg_2004} and CQP~\cite{gay_cqp_2004}.

\Umbrellatermpl{} can be divided into different categories according to various criteria.
A categorization of \umbrellatermpl{} was proposed based on their practicability, consisting of three distinct groups~\cite{zorzi_quantum_2019}. 
The first group comprises practical-minded tools that are designed to generate quantum circuit descriptions. 
Notable examples of such tools include Forest~\cite{smith_practical_2017}, Qiskit~\cite{qiskit}, and (Py)Quil~\cite{smith_practical_2017} which we classify differently.
The second category represents tools with a limited (or still underdeveloped) theoretical account.
This group includes Quipper~\cite{green_quipper_2013} and the Quantum Development Kit (QDK) which contains Q\#~\cite{svore2018q}. 
The third category comprises the theoretic languages which focus on semantics and denotational models. 
Notable examples are quantum lambda calculi~\cite{maymin_extending_1997,clairambault_full_2019,van_tonder_lambda_2004,tonder_lambda_2003,selinger_lambda_2006}, QML~\cite{altenkirch_qml_2005}, the Quantum IO Monad~\cite{gay_quantum_2009}, Proto-Quipper~\cite{fu_linear_2020,fu_tutorial_2020,ross_algebraic_2017,rios_categorical_2018}, QWIRE~\cite{paykin_qwire_2017}, qPCF~\cite{paolini_qpcf_2017}, and IQu~\cite{paolini_quantum_2019}.
In addition to the practicability, \cite{garhwal_quantum_2021} separates \umbrellatermpl{} according to five categories, including multi-paradigm, imperative, functional, quantum circuit, and quantum object languages. 
Moreover, recent trends of \umbrellaterm{} development are analyzed in this work as well. 

Another focus is quantum language design. 
Various requirements have been stated in several reviews~\cite{rudiger_quantum_2006, miszczak_models_2011, kesha_hietala_quantum_2016} based on \cite{bettelli_q_2003}, namely completeness (every quantum program can be coded within the language), extensibility (provide high-level classical constructs), separability (separation of classical and quantum programming), expressivity (provide high-level quantum constructs), and hardware independence (the design of language is independent from any particular quantum architecture). 
The design of \umbrellatermpl{} should address the aforementioned requirements with the help of an effective compilation toolchain, appropriate execution model, and related concepts.

Conceptual methods, including computational models, execution models, and formal and operational semantics, which fall under the general topic of programming language theory, were covered in various reviews~\cite{sofge_survey_2008,miszczak_models_2011,Valiron_2012,ying_quantum_2012}. 
Even more abstract classical topics, like software requirement analysis~\cite{zhao_quantum_2021} or software testing~\cite{zhao_quantum_2021, fingerhuth_open_2018}, have already been reviewed in the quantum context.
Some reviews covered the quantum toolchain, which consists of the full stack from the problem definition to the quantum device or simulator~\cite{chong_programming_2017, fingerhuth_open_2018, gill_quantum_2021}.
Some \umbrellatermpl{} such as Qiskit~\cite{qiskit}, Cirq~\cite{cirq_developers_cirq_2022}, XACC~\cite{mccaskey_xacc_2020}, or ProjectQ~\cite{steiger_projectq_2018} already contain the full quantum toolchain~\cite{fingerhuth_open_2018}.

While these studies have contributed significantly to the understanding of the \umbrellatermpl{} themselves, to the best of our knowledge, no research has 
reviewed existing \umbrellatermpl{} in terms of their suitability for \gls{HPC} applications or hybrid \gls{HPCQC} systems.

\section{Quantum Computing and Classical Computing Integration}\label{sec:qc_hpc_integration}

Although \gls{QC} has certain advantages over classical computing in specific scenarios, it is limited by the number of qubits available on current quantum computers and the lack of efficient ways to store, manipulate, and fetch classical information on quantum systems.
Therefore, the general way of making use of \gls{QC} is to treat \glspl{QPU} as additional, complex accelerators that are requested for certain computing tasks~\cite{ruefenacht2022ea}. 
Many near-term viable algorithms interleave classical and quantum information processing.
The execution of these \emph{hybrid} programs requires smooth integration of classical and quantum resources. 
Towards the integration of \gls{QC} into \gls{HPC}, some initial work was done:
\cite{humbleQuantumComputers2021} proposes three setups (``macroarchitecures'') to integrate \gls{QC} into \gls{HPC}. 
The first step is achieved by connecting a \gls{QC} device remotely with \gls{HPC} resources.
Then, quantum and classical resources are co-located and connected with a common network.
The last proposed expansion stage connects several quantum platforms additionally via a quantum network to send quantum data from one to the other. 
For the second stage, they detail a ``microarchitecure'' of such a quantum-accelerated node. 
Furthermore, \cite{wederProvenancePreservingAnalysisRewrite2023} concentrates on the detection of parts in a hybrid program where classical and quantum computations are tightly interlaced. 
To speed up the execution they propose to deploy those parts as one unit to a hybrid runtime. 
Besides their useful concretizations of the integration procedure the authors did not evaluate \glspl{QPT} regarding their suitability for the use in an \gls{HPC} environment. 
This section details the additional considerations when coupling a \gls{QPU} and an \gls{HPC} system:
the integration of \gls{QC} into the \gls{HPC} system needs to be accomplished on hardware level (\autoref{subsubsec:Hardware_Level}), as well as software level (\autoref{subsubsec:Software_Level}).  

\subsection{Hardware Level} \label{subsubsec:Hardware_Level} 

The hardware level considers the physical deployment of quantum hardware and \gls{HPC} resources. It focuses on the architecture of quantum and \gls{HPC} hardware components and the interactions among them.
We have identified four different \textit{scenarios} of \gls{HPCQC} integration~\cite{humbleQuantumComputers2021, ruefenacht2022ea, bartsch_valeria_2021_5555960}:
\begin{inparaenum}
  \item Loose integration --- Standalone, 
    \item Loose integration --- Co-located,
    \item Tight integration --- Co-located, and
    \item Tight integration --- On-node.
\end{inparaenum}
They represent an increasing level of complexity in integration, but also a timeline.
As a result, demands on  \umbrellatermpl{} rise in each scenario.

\emph{Loose integration --- Standalone}.
A standalone \gls{QPU} is the first step in developing a new quantum device.
In a lab setting, classical performance is neglected in favor of controlling the device correctly.
Problems are optimized manually by the users beforehand and undergo a minimal compilation procedure.
Access is granted through a web interface, which simply transfers circuits and measurement results.
Integration is minimal at this point.

\emph{Loose integration --- Co-located}.
Coupling a single quantum device to an \gls{HPC} system is the first step for integration. 
The device is typically co-located \emph{in a networking sense} and the \gls{QPU} is primarily accessed through the \gls{HPC} system.
This requires more sophisticated tools for deploying and managing the system.
As a result, the control over the complete system advances steadily.
Some \glspl{QPU} require extensive setups.
In these cases, integrating over the cloud instead of co-locating might be necessary but is usually not preferred.

\emph{Tight integration --- Co-located}.
Multiple devices, potentially from various vendors, are co-located \emph{physically}.
This opens up new possibilities to utilize existing classical hardware and properties of the quantum devices.
In this scenario, efforts for integration as well as expectations on performance increase drastically.
Multiple devices with different levels of control and functionality have to be supported in addition to the existing challenges of classical \gls{HPC}.
Due to hardware limitations, tight integration might still be too restrictive.
A weakened version could split some portion of the \gls{HPC} resources to tightly integrate the \gls{QPU}, while the main computation is located elsewhere.
In this way, computing resources can be accumulated (\emph{``snowballed''}) over multiple facilities.

\emph{Tight integration --- On-node}.
The ultimate goal of \gls{HPCQC} integration poses the highest technical challenges.
On-node integration, similar to \glspl{GPU}, enables tackling of major problems.
At fine granularity, existing solutions can shift to the quantum realm and thereby improve performance.
\glspl{QPU} are almost indistinguishable and tightly interconnected with classical components.

\begin{figure}
    \centering
    \includegraphics[width=0.8\textwidth]{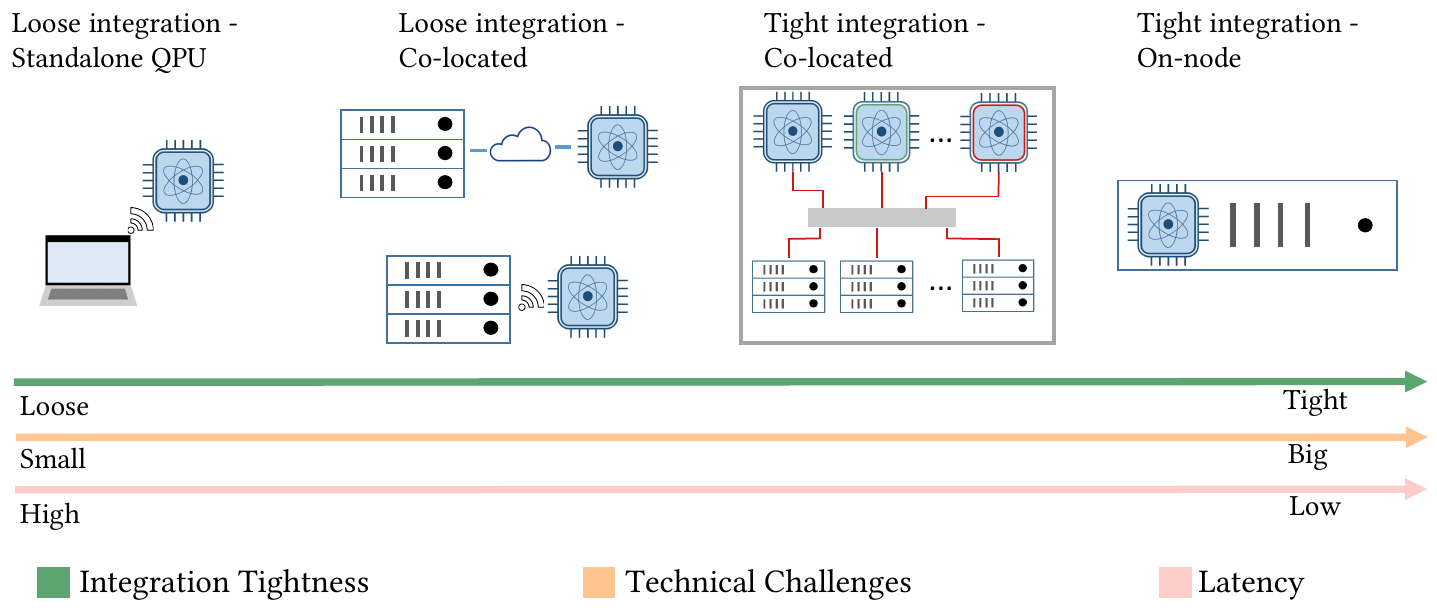}
    \caption{Trade-off between different \gls{QPU}-\gls{HPC} integration setups. From left to right, the demands of integration schemes change in different dimensions accordingly. 
    }
    \Description{The figure shows four integration schemes. From left to right those are:  Loose integration --- Standalone QPU,  Loose integration --- Co-located, Tight integration --- Co-located, and Tight integration --- on-node. 
    Scales referring to three dimensions indicate the increase in the integration tightness and technical challenges as well as the decrease in latency achieved by the tighter coupling.}
    \label{fig:IntegrationChallengesTightness}
    \vspace{-4mm}
\end{figure}

\autoref{fig:IntegrationChallengesTightness} shows the trade-off between different \gls{HPCQC} integration scenarios. The tighter the integration, the more efficient the interaction between \gls{HPC} nodes and \glspl{QPU}, but also the more challenging it is to realize and maintain the integrated system.

\subsection{Software Level} \label{subsubsec:Software_Level} 

\begin{figure}
    \centering
    \includegraphics[width=0.80\textwidth]{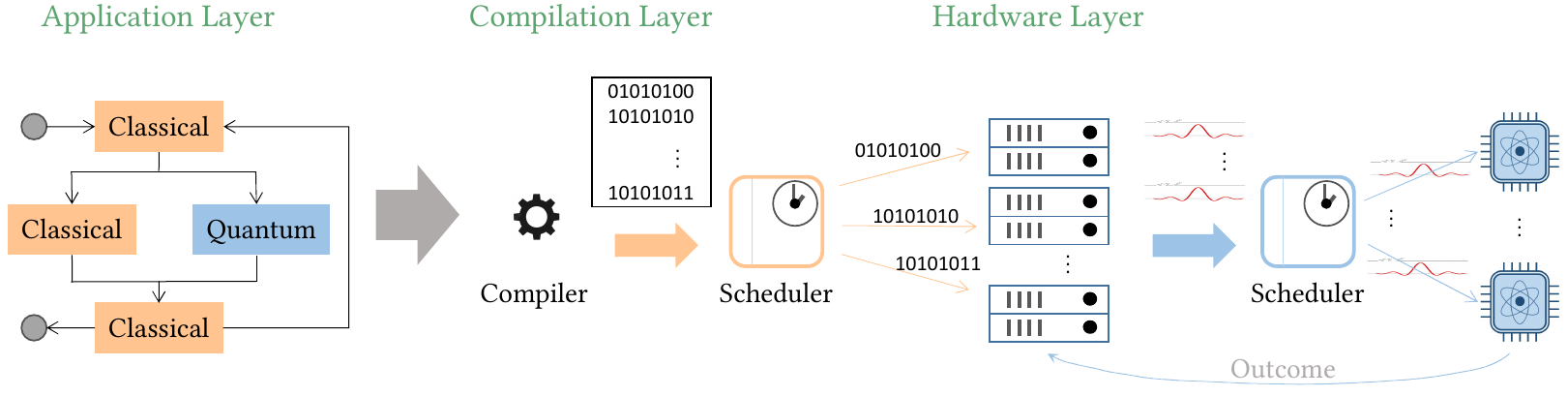}
    \caption{Software-level integration. Starting from the left, the first diagram illustrates integration in the application layer where algorithmic modules of classical computing and quantum computing are integrated by hybrid programming languages or workflow systems. The second diagram illustrates integration in the compilation layer where hybrid source code is compiled. 
    The final diagram illustrates integration in the hardware layer where \gls{HPC} binaries and \gls{QPU} pulse-level instructions are scheduled for execution on their target devices.}
    \Description{The figure describes the software-level quantum-classical integration. 
    In the application layer, it starts with a classical module, of which the output is split into a classical and a quantum module. The outputs of the two modules are then combined and fed to the classical module again. Then, a final output is returned or the output is used for another computation and the computation starts at the classical module again.
    During the compilation of the program, in the compilation layer, the quantum-classical source code is compiled by a compiler which supports quantum as well as classical features.
    In the hardware layer, the output of the compilation layer, i.e., the binaries for classical and pulses for quantum computations, are passed to the scheduler to determine the target devices for program execution. After the execution of the quantum part, the outcome is passed to the classical device to process the results.}
    \label{fig:SoftwareLevelIntegration}
    \vspace{-4mm}
\end{figure}

Software-level integration is important to the utilization of hybrid algorithms, as it facilitates the \emph{coordination} between \gls{HPC} and \gls{QC} resources.
More specifically, we define coordination as the allocation and utilization of computing resources (e.g., \gls{HPC} nodes and \glspl{QPU}) to efficiently complete jobs submitted by users while minimizing resource idle time.
Effective coordination can be achieved by understanding the underlying components or processes of the quantum software stack, from abstract application to compilation onto hardware, with an optimized scheduling strategy as the final goal.
As illustrated in \autoref{fig:SoftwareLevelIntegration}, software-level integration is realized at multiple layers of the quantum software stack. Given the current variability in quantum technologies (superconducting, ion-trapped, neutral atoms, etc.), all abstraction levels of the software stack need to be extensible and modular.
    
At the \emph{application layer}, the workflow is defined by a hybrid algorithm, specifying the interaction between classical and quantum systems to perform a task; at this level of abstraction, the quantum component is \umbrellaterm{}-agnostic~\cite{mccaskey_xacc_2020, humbleQuantumComputers2021}.
However, given the plethora of well-established classical computation tools on \gls{HPC} systems, most users will rely on existing technologies to perform their classical computation components. Hence, the \Umbrellatermpl{} used in the realization of quantum acceleration need to be compatible with these existing \gls{HPC} tools (compilers, libraries, parallel runtime environments).
This implies that the hybrid software tools provide system level language (e.g., C++) support with appropriate bindings to higher level languages (e.g., Python).

The \emph{compilation layer} connects the highly abstracted application layer and the underlying hardware. Considering the largely different quantum backends available, this layer must then be hardware-agnostic~\cite{mccaskey_xacc_2020}.
Building individual compilers for each specific quantum backend is not feasible, and a widely accepted alternative is a common \gls{IR} which facilitates the translation of domain-specific or standalone languages into hardware-specific instructions.
This provides an interface that frees programmers from rewriting boilerplate code for multiple backends and addresses common programming requirements, like scheduling, which is required for both \gls{HPC} and \gls{QC}.

However, designing and implementing such a standard is not trivial. Analogous to the way \glspl{ISA} operate in conventional computing, the foundational functions inherent to different technologies are essentially distinct. 
This means that toolchains must be able to account for hardware-specific optimizations, mappings, and routings when compiling quantum code.
Certainly, the current state-of-the-art requires scientists experimenting on quantum hardware to implement bespoke optimizations specific to their current test setup.
In addition, any generalized approach must be widely accepted to provide a stable \gls{API} for developers.
Currently, the ownership of this task is held almost solely by hardware or service providers who possess a monopoly on hardware-specific expertise required to implement such optimizations.
These responsibilities may shift when quantum technologies mature, but there is a high knowledge barrier to entry for the average programmer to perform such evaluations. More information is detailed in Section \ref{subsubsec:Cross_Platform_Execution}.

At the \emph{hardware layer}, schedulers allocate classical or quantum backends to run the compiled instructions from the previous layer and the hybrid nature of the overall task must be taken into account when assigning resources. 
Here, users can opt for existing \gls{HPC} programming models, such as the \gls{MPI} library, that support interfacing quantum and classical resources or rely on custom solutions if existing tools are not sufficient.

Finally, there are new challenges for \gls{HPCQC} system integration that affect multiple layers. 
Firstly, it is necessary that service providers accurately estimate the time and compute resource requirements of a quantum task or subroutine as these estimations do not only affect scheduling but also compilation.
This is especially the case when running hybrid algorithms as they often interleave quantum and classical computations.
However, accurately specifying the compute resource requirements of a given quantum task is still an open area of research.
Software-level integration must also consider the constraints imposed by the inherent properties of quantum systems. 
For example, it is not possible to partition an entangled state or to preempt and later restore the state of a quantum system, since there is no way to store and read out quantum states without destroying them.
The integration task is further complicated if error correction or error mitigation strategies are employed by the provider as the underlying implementation may not be transparent to users and care must be taken that any implemented strategies are fully verified and do not affect experiment results.

\section{Taxonomy}\label{sec:taxonomy} %

In this section, we introduce the criteria that allow us to characterize \umbrellatermpl{} from an \gls{HPC} perspective.
We distinguish the aforementioned four scenarios of quantum-classical integration, with increasing demands for our selected criteria. 
Moreover, we propose a weighting structure for these criteria that provide readers with insight about the possible benefits or drawbacks of adopting or using a particular \umbrellaterm{}.
We base our evaluation of \umbrellatermpl{} on these criteria, which follows in the next section.

\subsection{Criteria} \label{subsec:Criteria}

We propose a set of criteria consisting of three main categories inspired by a product life cycle.
The first category focuses on establishing a minimum viable product, encompassing foundational decisions crucial for any \umbrellaterm{}. These decisions are challenging to modify in the future and have far-reaching implications. Once a minimum viable product is developed, further enhancements are necessary to ensure competitiveness as the product evolves. The criteria falling under the second category significantly expand the capabilities and flexibility of the \umbrellaterm{}  after integration.
To achieve long-term sustainability and prevent regression, it becomes essential to adapt to changing requirements, such as new technologies or addressing legacy code.
Similar to classical \gls{HPC}, \umbrellatermpl{} should aim to establish standard, flexible solutions that can be extended for future use.
Criteria that cover these requirements fall into the third category.
The three categories and their respective criteria are the following:

\begin{enumerate}
    \item \Umbrellaterm{} Foundation, covered by Type, Host Language, and Execution Model
    \item \Umbrellaterm{} Maturity, covered by Toolchain Support and Scalability
    \item \Umbrellaterm{} Sustainability, covered by Cross-Technology Execution, Data Management, Resource Management, and Legacy Code Modernization
\end{enumerate}%

Each category is evaluated on an integer scale from $0$ to $10$, with the definitions for each rating defined in the respective sections.
Generally, a $0$ indicates no consideration of the criterion and a $10$ means full suitability in an \gls{HPC} environment or \gls{HPCQC} system.
We typically provide ranges which allow for some leeway.
A definite decision is dependent on the context in which the \umbrellaterm{} is evaluated, and we will provide useful guidelines how to make a decision.

\subsubsection{Type} \label{subsubsec:Type}

The first criterion is a categorical distinction between programming languages, libraries, and frameworks. This distinction is necessary to highlight implications on \emph{performance} and \emph{utilization}.
Programming languages follow a defined syntax and semantics, requiring a compiler or interpreter. \Glspl{DSL} (either embedded or external) are specialized languages designed for quantum operations only, such as QPCF~\cite{paolini_qpcf_2017}. All languages face challenges in integration, such as user effort, toolchain support, and performance. Extensions of existing languages, such as QCOR~\cite{mintz_qcor_2019}, can be used to build upon mature languages but still have limitations.

Libraries provide access to predefined functions and modules for expressing quantum programs, with an accessible and intuitive API. They are usually easier to use and provide faster prototyping, but may not support distributed computing. Frameworks, on the other hand, take over program execution and provide a structured option for customized control over quantum program execution. These frameworks can be beneficial for optimal control of quantum hardware.

\par We highlight the main challenges when utilizing a \umbrellaterm{} from either of the three categories. 
However, determining the suitability of a particular \umbrellaterm{} based on this categorization is highly subjective and is grounded in the needs of the user. Hence, we only provide a brief overview of the various challenges and do not assess this criterion.

\subsubsection{Host Language} \label{subsubsec:Host_Language}

Libraries and frameworks both require a \emph{host language} they are written or embedded in; the same holds for \gls{API} specifications like \textit{OpenMP}.
One example is Cirq~\cite{cirq_developers_cirq_2022}, where Python is the host language.
We additionally make the distinction between host languages and supported languages, which are made possible through language bindings. 
This is the case for Qiskit~\cite{qiskit}, a Python library with simulator backends that support C++ bindings.
In this paper, for (new) languages and \glspl{DSL}, the original or most influential language is identified as ``host language.''
For example, while IQu~\cite{paolini_quantum_2019} defines additional syntax, it adopts many features of Idealized Algol~\cite{Reynolds1997}; this determines its host language.

Given these definitions, there are two major considerations in this section: \emph{performance} and \emph{support}.
Our attention is directed towards the classical side language as we assume that the access to native quantum code is established.
Additionally, for hybrid programs, optimal performance on the \gls{HPC} side of the system is also crucial.
Performance is indicated by time to solution or performance benchmarking.
The lowest rated host languages in this category are those that are purely theoretical, such as the Lambda Calculus, the host language of the Quantum Lambda Calculus~\cite{maymin_extending_1997}. These require a complete implementation of a compiler or interpreter to function. 
Next are interpreted languages: purely measuring in execution time, memory, and energy, they are outperformed by compiled languages~\cite{pereira2021ranking}.
Some interpreted languages can support more performant compiled languages to remain competitive~\cite{qiskit}.
However, interpreted languages may require additional effort to run in an \gls{HPC} context and can face scalability issues which are detailed in \autoref{subsubsec:Scalability}.
Additionally, not all compiled languages are designed for performance, such as the Quantum IO Monad~\cite{gay_quantum_2009}, which is written in Haskell. 

Finally, we have \umbrellatermpl{} written in an \textit{\gls{HPC} language}, which we define as being well understood, performant, and widely used by the \gls{HPC} community. 
In addition to performance, these languages are \emph{well supported} in most \gls{HPC} infrastructures, which differentiates them from other performant, but less commonly used, compiled languages in an \gls{HPC} context.
This becomes relevant under the aspects of \emph{portability}, which is an active field of research. 
Examples of \gls{HPC} languages include C (despite manual memory management), C++, and FORTRAN.
We use the \emph{rating scheme} in \autoref{tab:concise_ratings_2-4} to rate the suitability of a \umbrellaterm{} based on its host language.

\subsubsection{Execution Model} \label{subsubsec:Execution_Model}
\begin{figure}[tb]
    \centering
    \begin{subfigure}{0.4\textwidth}
        \centering
        \includegraphics[scale=.65]{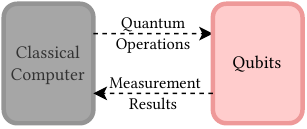}
        \caption{The \gls{QRAM} model.}
        \label{fig:qram}
    \end{subfigure}
    \hfill
    \begin{subfigure}{0.55\textwidth}
        \centering
        \includegraphics[scale=.65]{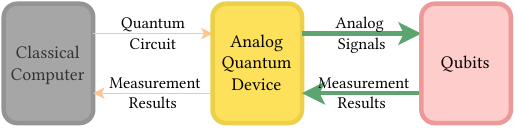}
        \caption{The restricted \gls{HQCC} model.}
        \label{fig:restricted_hqcc}
    \end{subfigure}
    \newline
    \begin{subfigure}{\textwidth}
        \centering
        \includegraphics[scale=.65]{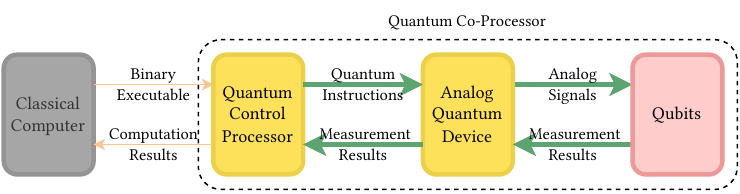}
        \caption{The refined \gls{HQCC} model.}
        \label{fig:refined_hqcc}
    \end{subfigure}
    \caption{Schematic representation of the three execution models referred to in this work. The arrows denote interaction between the two parties. Dashed arrows indicate interaction where hardware details are abstracted away. Thin arrows represent slow interaction and thick arrows fast interaction (based on Figure 2 in~\cite{fu2021quingo}).}
    \Description{A schematic representation of the three execution models referred to in this work. On the top left, the \gls{QRAM} model is shown: it consists of a classical computer and qubits. The classical computer sends quantum operations to the qubits to manipulate them, and measurement results are sent back from the qubits to the classical computer. 
    On the top right, the restricted \gls{HQCC} model is illustrated: it is extended with an analog quantum device between the classical computer and the qubits. The classical computer sends a quantum circuit to the analog quantum device, which then sends analog signals to the qubits to manipulate them. The measurement results are passed from the qubits to the classical computer via the analog quantum device. The interaction between classical computer and analog quantum device is slow (indicated by thin arrows) and the one between the analog quantum device and the qubits is fast (indicated by thick arrows). 
    On the bottom, the refined \gls{HQCC} model is depicted: it adds a quantum control processor between the classical computer and the analog quantum device and a quantum co-processor which comprises the quantum control processor, the analog quantum device, and the qubits. The classical computer sends a binary executable to the quantum processor, which then sends quantum instructions to the analog quantum device, which then sends analog signals to the qubits to manipulate them. The measurement results are passed from the qubits to the quantum control processor via the analog quantum device. The quantum control processor passes the computation results to the classical computer. The interaction from classical computer to quantum control processor is slow (indicated by thin arrows), the one within the quantum co-processor is fast (indicated by thick arrows.}
    \label{fig:execution_models}
    \vspace{-4mm}
\end{figure}

To execute quantum programs, instructions defining the quantum processes have to be sent from the classical device to the quantum hardware, which returns the computation result after the process is completed. 
Quantum execution models define this interaction between the quantum and classical hardware in a given quantum system.
An overview of the execution models considered in this work is illustrated in \autoref{fig:execution_models}. 
The quantum program starts within the classical computer and is executed on qubits.
After execution, results are sent back to the classical computer.
The rounded boxes denote the devices; the input and output are indicated by ingoing and outgoing arrows respectively.

The most studied execution model is the \emph{\gls{QRAM} model}~\cite{Knill_1996} introduced by Knill in 1996. 
It is similar to classical \gls{RAM}; the difference is that a restricted set of quantum operations, such as state preparation, unitary operations, and measurements, can manipulate a quantum state in the \gls{QRAM}. 
The structure of the \gls{QRAM} model is shown in \autoref{fig:qram}.
It assumes that the classical computer is able to control the quantum operations to be executed on the quantum computer in real-time.
However, due to the current limitations, implementing this execution model on a real device is impossible. 
This limitations arise from the nature of quantum states, which cannot be maintained indefinitely or at least until the classical computer initiates the next quantum operation or performs a qubit measurement.
As a result, the described integration scenarios do not accommodate this execution model, as near-term devices will not support such close coupling of classical and quantum computation. 
Nevertheless, there is a possibility of realizing this model in an on-node setting, thanks to the lower latency and greater control over the interaction between node and accelerator.

In reality, most quantum analog devices capable of executing quantum operations on real quantum hardware cannot execute classical operations, and the other way around as well~\cite{fu2021quingo}.
This means that an additional device for the execution of quantum operations is necessary.
Hence, another execution model, the \emph{\gls{HQCC} model}, seems to be more promising and viable~\cite{mintz_qcor_2019}. %
The \gls{HQCC} model can be separated in two categories~\cite{fu2021quingo}: the \emph{restricted HQCC model} and the \emph{refined \gls{HQCC} model}, illustrated in \autoref{fig:restricted_hqcc} and \autoref{fig:refined_hqcc}.

\Umbrellatermpl{} following the \emph{restricted \gls{HQCC} model} do not require real-time feedback from the classical computer during circuit execution time, which means that only offline quantum-classical interaction is possible. 
For instance, OpenQASM~\cite{cross_open_2017} and Cirq~\cite{cirq_developers_cirq_2022} implement the restricted \gls{HQCC} model.
In comparison to the \gls{QRAM} model, an analog quantum device, which controls the execution of quantum operations, is added.
The \emph{refined \gls{HQCC} model} adds a quantum control processor into a quantum co-processor, which enables controlling the application of quantum operations and the execution of a limited set of classical instructions. 
\Umbrellatermpl{} that implement the refined \gls{HQCC} model include Qiskit~\cite{qiskit}, XACC~\cite{mccaskey_xacc_2020}, (Py)Quil~\cite{smith_practical_2017} and Q\#~\cite{svore2018q}.
Most common is a \emph{total job} approach, in which classical code and quantum code are bundled and executed simultaneously on co-located hardware.
Instead of fixed quantum circuits, the classical computer sends binary executables to the quantum co-processor. 
The quantum control processor then sends quantum instructions to the analog quantum device that applies the operations on the qubits.
The measurement results of the computations are passed back to the quantum control processor via the analog quantum device, and classical registers may be updated.

The assumed \textit{Execution Model} poses different kinds of constraints on the \umbrellaterm{}.
Direct control, as required in the \gls{QRAM} model, has only been realized in standalone hardware in lab setups.
As a short-term solution, for a loose integration scenario, the quantum system can be treated as a black box, which moves the responsibility to the provider of the \gls{QPU}.
In return, the user loses out on optimizing the problem themselves.
For short-term applications, the \gls{HQCC} models are more realistic.
If the necessary tooling is provided, all models are possible, but the refined \gls{HQCC} model is the most suitable for tight integration scenario. The rating scheme to assess the suitability of each execution model for the \gls{HPCQC} integration is shown in \autoref{tab:concise_ratings_2-4}.

\subsubsection{Toolchain Support} \label{subsubsec:Toolchain_Support}

The compilation toolchain is a high-level abstraction of the process of compiling quantum programs into executable entities for heterogeneous hardware.
The toolchain hides details specific to actual compilation processes, such as the following: the choice of \glspl{IR} embedded in the compiler, optimization methods performed on the quantum circuits, or scheduling algorithms running on various types of quantum hardware platforms, etc.
The \emph{Toolchain Support} criterion focuses on the range of modules in the toolchain covered by a \umbrellaterm{}.

\begin{figure}[tb]
    \centering
    \includegraphics[width=0.75\textwidth]{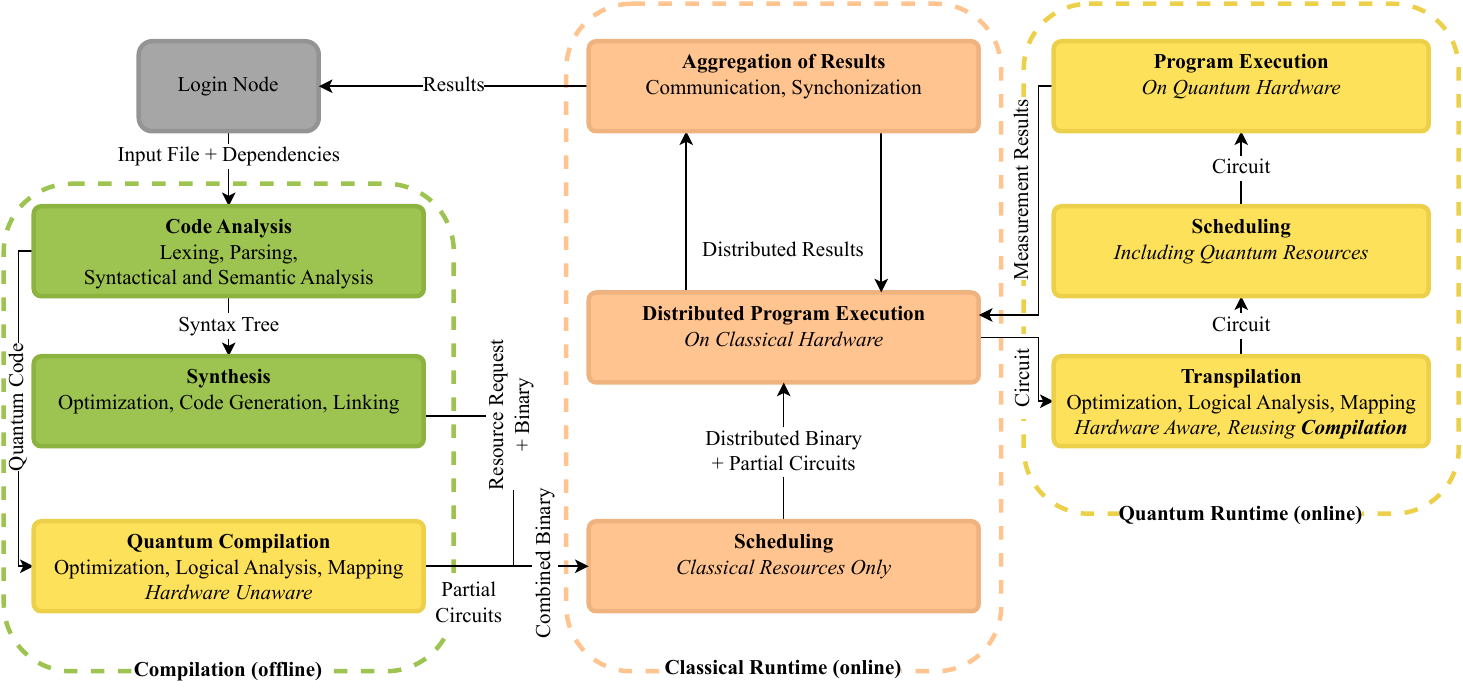}
    \caption{The hybrid toolchain in high abstraction.
    It enriches the original toolchain by a quantum compilation step and a quantum runtime (yellow). 
    The quantum compilation can pre-compile certain circuits as part of the binary and is then reused during run time.
    \gls{QPU} candidates can be selected based on these circuits and finalized after the full transpilation procedure.
    The Compilation Toolchain starts from the Login Node and follows the directions of arrows. Compared to \autoref{fig:Toolchain_Workflow_HPC}, it splits the runtime into a classical runtime and a quantum runtime.  
    }
    \Description{The figure describes the hybrid compilation toolchain. The desciption goes from left to right. The toolchain starts at the login node of the HPC system. The parser and lexer take the input files and dependencies as an input. They recognize the quantum kernels within the code and pass them to the quantum compilation toolchain for further processing. They also generate the corresponding syntax tree.  Then the code generation step takes the tree as an input and generates the binary executable files. The executable files and partially compiled quantum circuits are passed to the classical scheduling system, which allocates the compute resources during classical run time. During the execution, quantum transpilation is invoked to optimize the quantum circuits and map them to the available quantum hardware. Finally the quantum scheduler allocates the quantum resources and the results are passed back to the classical computing resources.}
    \label{fig:runtime_compilation_time}
    \vspace{-4mm}
\end{figure}

With the help of the toolchain, the comparison among \umbrellatermpl{} becomes more substantial. 
For instance, comparing the performance of two \umbrellatermpl{} is \emph{inconclusive} when they cover \emph{different} modules in the toolchain, or focus on \emph{different} stages in the compilation.
If one manages to identify some overlapping modules of both \umbrellatermpl{}, then comparing the performance of this module is more meaningful because the input, output, and the task of this module is clearly defined.
The hybrid toolchain, illustrated in \autoref{fig:runtime_compilation_time}, broadly consists of three parts: the left part corresponds to the \emph{compilation}, the middle part corresponds to the \emph{classical runtime}, and the right part corresponds to the \emph{quantum runtime}. 

During the \emph{compile time}, the source code is compiled into binaries from which quantum circuits are generated.
In the code analysis stage, user-written programs are going through lexical analysis, syntactical analysis, and semantic analysis, %
generating an \gls{IR} (e.g., syntax trees) along with the quantum part of the source code. 
The \gls{IR} is then fed to the synthesis stage, where optimization and code generation are performed. 
Afterwards, the linker combines all object and library files into one binary. The quantum part of the source code can be used for hardware-independent analysis and optimizations where supplementary information on circuits is produced.

At \emph{classical run time}, the binary combined with supplementary circuit information is scheduled to suitable \gls{HPC} nodes as distributed tasks. 
On \gls{HPC} nodes, quantum circuits are generated and their execution initiated. This marks the beginning of \emph{quantum run time}, during which the circuits are transpiled and executed.  %
The circuits undergo a series of transpilation steps, which include tasks like restricted architecture-independent optimization, error correction, error mitigation, transpilation to hardware-specific gate sets, qubit mapping, or qubit routing. %
The transpiled circuits are scheduled to a quantum hardware for execution. 
Measurement outcomes from the quantum hardware are handed back to the classical runtime and are synchronized, accumulated, and communicated to the users.
In classical \gls{HPC}, providing data to another device is usually called \emph{offloading}. 
At this time there is no uniform standard for offloading in the quantum realm, and it is not clear whether this responsibility lies in the \umbrellaterm{} or the hardware.

Hence, it is crucial to have a toolchain that maps high-level algorithms to executable instructions.
When surveying the current landscape, many \umbrellatermpl{} lack actual \textit{Toolchain Support}. 
Others provide only a limited implementation of the toolchain, which is not comprehensive enough for real applications. 
Hence, we adopt this criterion to judge the maturity of available \umbrellatermpl{} and to measure their ability to support the upcoming phase of hybrid quantum-classical applications.
A compiler or interpreter is the fundamental piece of any toolchain; therefore, this is the focus in our rating scheme for this category. %
We have to distinguish between the classical and the quantum domain, as they have different characteristics.
From an advanced compiler we expect \emph{good performance}, which depends on the given domain, typically speed in \gls{HPC} or gate depth for quantum circuits (longest path from input to output).
As an indicator for performance we recommend the tightness of integration with the \umbrellaterm{}, but other indicators are also conceivable.
Common classical aspects include resulting execution speed, compile time, memory and energy efficiency, and the level of optimization which is also crucial for \gls{QC}.
Additionally, \gls{QC} has to solve the additional challenges mentioned in \autoref{sssec:challenges_qc}.
Compilers should also maintain semantic correctness to some extent.
For additional tools, we judge their performance and how well they can be adapted to the users needs. 
In the rating scheme given in \autoref{tab:concise_ratings_2-4}, higher ranges build upon the previous ones.

\subsubsection{Software Scalability} \label{subsubsec:Scalability}

Traditionally, the scalability of quantum systems is hardware-focused; namely, how many qubits a quantum system can provide at sufficient readout fidelity.
However, in the context of this work, we focus on \emph{software scalability}.
Several aspects of quantum algorithm scaling need to be considered when integrating a quantum system with a \gls{HPC} framework with increasingly large problem sizes.
We judge how well a given \umbrellaterm{} handles quantum hardware with increasing qubit numbers and the associated consequences.

The first aspect is \emph{transpilation}; this converts a given algorithm to a series of instructions that can be executed on hardware with the highest fidelity.
This process can be further divided into multiple steps, which include circuit optimization~\cite{nam_automated_2018}, mapping and routing~\cite{cowtan2019, willeMappingQuantumCircuits2019}, and decomposition to the native gate set, each of which are hard problems in their respective communities.

While the main focus of these steps is performance (i.e., how much is circuit depth reduced or how is fidelity improved), the complexity of such algorithms must also be considered. 
Many existing algorithms, such as QAOA~\cite{farhi2014quantum}, quantum chemistry simulations~\cite{Knizia2012}, or QSVT~\cite{Dong2022}, require deep circuits when implemented on native gate-sets.
Scalability in terms of qubit number is also a concern, especially as the number of qubits on currently existing quantum hardware exceeds 400 qubits, with further plans to surpass 1,000 qubits this year and to exceed 10,000 qubits by 2026~\cite{ibm_roadmap} for superconducting devices.
In other technologies, exponential growth of all-to-all connectivity might become a bottleneck.
Hence, \umbrellatermpl{} must be able to support the projected problem sizes throughout the different stages of \gls{HPCQC} integration either by offering such services natively or incorporating external tools that do so.

\par
The second aspect is the \emph{communication} between control hardware and the underlying quantum machinery. 
The authors of the quantum \glspl{IR} eQASM and QuMIS~\cite{fu_eqasm_2019, Microarchitecture_Fu} have commented on the necessity of high information density in quantum \glspl{ISA}. 
This conclusion was reached after experimentation with QuMIS, in which the limited bandwidth of existing hardware required a denser encoding of the quantum \glspl{ISA} with an increasing number of qubits.

Finally, the \emph{scalability of the host language} of a given \umbrellaterm{} should also be considered. The language should run easily in an \gls{HPC} environment.
Besides the challenges for quantum toolchains, classical \gls{HPC} still remains complex.
Moving classical applications to the exa-scale regime requires scalable software on the classical side as well.
The most common tools are \textit{OpenMP} for on-node parallelism and \textit{MPI} for cross-node parallelism.
They offer programming interfaces in multiple \gls{HPC} languages and are the quasi-standard in \gls{HPC}.
\par
We note that the concept of software-side scalabilty for quantum computing systems is not a mature field of research as, at the time of writing, hardware-side scalability is generally accepted as the dominating bottleneck.
Therefore, we are unable to provide exact numbers or scaling complexities to be used as a benchmark. 
Instead, we request that users of the blueprint tailor the rating according to the use case at hand.
As a result, \autoref{tab:concise_ratings_2-4} has only five distinct categories; we have included a rating from 0 to 10 to convey the relation to the other criteria.

\begin{table}[bt]
    \vspace{-2mm}
    \tiny
    \centering
    \caption{Rating schemes for \emph{Host Language}, \emph{Execution Model},\emph{Toolchain Support} and \emph{Scalability} }
    \label{tab:concise_ratings_2-4}
    \begin{tabular}{llr}
    \toprule
        Range &  Implementation  & Rating\\
    \toprule
    \multicolumn{3}{c}{\emph{Host Language}} \\
    \midrule
        Purely theoretical language & Concept only & 0 \\
                                    & Basic syntax, Basic semantics & 1 \\
                                    & Advanced syntax, Advanced semantics & 2 \\ 
        \midrule
        Interpreted language without bindings & Low Expressiveness, Limited standard libraries, Low community and ecosystem support & 3 \\
                                              & High Expressiveness, Comprehensive standard libraries, High community and ecosystem support & 4 \\
        \midrule
        Interpreted language with bindings &  Low Expressiveness, Limited standard libraries, Low community and ecosystem support  & 5 \\
                                           &  High Expressiveness, Comprehensive standard libraries, High community and ecosystem support & 6 \\
        \midrule
        Compiled language & Functional, Minimal memory management, Low community and ecosystem support & 7 \\
                          & Imperative, Efficient memory management, High community and ecosystem support & 8 \\ 
        \midrule
        Compiled language, HPC language & Imperative, Efficient memory management, Performant, Limited legacy code support  &  9\\ 
                               & Imperative, Efficient memory management, Performant, Legacy code support   & 10\\
    \toprule
   \multicolumn{3}{c}{\emph{Execution Model}} \\
    \midrule
        No assumed execution model  & No consideration  & 0 \\
                                    & No consideration: Justified & 1 \\
                                    & No consideration: Implied &  2 \\ 
        \midrule
        QRAM: Abstract  & Assumption    & 3 \\
                          & Specification & 4 \\
        \midrule
        Restricted HQCC &  Interaction protocol: Not well defined & 5 \\
                        &  Interaction protocol: Well defined & 6 \\
        \midrule
        Refined HQCC & Interaction protocol: Not well defined  & 7  \\
                     & Interaction protocol: Well defined & 8 \\ 
        \midrule
        QRAM: Realized & Controlability: Qubit manipulation    &  9\\ 
                         & Controlability: Instruction streaming & 10\\
        \toprule
        \multicolumn{3}{c}{\emph{Toolchain Support}} \\
        \midrule
        Support level: No support  & Availability: Non existent & 0\\
                                     & Availability: Assumed/Mentioned  & 1\\
                                     & Availability: Classical only     & 2\\ 
        \midrule
        Support level: Compilation, Classical optimization & Loosely Integrated & 3\\
                                                              & Tightly Integrated & 4\\
        \midrule
        Support level: Compilation, Quantum optimization & Loosely Integrated & 5\\
                                                          & Tightly Integrated  & 6\\
        \midrule
        Support level: Additional tooling & Proprietary or fixed & 7\\
                                            & Customizable & 8\\ 
        \midrule
        Support level: Full toolchain, SDK & Proprietary or fixed &  9\\ 
                                               & Customizable & 10\\
        \toprule
         \multicolumn{3}{c}{\emph{Scalability}} \\
        \midrule
        No scaling considerations &  & 0 \\
        \midrule
        Supports insufficient qubits and gates & Transpilation of qubit- and gate-restricted use cases possible & 3 \\
        \midrule
        Supports sufficient qubits or gates & Transpilation of qubit- or gate-restricted use cases possible & 4 \\
        \midrule
        Supports sufficient qubits and gates & Transpilation of unrestricted use cases possible  & 7 \\
        \midrule
        Future proof            & Supported qubits and gates already exceeds projected future scalability requirements & 10 \\ 
        \bottomrule
    \end{tabular}
    \vspace{-4mm}
\end{table}

\subsubsection{Cross-Technology Execution} \label{subsubsec:Cross_Platform_Execution}

There is currently active research into what may be viable quantum computing technologies, with no implementation clearly superior to all others. 
Current state-of-the-art implementations include superconducting qubits~\cite{Bravyi2022, Krantz2019}, Rydberg neutral atoms~\cite{Cohen2021}, ion traps~\cite{Ladd2010}, diamond vacancies~\cite{diamond_qubit}, and photons~\cite{AspuruGuzik2012}.
These various hardware technologies (often called platforms or modalities) have different strengths and shortcomings which are relevant when selecting a platform for a given quantum algorithm.
One real-world example where such flexibility is required is Amazon Braket. It offers four different technologies to be used for \gls{QC}~\cite{AmazonBraKet}.
This then raises the question which stakeholder is responsible for choosing a specific \gls{QPU} (at compile or run time) and who is responsible for hardware-specific compilation.
During the development process, users are typically expected to select the \gls{QPU} themselves as well as to provide optimizations specific to the hardware.
However, it is also feasible that service providers take care of those hardware-specific optimizations (during circuit optimization, mapping, and routing).
Especially, if multiple backends are available, the optimal target technology might change during optimization.

Recent and commonly used \umbrellatermpl{}~\cite{gill_quantum_2021} generally support hardware agnostic gate-based \gls{QC}.
This abstracts away the hardware specifics required to compile to a chosen \gls{QPU} technology.
Based on this process, a backend- and hardware-aware toolchain is triggered. This part of the toolchain is responsible for generating executable QASM code corresponding to different specific platforms~\cite{khammassi_openql_2022}.

We also see that the variety of quantum platforms implies a split between hardware-aware and hardware-agnostic tools and significant support is required for smooth, heterogeneous operations. 
This adds a new requirement to the unified toolchain.

Heterogeneous systems allow the users to optimize for a specific technology.
For example, a chemistry simulation might be more suited for ion traps in one configuration and superconducting qubits in another.
This is not relevant for single \gls{QPU} systems, but it is highly useful for the co-located integration case and possibly mandatory for on-node integration applications which depend on multiple \gls{QPU} accelerators.
Quantum parallelism, in the form of running a program on multiple \glspl{QPU} simultaneously, is interesting once communication between devices becomes relevant for computation, but for now, we consider it in other parts of this review. %
We judge how easily programs can be deployed to heterogeneous quantum platforms and how this is supported by the toolchain.
The rating scheme is stated in \autoref{tab:concise_ratings_5-8}.

\subsubsection{Data Management} \label{subsubsec:Memory_Model_and_Data_Mapping}

As described in \autoref{subsec:HPC_World}, \gls{HPC} architectures support algorithms intractable on single machines by parallelizing and distributing the workload to multiple machines.
To realize this, communication standards, such as \gls{MPI}, have been developed to support a distributed memory model.
While there is active research into developing similar approaches for quantum applications~\cite{Gyongyosi2019,Cuomo2020,Caleffi2022}, this concept is not yet sufficiently mature to be integrated into any \umbrellaterm{} that we are aware of. 
Most \umbrellatermpl{} use a conventional data model, where shared registers between \gls{QPU} and \gls{CPU} are assumed.
This means that the classical CPU and the QPU communicate via shared memory buffers or registers. %
In this model, the CPU can load data into the shared register, the QPU can perform quantum operations on the data, and the results can be retrieved by the CPU from the same shared registers. Yet, it may also be limiting, as it imposes certain restrictions on the types of quantum circuits that can be implemented and the speed of communication between the classical and quantum domains. For instance, the shared register model may not be suitable for quantum circuits that require complex data structures (such as graphs or tensor networks) or non-local interactions between qubits.
To respect the \emph{no-cloning theorem}~\cite{Wootters1982}, some \umbrellatermpl{}  use additional write-only registers to prevent copying of quantum variables, for example LanQ~\cite{mlnarik2006introduction}.
Not covered is the possibility of shared \gls{QPU} memory, 
where various classical nodes can access the same QPU memory space to store and retrieve quantum data. Shared QPU memory can be realized using multiple hardware and software techniques, such as shared memory buffers, caches, or interconnects, depending on the QPU architecture and the communication protocols adopted.
They assume that data consistency and synchronization is either handled by the \gls{QPU} itself or the user.
Similar to developments in classical computing, this is not sufficient for large-scale applications in a post-\gls{NISQ} era.
This presents a possible gap that could be addressed by a well-designed quantum toolchain, especially if service providers possess multiple quantum accelerators.
As of now, shared memory is not the limiting factor for quantum-classical applications.

On a different note, data mapping and communication overhead are influenced by the adopted memory model.
A memory model specifies the mechanisms to manage memory, to exchange data, and to quantify how much responsibility is left to the programmer.
The availability of data at the correct node has a high impact on I/O operations.
A dedicated memory concept is one of the requirements for the \gls{HPCQC} tight integration scenario. %
This affects heterogeneous data, with different contexts and localities.
The runtime environment handles \textit{data mapping} for files which contain quantum kernels.
In the tight integration scenario, instruction streaming~\cite{nguyen2022extending}, which is supported in QCOR~\cite{mccaskey_extending_2021}, or more sophisticated models might be necessary to reduce memory overhead.
\autoref{tab:concise_ratings_5-8} contains the rating scheme.

\subsubsection{Resource Management} \label{subsubsec:Resource_Management}

Resource management is currently a complex problem in classical computing.
Adding quantum resources further increases the complexity. That is applicable on three levels. 
The first level concerns the availability of quantum-enabled nodes. This is subject to the runtime environment. The importance of scheduling in this context has been illustrated before~\cite{ravi2021adaptive}. 
The second one focuses on parallel execution of multiple quantum programs on a single machine. Presently, managing individual qubits might no be relevant, but it is possible for future applications. For this goal, safely freeing and allocating extra qubits is a desirable feature. 
The third level looks into the possibility of defining \textit{where} to run specific instructions when combining quantum and classical instructions.
Instructions can be executed on the controller of the \gls{QPU} or the \gls{HPC} node.
We refer to this feature as \textit{instruction pinning}.

\textit{Managing} the available resources on the hardware level and the software level is a desirable feature. 
Using quantum resources efficiently reduces cost and improves quality.
The downtime should be minimized and, in turn, utilization maximized.
Giving the user fine-grained control over classical and quantum resources is an indicator for viable tight integration scenario solutions. 
Still, the performance should not suffer if resources are not explicitly managed by the user. The rating scheme for this criterion is shown in \autoref{tab:concise_ratings_5-8}.

\subsubsection{Legacy Code Modernization} \label{subsubsec:Legacy_Code_Support}

In \gls{HPC}, most simulations depend on a legacy code base.
From this, additional challenges arise for \umbrellatermpl{}.
In the best case, users can extend existing programs with quantum functionality.
This is especially true for the area of simulation software.
Similar to \glspl{GPU}, \glspl{QPU} promise speedup compared to classical devices. 
To leverage this, \umbrellatermpl{} should follow proven practices and enable the users to harness the quantum advantage.
A proposed method ~\cite{wang2014exploitation} for automatically detecting and enhancing sequential code serves as an example of efforts to modernize legacy code through GPU integration.
Our main purpose for this criterion is the integration with legacy code and how easily a programmer can enrich existing code with quantum programs.

We already emphasized the benefit of using one of the traditional \gls{HPC} languages.
Providing support for \textit{Legacy Code} is closely connected.
The integration process for legacy code is tedious.
One intermediate requirement is feature stability and backwards compatibility.
While a \umbrellaterm{} undergoes regular changes, this cannot be realized.
Nonetheless, this can be a guiding principle for the long term; see \autoref{tab:concise_ratings_5-8} for the proposed rating scheme.

\begin{table}[tb]
    \vspace{-2mm}
    \tiny
    \centering
    \caption{Rating schemes for \emph{Cross-Technology Execution}, \emph{Data Management},\emph{Resource Management} and \emph{Legacy Code Modernization}}
    \label{tab:concise_ratings_5-8}
    \begin{tabular}{llr}
    \toprule
        Range &  Implementation  & Rating\\
    \toprule
        \multicolumn{3}{c}{\emph{Cross-Technology Execution}} \\
    \midrule
        Backend support: Only simulators   & No implementation & 0 \\
                                            & State vector based, Noise not considered  & 1 \\
                                            & Density matrix based, Noise considered  & 2 \\ 
    \midrule
        Backend support: Single QPU    &  Hardware technology coupled & 3  \\
                                        &  Sequential access (QPU can change in between) & 4  \\
    \midrule
        Backend support: Multiple QPUs, Single technology  &  Parallel access: Same quantum program & 5 \\
                                                             &  Parallel access: Different quantum programs  & 6 \\
    \midrule
        Backend support: Multiple QPUs, Multiple technologies &  Parallel access: Same quantum program & 7  \\
                                                                &  Parallel access: Different quantum programs  & 8  \\ 
    \midrule
        Backend support: Arbitrary QPUs & Arbitrary parallel access & 9  \\ 
                                         & QPU data parallelism  & 10 \\
    \toprule
        \multicolumn{3}{c}{\emph{Data Management}} \\
    \midrule
    Memory Model: Not considered & No consideration & 0  \\
                                   & Implicit: Identical classical and QPU memory & 1  \\
                                   & Implicit: Distinct classical and QPU memory  & 2  \\ 
    \midrule
    Memory Model: Shared Memory & Explicit: Identical classical and QPU memory & 3  \\
                                  &  Explicit: Distinct classical and QPU memory & 4  \\
    \midrule
    Memory Model: Distributed classical memory & QPU access: On-node only & 5  \\
                                                 & QPU access: Across-node  & 6  \\
    \midrule
    Memory Model: Distributed classical memory and data mapping & Data mapping: Classical only & 7  \\
                &  Data mapping: Access to across-node QPUs & 8  \\ 
    \midrule
    Memory Model: Distributed classical and quantum memory & Distributed: On-node QPU memory & 9  \\ 
                &  Distributed: Global QPU memory & 10 \\
    \toprule
       \multicolumn{3}{c}{\emph{Resource Management}} \\
    \midrule
    Resource accessibility level: No resource management  & None & 0  \\
                            & Implicit: Resources provided by system & 1  \\
                            & Implicit: Separate classical and quantum resources & 2  \\ 
    \midrule
    Resource accessibility level: On-Node     & Cores, Local QPU access  & 3  \\
                & Cores, Local Qubit allocation & 4  \\
    \midrule
    Resource accessibility level: Across-Node  & Nodes, Local QPU   & 5  \\
                & Nodes, Global QPU access  & 6  \\
    \midrule
    Resource accessibility level: Instruction Pinning, QPU & Bundled  & 7  \\
                &  Single instructions & 8  \\ 
    \midrule
    Resource accessibility level: Qubit Control   &  Global Qubit access & 9  \\ 
                    &  Global Qubit instruction pinning & 10 \\
    \toprule
        \multicolumn{3}{c}{\emph{Legacy Code Modernization}} \\
    \midrule
            Legacy code support: Not possible & No implementation & 0  \\
                                                & Language incompatibility  & 1  \\
                                                & Execution model incompatibility & 2 \\ 
    \midrule
            Legacy code support: Quantum routines in binaries  & Not optimized & 3 \\
                                                                 & Optimized & 4  \\
    \midrule
            Legacy code support: Quantum routines in other languages & Interpreted languages & 5  \\
                & Compiled languages & 6  \\
    \midrule
            Legacy code support: Quantum routines in the same language & Binding code required & 7  \\
                                                        &  No binding code required  & 8\\ 
    \midrule
            Legacy code support: Quantum routines & Offloading API & 9  \\ 
                                    & Automatic replacement & 10 \\
    \bottomrule
    \end{tabular}
    \vspace{-4mm}
\end{table}

\subsection{Scenario Thresholds} \label{subsec:Weighting}

In \autoref{tab:tag_weights}, we have summarized a recommendation for the thresholds in each scenario we introduced in \autoref{subsubsec:Hardware_Level}, based on our hands-on experience with the \gls{HPCQC} integration process.
The weights provide a set of requirements that we deem necessary for a smooth \gls{HPCQC} interaction that serves the technical demands within a given scenario.
First, we note that the thresholds are increasing for each scenario.
This is warranted, as our scenarios also represent a timeline of increasing integration.
Consequently, requirements transfer from one scenario to the next, typically including additional aspects.

\textit{Type} and \textit{Host language} behave similarly.
They dictate the ease of use during the early stages of the technology.
In later scenarios, performance is more critical, where one might want to switch to a classical \gls{HPC} language compatible with the existing infrastructure.
This is also desirable to enable extensibility for \textit{legacy} projects, because they are more prominent in the exa-scale scenario.
The \textit{execution model} expands on this, but has a more crucial role.
The restricted HQCC model is already insufficient for small-scale operations due to its overhead.
With increasing problem size, the supporting infrastructure has to become more sophisticated.
This affects the complete \textit{toolchain} and \textit{scalability}, where tool support is preferred for the initial coupling.
Multiple \textit{technologies} become relevant once single \gls{QPU} boundaries are crossed, which we deem necessary for small-scale \gls{HPC} integration.
The need for \textit{resource management, memory model, and data mapping} rises steadily, similar to the classical domain. 
\begin{table}[tb]
    \tiny
    \newcolumntype{P}[1]{>{\centering\arraybackslash}p{#1}}
    \caption{Minimum requirement of each criteria in different integration phases. Weight range is [0,10].}
    \label{tab:tag_weights}
    \begin{tabular}{p{3.0cm}P{2.0cm}P{1.8cm}P{2.7cm}P{2.5cm}}\toprule
        \textbf{Criterion} & \textbf{Loose integration - Standalone QPU} & \textbf{Loose integration - Co-located} & \textbf{Tight integration - Co-located} & \textbf{Tight integration - On-node} \\ \midrule
        Host Language & 1 & 2 & 4 & 7\\
        Execution Model & 1 & 4 & 8 & 10\\
        Toolchain Support & 3 & 5 & 7 & 10 \\
        Software Scalability & 3 & 4 & 7 & 10 \\
        Cross-Technology Execution & 0 & 1 & 6 & 8 \\
        Data Management & 1 & 3 & 6 & 9 \\
        Resource Management & 1 & 3 & 6 & 9 \\
        Legacy Code Support & 0 & 2 & 4 & 7 \\
    \bottomrule
    \end{tabular}
    \vspace{-4mm}
\end{table}

\section{QPT Analysis}\label{sec:comparative_analysis} %
In this section, we provide an overview on existing \umbrellatermpl{}.
We collected a multitude of scientific publications and open source projects relevant to this topic, examining their methodologies, results, and conclusions. We also studied their innovative approaches and solutions to the \gls{HPCQC} integration problem. 
For this we developed an \textit{analysis blueprint} that can be applied to evaluate existing \umbrellatermpl{}.
\autoref{tab:qpts} gives an overview of \umbrellatermpl{} with compilation toolchain support, whereas the more theoretical \umbrellatermpl{} are not included. %

A large fraction of the \umbrellatermpl{} were not incorporated into this segment due to various considerations.
Several of these \umbrellatermpl{} are merely theoretical concepts, others are still in a developmental stage, or they fail to satisfy the requisite criteria upon initial inspection.
From $44$ collected \umbrellatermpl{}, we select six diverse candidates to demonstrate the application of our proposed methodology.
Additionally, exclusion of a specific \umbrellaterm{} might be attributed to its similarity with one of the shortlisted notable candidates, whereby the selected candidate showcases marginally enhanced suitability or significance.

\subsection{Analysis Blueprint}\label{ssec:blueprint}
\begin{figure}[t]
    \centering
    \includegraphics[width=\textwidth]{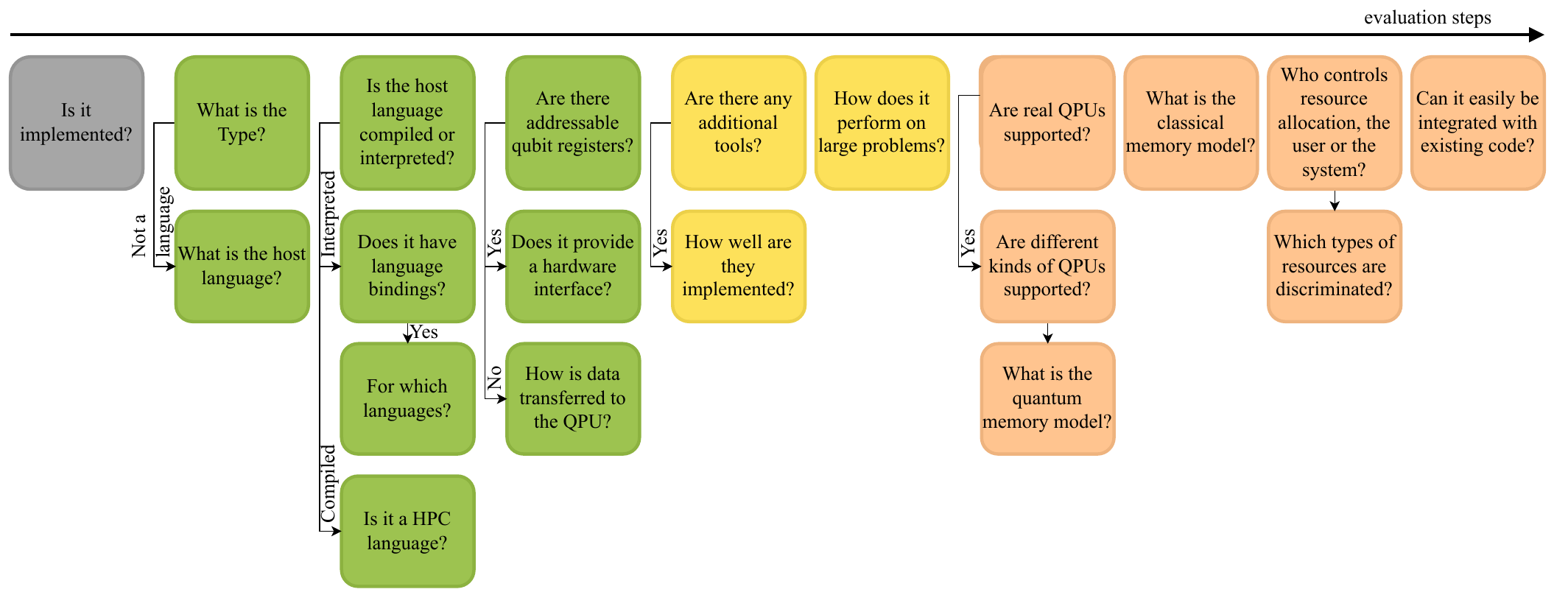}
    \caption{Flowchart of questions for the analysis blueprint. Answering the questions serves as a guidance to judge the tool based on our criteria. The three main categories from \autoref{subsec:Criteria} are color-coded and each column represents one criterion. Unlabeled arrows indicate follow-up questions and labeled arrows conditions. One should go from left to right, at each column of questions pick and record the one led to by the answer to the previous question or the only one the column contains, and eventually a list of questions is produced upon reaching the rightmost endpoint of the horizontal axis, each from one of the column.}
    \Description{This figure introduces a flow chart that represents the analysis blue print. The analysis blue print asks the readers questions to enable them to judge the quantum programming tool (QPT) at hand. The questions goes as follows in order: 
    1- Is the QPT implemented or not?
    2- What is the type of the QPT?
        2.1- In case it is not a standalone language, what is the host language?
    3- Is the host language compiled or interpreted?
        3.1- In case of interpreted host language, does it support bindings or not?
        3.2- in case of compiled language, is it an HPC language?
    4- Are there addressable qubit registers?
        4.1- In case of yes, Does it provide a hardware interface?
        4.2- In case of no, How is data transferred to the QPU?
    5- Are there any additional tools?
        5.1- How well are they implemented?
    6- How does it perform on large problems?
    7- Are real QPUs supported?
        7.1- In case of yes, Are different kinds of QPUs supported?
            7.1.1- What is the quantum memory model?
    8- What is the classical memory model?
    9- Who controls resource allocation, the user of the system?
        9.1- Which types of resources are discriminated?
    10- Can it easily be integrated with existing code?}
    \label{fig:analysis_flowchart}
    \vspace{-4mm}
\end{figure}

The analysis of \glspl{QPT} is largely based on the associated publications, but also examines supplementary material, such as documentation or implementation, if available.
In the previous section, we have established our criteria and their respective rating schemes.
We use them with a simple flowchart as the basis for our \textit{analysis blueprint}, wherein we systematically evaluate a tool by %
assessing criteria with respect to the \umbrellaterm{}.
The objective is to streamline the process of evaluating the compatibility of a \gls{QPT} for \gls{HPCQC} integration for any potential user.
The complete procedure is depicted in \autoref{fig:analysis_flowchart}, it is meant to guide users through the evaluation process.
To help classify certain properties like \textit{Host Language} or \textit{Execution Model}, we included follow-up questions and their implication.
A sample evaluation is given in the supplementary material. %
The goal of this blueprint is to provide means to evaluate the suitability of a \umbrellaterm{} for \gls{HPCQC} integration following abstract and quantitative criteria. 

\subsection{Emerging Insights}

We have conducted a comprehensive survey to find promising candidates for \gls{HPCQC} integration. 
Upon analysis of all \umbrellatermpl{}, we identified four main overlapping categories of \umbrellatermpl{}.
Each category targets a specific challenge, while compromising other areas.
In an ideal \umbrellaterm{}, all of the properties would be combined. 
We roughly summarize our findings for these categories in this section.
It is worth mentioning that most of the \umbrellatermpl{} provide different flavored solutions to the challenges. 
Typically \gls{HPC} is not covered by these challenges which is not ideal for \gls{HPCQC} integration.

\subsubsection{Correctness} 

\Umbrellatermpl{} within the \emph{correctness} category aim to capture general quantum operations in a verifiable, correct manner.
It consists mainly of functional languages like Qunity~\cite{voichick_qunity_2022}.
Quantum operations are expressed in a formal language, usually above circuit level, and combined with classical logic.
The \umbrellatermpl{} offer a rich set of expressive features and employ elegant language constructs to model quantum properties.
Through formal analysis, they ensure correctness or the scope of the computation in a theoretical sense.
However, most lack an implementation beyond operational semantics and usually perceive the \gls{QPU} as a black box system.
As a result, quantum-classical interfaces are left undefined, which renders this approach impractical for \gls{HPCQC} integration. 

\subsubsection{Generation}

The majority of libraries, for example Cirq~\cite{cirq_developers_cirq_2022}, belong to the category \emph{generation}.
These \umbrellatermpl{} simplify development of quantum algorithms.
Generating quantum algorithms with libraries, primarily by the means of circuits, allows using existing classical infrastructure. 
Additionally, the complexity of quantum physics is moved to an existing environment, lowering the barriers to entry.
Using established host languages addresses a large user base and enables fast prototyping.
At the same time, this limits choices when developing new tools as they need to be compatible with existing tools.
Performance and scalability are typically secondary or tertiary targets only.

\subsubsection{Toolchain}

Contrary to the previous category, certain \umbrellatermpl{}, for instance t|ket⟩~\cite{sivarajah_tket_2020}, focus on providing powerful tools for circuit compilation.
Usually, they outperform other \umbrellatermpl{} (along with their respective tools) significantly in terms of relevant metrics such as gate depth for circuit optimization.
The language part of the \umbrellaterm{} provides the entry point and is tailored to enable specific compilation techniques.
While integration interfaces are well defined, the adoption is limited due to the tools' specific nature.
Combining tools from multiple \umbrellatermpl{} is also nearly impossible or is connected with significant overhead.
In our vision, specialized tools should seamlessly interact to enable a flexible unified toolchain.

\subsubsection{Execution}

The last category concentrates on the \emph{execution} of hybrid problems. 
One key component is explicit addressing of (physical) qubits.
Frameworks like Quingo~\cite{fu2021quingo} offer full access to the hardware capabilities in a generic manner.
As hardware interfaces rapidly evolve, execution-focused \umbrellatermpl{} provide a high-level abstraction while retaining some low-level details.
In our opinion, this is the most promising approach, especially when incorporating aspects of the previously mentioned categories.
Currently, the limiting factor is access to quantum hardware and vendor-restricted interfaces, which prevents technology-specific operations.
Upon overcoming this hurdle, the emphasis should shift towards integrating with existing \gls{HPC}  processes.
During the maturing process, it is crucial to ensure that control over \glspl{QPU} does not adversely affect established classical solutions.

\begin{table}[t]
    \tiny
    \newcolumntype{P}[1]{>{\raggedright\arraybackslash}p{#1}}
    \begin{minipage}{\linewidth} %
    \caption{Overview of \umbrellatermpl{} with toolchain support.}
    \label{tab:qpts}
    \centering
    \begin{tabular}{cP{2.9cm}P{1.6cm}P{2cm}P{1.9cm}P{1.5cm}}
    \toprule
        Year & Name & Type & Host Language & Paradigm & Execution Model\\
    \midrule
        1998 & QCL~\cite{bernhard_omer_qcl_1998,bernhard_omer_qcl_2000,bernhard_omer_qcl_2003} & Language & based on C/C++ & Imperative & QRAM \\
        2004 & QFC/QPL~\cite{selinger_qfc_2004} & Language &  & Functional & QRAM \\
        2005 & cQPL~\cite{mauerer_cqpl_2005} & Extension & QPL & Functional & QRAM \\
        2005 & QML~\cite{altenkirch_qml_2005} & Language &  & Functional &  \\
        2006 & LanQ~\cite{mlnarik2006introduction} & Language & C & Imperative &  \\
        2008 & NDQFP~\cite{xu_ndqjava_2008} & Language & FP, compiles to C++ & Functional & QRAM \\
        2008 & NDQJava~\cite{xu_ndqjava_2008} & Language & Extends Java & Imperative & QRAM \\
        2009 & Cove~\cite{purkeypile_cove_2009} & Library & C\# & Imperative & Restricted \\
        2009 & Quantum IO Monad~\cite{gay_quantum_2009} & Library & Haskell/Agda & Functional &  \\
        2012 & Scaffold~\cite{abhari_scaffold_2012,javadiabhari_scaffcc_2015} & Language & C/C++ & Imperative & QRAM \\
        2013 & QuaFL~\cite{lapets_quafl_2013} & Language &  & Functional & Restricted \\
        2017 & Quil~\cite{smith_practical_2017} & Library & Python & Imperative & Refined \\
        2017 & OpenQASM~\cite{cross_open_2017} & Language &  & Imperative & Refined \\
        2017 & Qiskit~\cite{qiskit} & Library & Python & Imperative & Refined \\
        2017 & QWIRE~\cite{paykin_qwire_2017} & Language & Any, Coq & Circuit & QRAM \\
        2018 & Cirq~\cite{cirq_developers_cirq_2022} & Library & Python & Imperative\footnote{with functional elements\label{fn:gen_1}} & Restricted \\
        2018 & cQASM~\cite{khammassi_cqasm_2018} & Language &  & Imperative & Refined \\
        2018 & ProjectQ~\cite{steiger_projectq_2018} & Extension & Python & Imperative\footref{fn:gen_1} & Refined \\
        2018 & Q\#~\cite{svore2018q} & Language & C\#, F\# & Imperative & Refined \\
        2018 & QuantumOptics.jl~\cite{kramer_quantumopticsjl_2018} & Framework & Julia & Imperative &  \\
        2019 & Blackbird~\cite{killoran_strawberry_2019} & Language & Python & Imperative\footref{fn:gen_1} &  \\
        2019 & ReQWIRE~\cite{rand_reqwire_2019} & Extension & QWIRE & Circuit & QRAM \\
        2019 & Strawberry Fields~\cite{killoran_strawberry_2019} & Library & Python, Blackbird & Imperative\footref{fn:gen_1} &  \\
        2020 & QCOR~\cite{mintz_qcor_2019} & Extension & C++ & Imperative & Refined \\
        2020 & staq~\cite{amy_staq_2020} & Toolkit & C++ &  &  \\
        2020 & t|ket⟩~\cite{sivarajah_tket_2020} & Framework & C++ &  & Restricted \\
        2020 & XACC~\cite{mccaskey_xacc_2020} & Framework & C++ & Imperative & Refined \\
        2021 & OpenQL~\cite{khammassi_openql_2022} & Framework & C++/Python & Imperative & Refined \\
        2021 & Quingo~\cite{fu2021quingo} & Framework \& Language & Python and can be coupled with any classical language & Imperative & Refined \\
        2022 & Twist~\cite{yuan_twist_2022} & Language &  & Functional & Simulator \\
        2022 & OpenQASM 3~\cite{cross_openqasm_2022} & Language &  & Imperative & Refined \\
    \bottomrule
    \end{tabular}
    \end{minipage}
    \vspace{-4mm}
\end{table}

\begin{figure}[t]
    \centering
    \includegraphics[width=0.8\textwidth]{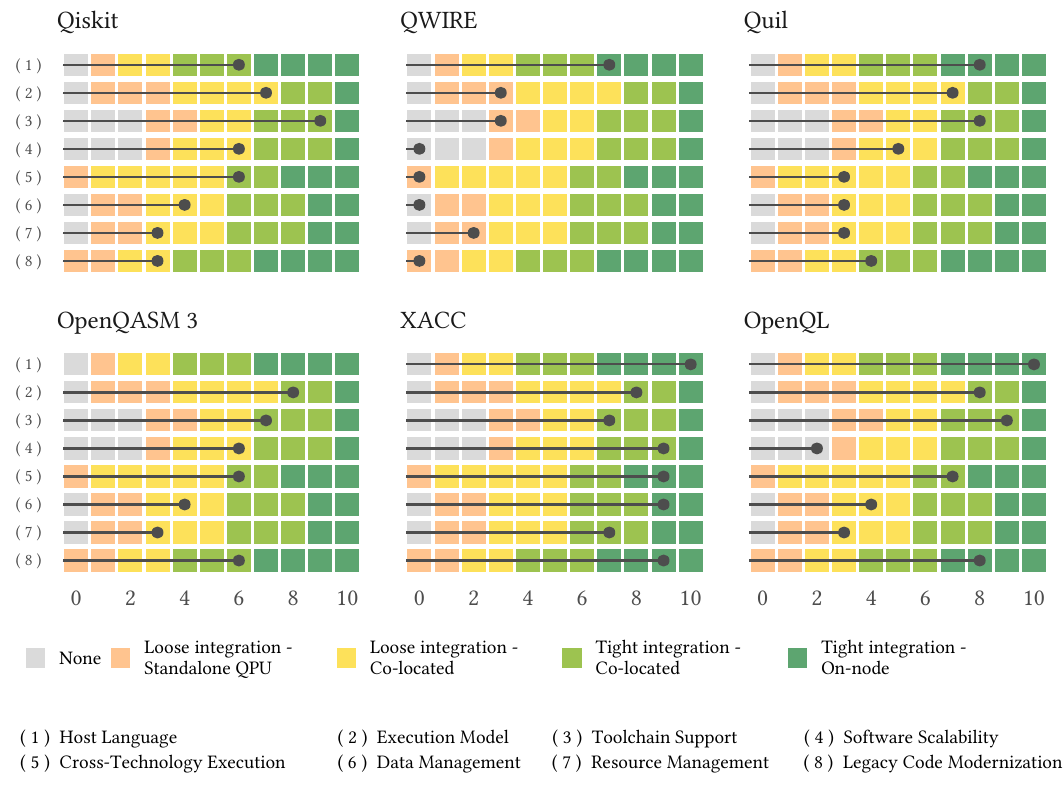}
    \caption{The rating of noteworthy \umbrellatermpl{} according to their potential in various \gls{HPCQC} integration scenarios.}
    \Description{This figure shows how the noteworthy candidates (i.e. OpenQL, QASM 3, Qiskit, Quil, Quingo, Qunity, QWIRE, and XACC) perform according the chosen criteria (i.e. Host Language,  Execution Model,  Toolchain Support,  Scalability,  Cross-Technology Execution, Memory Model and Data Management,  Resource Management and  Legacy Code Support). The ratings are as follows:
    OpenQL: 
    Host language: 10
    Execution Model: 8
    Toolchain Support: 9
    Scalability: 3
    Cross-Technology Execution: 7
    Data Mangement: 4
    Resource Management: 3
    Legacy Code Support: 8
    QASM 3: 
    Host language: None
    Execution Model: 8
    Toolchain Support: None
    Scalability: None
    Cross-Technology Execution: 5
    Data Mangement: 2
    Resource Management: 2
    Legacy Code Support: 7
    Qiskit: 
    Host language: 6
    Execution Model: 7
    Toolchain Support: 9
    Scalability: 6
    Cross-Technology Execution: 4
    Data Mangement: 4
    Resource Management: 3
    Legacy Code Support: 3
    Quil: 
    Host language: 8
    Execution Model: 7
    Toolchain Support: 8
    Scalability: 5
    Cross-Technology Execution: 3
    Data Mangement: 3
    Resource Management: 3
    Legacy Code Support: 4
    Quingo: 
    Host language: 4
    Execution Model: 7
    Toolchain Support: 4
    Scalability: 3
    Cross-Technology Execution: 2
    Data Mangement: 4
    Resource Management: 4
    Legacy Code Support: 5
    Qunity: 
    Host language: 1
    Execution Model: 3
    Toolchain Support: 5
    Scalability: 0
    Cross-Technology Execution: 0
    Data Mangement: 0
    Resource Management: 0
    Legacy Code Support: 0
    QWIRE: 
    Host language: 7
    Execution Model: 3
    Toolchain Support: 3
    Scalability: 0
    Cross-Technology Execution: 0
    Data Mangement: 0
    Resource Management: 2
    Legacy Code Support: 0
    XACC: 
    Host language: 10
    Execution Model: 8
    Toolchain Support: 7
    Scalability: 9
    Cross-Technology Execution: 9
    Data Mangement: 9
    Resource Management: 7
    Legacy Code Support: 9
    }
    \label{fig:rate_qpts}
    \vspace{-4mm}
\end{figure}

\subsection{Noteworthy Candidates} \label{ssec:candidates}

In this section, we want to discuss selected \umbrellatermpl{} in more detail.
To give as much information as possible, we use the analysis blueprint introduced in \autoref{ssec:blueprint} for evaluation.
Our selection fulfills two purposes. Firstly, we try to cover a diverse lineup with respect to our criteria. This allows us to fully demonstrate the blueprint and displays the multi-faceted nature of designing a \umbrellaterm{}. Secondly, the noteworthy candidates provide unique solutions to one or more challenges under the integration aspect. We believe that a viable long-term solution should include most of these approaches as features. 
The result of the analysis study is summarized in \autoref{fig:rate_qpts} %
that illustrates the ratings given for each \umbrellaterm{}. 
Moreover, it shows for which \gls{HPCQC} integration scenario(s) a \umbrellaterm{} could be used. 
The motivation for the ratings is explained in the following sections.   
The overall rating of a \umbrellaterm{} is given in the corresponding subsection in the following form: [host language, execution model, toolchain support, software scalability, cross-technology execution, data management, resource management, legacy code modernization].
At the time of writing, new proposals have been put forward by NVIDIA (\emph{CUDA-Quantum}~\cite{cuda_quantum}) and PennyLane (\emph{Catalyst}~\cite{pennylane_catlyst}).
Both seem promising as practical \umbrellatermpl{}, but are too recent to be considered here.

\subsubsection{Qiskit [6,7,9,6,6,4,3,3]}
Qiskit~\cite{qiskit} is a Python development kit that includes a software library and additional tools which grant access to IBM's quantum hardware.
Additionally, it also includes a simulator written in C++.
The main use case is to construct circuits and submit them as a job in an accessible way which is part of the core library.
In the original proposal, Qiskit implemented a restricted \gls{HQCC} model, but recently it has also followed the refined \gls{HQCC} model.
For the user, this changes the interaction with the quantum system, as the refined \gls{HQCC} model is coupled to a different programming model.

Qiskit includes a full toolchain as part of its core module.
The transpilation procedure is comprised of so-called \textit{passes}, which work on a specific level of circuit refinement.
It does not use any specific \glspl{IR} and all transformations are performed directly on the circuit represented as acyclic graph.
The transformations, including mapping, routing, and synthesis, are implemented in Python.
From a performance standpoint, this is not sufficient.
Compared to other tools, scalability is a big issue because the transpiler is usually outperformed by other tools~\cite{sivarajah_tket_2020}.
To combat this, tools are increasingly ported to more suitable languages and included with Python bindings.

This procedure makes it harder to directly add Qiskit into existing legacy applications, but the quantum hardware is also accessible without using the library:
one can submit jobs by calling an \gls{API} with a (compiled) circuit as input, and then they are scheduled internally; the choice of hardware is left to the user.
The hardware specification is made available, so the responsibility to optimize more than what is available by default lies with the user.
Recently, the so-called \emph{Qiskit Runtime} was included. %
Instead of making slow calls to a remote accelerator, one can upload a fully specified hybrid program; the execution follows the refined \gls{HQCC} model.
The inner workings are not open source, so we cannot judge them properly.
We assume that resource management, memory model, and data mapping are cared for, but not in which quality.

Qiskit showcases the significance of having a fully defined toolchain which warrants the score of 9 points.
It is by far the most popular tool used in \gls{QC} and has been implemented in different flavors \cite{cirq_developers_cirq_2022, AmazonBraKet}.
This is due to the fact that one can write and evaluate quantum programs easily. 
Furthermore, it has an exceptional usability and provides a wide range of functions.
Layers of abstraction from high-level quantum algorithms to low-level pulses are possible.
In the short term, where performance is not critical, Qiskit is a prime example of a \umbrellaterm{}.
Scaling and future-proofing are not sufficient yet and depend on further development.

\subsubsection{QWIRE [7,3,3,0,0,0,2,0]}
In addition to many approaches that mainly follow an imperative paradigm, researchers also proposed functional languages~\cite{green_quipper_2013,yuan_twist_2022,paykin_qwire_2017,rand_reqwire_2019}.
Indeed, some aspects of functional languages are a perfect fit for quantum programs: in the purely functional perspective, a function receives an input and computes the output only based on the given input.
It neither uses any environment variables nor modifies them (side effects).
Quantum gates resemble this behavior: they manipulate a quantum state and produce identical outcome for the same input.

QWIRE~\cite{paykin_qwire_2017} is designed as a language that can be embedded in any host language.
The only thing the host language must provide is a static type system and support for the types \lstinline|unit|, \lstinline|bool|, \lstinline|tuples|, and a pattern type \lstinline|Circ|.
Circuits constructed in the embedded language are boxed and become patterns in the host language.
For QWIRE, an implementation exists in coq~\cite{coq_2022}. 
Qubits are treated as linear resources, i.\,e., they can only be used exactly once.
This is a typical choice in functional languages to enforce the no-cloning theorem.

QWIRE provides a clear separation of classical data in the host language and quantum data in the embedded language.
However, the required boxing of circuits for the host and unboxing of circuits to be modified in the embedded language introduces some overhead.
Nevertheless, this approach circumvents the problem of a complicated type system that ensures the no-cloning theorem when dealing with quantum and classical data in one language.

Regarding its execution model, the implementation of QWIRE is based on \gls{QRAM}, but with no hardware interface. 
QWIRE can return a circuit as QASM code that can be sent to a quantum computer.
As QWIRE focuses on the verification of quantum algorithms, it is not surprising that it does not provide a wide toolchain support.
The export of QASM code already defines the capabilities when it comes to toolchain support; QWIRE does not implement any optimization on the circuit or other compilation steps in the toolchain. 
Hence, it only covers the box \emph{Code Analysis} in \autoref{fig:runtime_compilation_time}.
The scalability cannot be judged in a meaningful way because QWIRE does not integrate any quantum platform neither provides a simulator.
This also means that neither data- and resource-management nor legacy code modernization is considered; yet, proofing statements on ancillary qubits is possible.

\subsubsection{Quil [8,7,8,5,3,3,3,4]}
As an instruction set introduced for an abstract machine architecture for hybrid quantum-classical computation, Quil~\cite{smith_practical_2017} is an assembly-style low-level language equipped with a set of fine-grained control instructions, provided for the \umbrellaterm{} of Rigetti's quantum platform.
While scaling could be a big issue, Quil is well suited for program analysis and compilation. 
The refined \gls{HQCC} model is considered as Quil's execution model since Quil has an instruction to suspend the execution of the program until some condition is satisfied, which is mainly used to synchronize classical and quantum parts of the program.
For example, to implement variational algorithms in Quil, one could use this instruction to suspend the execution of the program until all measurement results are obtained.

For toolchain support, Rigetti Forest Software Development Kit contains (i) PyQuil, a Python library that generates and executes Quil instructions, (ii) the Rigetti Quil Compiler (quilc), which compiles Quil programs to native gate sets and performs optimization, and (iii) the Quantum Virtual Machine (qvm), a simulator for Quil programs. 
PyQuil may be helpful to import Quil applications to legacy projects, with the help of well-established bindings to and from Python. PyQuil faces similar situation to Qiskit with regard to software scalability: The core of PyQuil is a toolchain that compiles quantum source code to backend-executable entities, so its intermediate transpiling processes can come inefficient as the scale of quantum circuits grows. As mentioned, however, Python bindings can be used to integrate more efficient tools. 

\subsubsection{OpenQASM~3 [-,8,7,6,6,4,3,6]}

As a quantum assembly language, OpenQASM~3~\cite{cross_openqasm_2022} is used to describe quantum programs on a low-level circuit or even at pulse level. 
It is the successor of OpenQASM~\cite{cross_open_2017} and extends the language with real-time computation capabilities and pulse-level control. 
OpenQASM, which itself is based on various dialects of quantum assembly languages~\cite{balensiefer2005qale,qasm2circ, dousti2015squash, svore_layered_2006}, served as a de-facto standard for circuit description, used and supported by many tools, for instance~\cite{qiskit,mintz_qcor_2019,amy_staq_2020, javadiabhari_scaffcc_2015}. 
With OpenQASM~3, IBM aims to provide a new standard for future quantum applications that gives programmers more control over more capable quantum control hardware. 
Unique features include the ability to perform classical computations with the option to include arbitrary classical functions, have precise control over gate timings and scheduling, and the option to include hardware vendor-specific language grammars to describe pulse implementations for gates~\cite{cross_openqasm_2022}.
While the paper itself does not describe a compilation infrastructure, OpenQASM~3 is supported in parts by IBM~Quantum~\cite{ibm_qasm3_features} and Amazon~Braket~\cite{aws_qasm3_features}. We refer our rating to the implementation of IBM.

OpenQASM~3 is an external \gls{DSL}, and hence independent of any host language, which is why there is no rating for this category. 
The language follows the refined \gls{HQCC} execution model with a quantum controller capable of performing limited classical computations including real-time feedback within the coherence time of the qubits. 
Since OpenQASM~3 is being integrated into the Qiskit toolchain, a few ratings are the same as for Qiskit (scalability, cross-technology execution, data management and resource management). 
The toolchain support rating is lower than for Qiskit, as not all features are implemented yet. OpenQASM~3 is a compiled external \gls{DSL} which allows for easier integration into legacy applications.
Overall, this renders OpenQASM~3 a promising candidate for an expressive quantum assembly language.

\subsubsection{XACC [10,8,7,9,9,9,7,9]}
XACC~\cite{mccaskey_xacc_2020} is a system-level software framework which is implemented in C++. 

It provides a programming model which brings quantum acceleration within \gls{HPC} software workflows. 
XACC offers full toolchain support and is comprised of a three-layered (front-, middle-, and backend) software architecture, each layer providing different extension points. 
The frontend provides extensible compiler infrastructure, which enables the specification of quantum kernels in a language agnostic fashion by transforming it to a low-level polymorphic \gls{IR}. 
The middleend layer exposes an \gls{IR} object model. 
Transformation on \gls{IR} is a critical concept of the middle layer, which provides an extension point of XACC for performing standard quantum compilation routines. 
The \gls{IR} design provides the core integration mechanism bridging multiple \umbrellatermpl{} with multiple quantum backends.
The extensible design of the backend layer provides communication protocols for an abstract type of quantum backend, which makes it feasible for cross-technology execution. 

XACC treats the \gls{QPU} as a composition of quantum registers, quantum control units, and classical memory space for storing quantum execution results, which resembles a refined \gls{HQCC} execution model~\cite{fu2021quingo}. 
The quantum-classical interaction is realized via the \textit{client-server} model, with the classical computing system on the client-side and the quantum accelerator on the server-side. 
General quantum programs are modeled as operations of these quantum registers. 
XACC puts forward an abstraction of \textit{accelerator buffer} that models underlying quantum registers. 

C++ is a commonplace language when it comes to portability, performance, and efficiency in the HPC environment. 
This makes it possible for XACC to integrate legacy \gls{HPC}-code for hybrid computing by leaving implementation details to the programmer. 
Hence, XACC is rated high in the respective category of \autoref{fig:rate_qpts}.
Having C++ as the host language, XACC lays the groundwork for tighter quantum-classical computing. 
Besides serial quantum-classical computing, XACC also facilitates massively parallel distributed quantum-accelerated \gls{HPC}. 

XACC is designed in such a way that it is quite convenient to extend its core modules. 
For instance, the frontend layer can be extended to parse nascent \umbrellatermpl{} by extending its compiler interface. 
Because of the support of MPI, resource scaling and data management are quite efficient and convenient via XACC, hence these categories are rated high in \autoref{fig:rate_qpts}. 

\subsubsection{OpenQL [10,8,9,2,7,4,3,8]}

OpenQL~\cite{khammassi_openql_2022} includes a complete compilation toolchain that consists of a high-level programming language and a standard interface for extendable \gls{QPU} backends. 
The \gls{API} offers support for two distinct programming languages, Python and C++. 
Algorithm developers are free to select their preferred language to describe algorithms. 
As a framework, it incorporates two separated compilation procedures, each offering its own optimization and scheduling process. 
The first process is the hardware-agnostic compilation flow that compiles the program to an \gls{IR}. 
Then follows the second part functioning as hardware-dependent that can further compile and optimize the program based on specific technologies. 
Based on different hardware specification, the compilation process generates an executable QASM that can be directly executed on the control hardware. 
This separated compilation process facilitates the extension of the \gls{QPU} backend and enables better optimization of the instruction sequence. 

In addition to the compilation toolchain, OpenQL also provides an interface to a quantum simulator. 
This feature assists the developer to verify the algorithm and transfer it to the backend implementation without additional efforts. 
The scalability of this firmware is mainly restricted by the optimization mapping and routing algorithms. %
Large circuits will exponentially increase the computational complexity and lead to an unreliable algorithm. 
Additionally, when interfacing the backends in this firmware, scalability also depends on the backend design of the quantum \gls{ISA} and microarchitecture. 
OpenQL uses a backend from~\cite{fu_eqasm_2019}, which constrains the scalability of the firmware because of the \gls{ISA} encoding limitation.
But its performance can be enhanced through a novel and robust optimization, mapping algorithm, and backend design. 
Due to an absence of relevant information in the paper, comprehensive analysis of the resource management, memory model, and data mapping is hindered.

Overall, OpenQL is a comprehensive framework that provides relatively full toolchain support, covering different stages with optimizations from quantum programs to QPU interfaces. Furthermore, it features the refined \gls{HQCC} model and quantum co-processing computational module. With these unique characteristics, OpenQL receives high ratings regarding the \textit{toolchain support} and \textit{execution model} criteria and makes itself a noteworthy candidate for further \umbrellaterm{} research.

\section{Road Map}\label{sec:discussion}

It is clear from the previous section that \gls{HPCQC} integration is still in an initial state. 
In this section, we highlight the gaps which inhibit \gls{HPCQC} integration, and motivate our future work based on the outcome of the analysis above. 
First, we highlight the need of a unified quantum-classical toolchain.
Then we illustrate the main shortcomings of the currently existing toolchains with regard to \gls{HPC} integration.

\subsection{Hybrid Quantum-Classical Toolchain} \label{subsec:Hybrid_Q_C_tlc}
High performance for large-scale problems is the origin of \gls{HPC}.
For classical problems, a well-established way of tackling them exists.
New methods can be integrated into existing workflows and the applications can be scaled on different architectures.
In contrast, the progress in \gls{QC} is scattered and unconnected pieces are driven by individual efforts.
A unified \emph{hybrid classical-quantum toolchain} as a guiding structure for specialized tools to fit in is necessary. 
Specifically, a unified hybrid classical-quantum toolchain is a software framework that combines both classical and quantum computing methods into a single, integrated system. This toolchain aims to streamline the development, implementation, and scaling of computational solutions by providing a consistent and coherent structure where specialized tools for classical and quantum computations can be seamlessly combined. Such a toolchain would enable researchers and developers to harness the strengths of both computing approaches while addressing the limitations of current \glspl{QPT}.

Some proposals have been put forward, but they are not sufficient in the \gls{HPC} domain.
In an upcoming publication, we will present our modular and scalable quantum-classical software toolchain. 
It is built to integrate into existing \gls{HPC} workflows, while still maintaining the necessary flexibility.

One of the key challenges is to provide full toolchain support, but also to guarantee scalability.
It is clear from our results that most \umbrellatermpl{} focus on only one of those.
For example, Qiskit provides an easy way to add new tools, but it comes short in scalability.
Additionally, it focuses on specific areas to provide a unique feature compared to other \umbrellatermpl{}.
These features do not explicitly target \gls{HPC} yet, but they can be used in this context.
Circuit optimization and related tasks that solve \gls{NISQ}-specific problems are also actively researched but rarely integrated in a toolchain.
We think it is crucial to provide a uniform method to provide these tools to a broader audience. 

As a guiding principle, we advocate the analysis blueprint.
It highlights the evolution of requirements up to the on-node integration in the future.
Strategically placing emphasis on the key obstacles allows us to find the most useful tools for a task.
Also, it becomes quite clear which areas need most improvement.
We bring attention to these shortcomings in the next section.

\subsection{Gaps in \glsentryshort{HPCQC} Integration Tools}
\label{subsec:Gaps_in_QC_HPC_Integration_Tools}

As stated previously, the hybrid quantum-classical toolchain plays a crucial role for incorporating quantum devices into existing classical \gls{HPC} systems.
According to our analysis, none of the analyzed \umbrellatermpl{} fulfills all the listed criteria.
Most tools focus on the current challenges, namely restricted access to restricted hardware.
However, in the long term, these tools are not suitable for exascale \gls{HPC} integration individually.
In this area, we can learn from classical problems and have identified key obstacles for quantum-enabled compute nodes.
With this section, we point out interesting topics and inspire future research.

Classical resource managers are employed in any modern \gls{HPC} system.
So far, quantum resources are not supported because they have different requirements on scheduling.
Some ideas have already been explored, but not with \gls{HPC} as their target.
The quantum-classical interaction in hybrid problems also poses new challenges.
For instance, the resource manager must monitor the performance of the quantum hardware and dynamically adjust run time parameters like allocation of resources to achieve goals like error mitigation or optimization of computing resource utilization efficiency.
Per-machine scheduling is necessary; eventually one can run multiple problems on one machine if they fit.
Scheduling can also be done on a qubit level over multiple machines once communication becomes available.
Furthermore, it is conceivable that various technologies are used simultaneously.
To our knowledge, this problem has not been considered so far.

Utilizing the quantum advantage aims to provide viable solutions to classically hard problems.
For many problems, classical solutions already exist.
No effort has been made to integrate quantum solutions with legacy applications.
This task requires work from various communities.
As a first step, such applications which would benefit from quantum acceleration need to be identified.
Afterwards, integration becomes relevant: the foremost requirement is the compatibility with \gls{MPI}.
Distributed memory systems are essential for parallelized applications.
\gls{HPC} systems cannot afford to synchronize access to a single shared \gls{QPU}.
Offloading tasks which end up in a job queue are a major bottleneck.
If the tasks are assembled by multiple classical nodes, even data mapping might become relevant.
This is still limited by hardware availability, but has to be solved nonetheless.

To ensure the validity of future development, two additional requirements are necessary.
Tools should adhere to a standard way of communicating their in- and output as well as their transformation of the problem instance.
A \gls{SQIR} can streamline the integration of quantum computers and HPC systems.
A common language and protocols enable seamless communication and interoperability, simplifying hybrid quantum-classical algorithm development.

We are not the first to recognize the need for standardization.
The \gls{QIR}~\cite{QIRSpec2021} is an LLVM-based \gls{IR} for quantum programs.
It acknowledges the three critical phases: language-specific, generic, and target-specific and furnishes an interface for the generic stage.
The proposed strategy involves employing a suitable \umbrellaterm{} to express the program, executing high-level optimization, and translating it into the universally accepted QIR. 
At this juncture, numerous generic optimizations can be implemented prior to the final translation step to hardware, facilitated by a target-specific compiler.
For our purposes, this proposal is not extensive enough.
It shifts some of the problems we uncovered in this paper to the hardware implementation.
Scalability is achieved via the utilization of the LLVM framework.
We believe this is the most reasonable way to tackle the problem, but more research has to be done to adopt similar techniques to the runtime environment.
Not all classical challenges like communication, resource management and data management can be resolved by using \gls{QIR} alone.
Still, \gls{QIR} is an actively evolving standard and we hope that it or one of  its subsequent iterations fulfill our criteria.

Any unified representation facilitates robust and efficient toolchains, allowing easy code porting and optimization across various platforms.
Moreover, a \gls{SQIR} fosters collaboration among researchers, developers, and manufacturers, promoting resource sharing and best practices. This accelerates the maturation of \gls{QC} technologies, enhancing integration with HPC systems and unlocking their full potential.
An extended standard will be part of our toolchain proposal, but it also has to be accepted by the community.
Progress should then be tracked by common benchmarks.
In classical \gls{HPC}, it is straightforward how to evaluate the performance of a system.
For hybrid systems, such benchmarks are not yet available.

Once all necessary tools exist we plan to unify them and deploy our toolchain to an \gls{HPC} system.
There, it will be tested and benchmarked in real-world scenarios.

\section{Conclusion and Future Work}\label{sec:conclusion}

Most previous reviews have investigated the language aspect of \umbrellatermpl{}.
In this work, we have reviewed multiple \umbrellatermpl{} under the unique aspect of \gls{HPC} integration.
We look at four increasingly complex scenarios of integration and provide the relevant criteria to judge \umbrellatermpl{} accordingly.
Each criterion belongs to one of three categories, with the most important one being \textit{product maturity} in \gls{HPC}.
It contains the main essence of classical \gls{HPC} problems, \textit{scalability} and \textit{toolchain support}, transferred to the quantum domain.
With our \emph{analysis blueprint}, we provide a way to rate the integration potential of \umbrellatermpl{} by asking a series of questions.
We have used it to analyze six \umbrellatermpl{} in-depth, showing the various focuses and advances.
As a result, we concluded that existing \umbrellatermpl{} only partially cover the requirements needed for full \gls{HPCQC} integration and that future work on \umbrellatermpl{} must consider integration as a crucial aspect.

This only marks the first step in an continuous integration effort.
In future work, we plan to move forward from analyzing the existing tools to provide a straightforward way to use them in an \gls{HPC} context.
For this, we propose the approach of a \textit{unified} \gls{HPCQC} toolchain.
By defining the interfaces and interactions of tools, we want to enable researchers to easily tackle \gls{HPC} problems.
In the best case, we can extend our blueprint to not only cover analysis, but also to guide the integration process of \umbrellatermpl{}.
We deem this a necessary step for scaling applications, due to the different requirements compared to traditional toolchains.

Another problem is the comparability of results.
In the \gls{HPCQC} domain, no straightforward benchmarks exist.
Existing \gls{NISQ} problems emphasize quantum challenges but rarely exhaust classical resources.
To gain a better understanding of the requirements, both sides have to be tested equally.
Thus, our final concept includes phase analysis, integration, and evaluation.

\begin{acks}
The research is part of the Munich Quantum Valley~(MQV), which is supported by the Bavarian state government with funds from the Hightech Agenda Bayern Plus.
Moreover, this project is supported by the Federal Ministry for Economic Affairs and Climate Action on the basis of a decision by the German Bundestag through project QuaST, as well as by the Bavarian Ministry of Economic Affairs, Regional Development and Energy with funds from the Hightech Agenda Bayern.
Further funding comes from the German Federal Ministry of Education and Research~(BMBF) through project MUNIQC-SC.
Finally, this work is also supported by BMW.  
\end{acks}

\bibliographystyle{ACM-Reference-Format}
\bibliography{references}


\begin{thebibliography}{142}


\ifx \showCODEN    \undefined \def \showCODEN     #1{\unskip}     \fi
\ifx \showDOI      \undefined \def \showDOI       #1{#1}\fi
\ifx \showISBNx    \undefined \def \showISBNx     #1{\unskip}     \fi
\ifx \showISBNxiii \undefined \def \showISBNxiii  #1{\unskip}     \fi
\ifx \showISSN     \undefined \def \showISSN      #1{\unskip}     \fi
\ifx \showLCCN     \undefined \def \showLCCN      #1{\unskip}     \fi
\ifx \shownote     \undefined \def \shownote      #1{#1}          \fi
\ifx \showarticletitle \undefined \def \showarticletitle #1{#1}   \fi
\ifx \showURL      \undefined \def \showURL       {\relax}        \fi
\providecommand\bibfield[2]{#2}
\providecommand\bibinfo[2]{#2}
\providecommand\natexlab[1]{#1}
\providecommand\showeprint[2][]{arXiv:#2}

\bibitem[Aaronson and Chen(2017)]%
        {Aaronson2017}
\bibfield{author}{\bibinfo{person}{Scott Aaronson} {and} \bibinfo{person}{Lijie
  Chen}.} \bibinfo{year}{2017}\natexlab{}.
\newblock \showarticletitle{Complexity-Theoretic Foundations of Quantum
  Supremacy Experiments}. In \bibinfo{booktitle}{\emph{Proceedings of the 32nd
  Computational Complexity Conference}} (Riga, Latvia)
  \emph{(\bibinfo{series}{CCC '17})}. \bibinfo{publisher}{Schloss
  Dagstuhl--Leibniz-Zentrum fuer Informatik}, \bibinfo{address}{Dagstuhl, DEU},
  Article \bibinfo{articleno}{22}, \bibinfo{numpages}{67}~pages.
\newblock
\showISBNx{9783959770408}
\urldef\tempurl%
\url{https://doi.org/10.48550/ARXIV.1612.05903}
\showDOI{\tempurl}


\bibitem[Abelson and Sussman(1996)]%
        {Abelson1996-qw}
\bibfield{author}{\bibinfo{person}{Harold Abelson} {and}
  \bibinfo{person}{Gerald~Jay Sussman}.} \bibinfo{year}{1996}\natexlab{}.
\newblock \bibinfo{booktitle}{\emph{Structure and interpretation of computer
  programs} (\bibinfo{edition}{2} ed.)}.
\newblock \bibinfo{publisher}{MIT Press}, \bibinfo{address}{London, England}.
\newblock
\urldef\tempurl%
\url{http://library.oapen.org/handle/20.500.12657/26092}
\showURL{%
\tempurl}


\bibitem[Abhari et~al\mbox{.}(2012)]%
        {abhari_scaffold_2012}
\bibfield{author}{\bibinfo{person}{Ali~Javadi Abhari}, \bibinfo{person}{Arvin
  Faruque}, \bibinfo{person}{Mohammad~Javad Dousti}, \bibinfo{person}{Lukas
  Svec}, \bibinfo{person}{Oana Catu}, \bibinfo{person}{Amlan Chakrabati},
  \bibinfo{person}{Chen-Fu Chiang}, \bibinfo{person}{Seth Vanderwilt},
  \bibinfo{person}{John Black}, \bibinfo{person}{Fred Chong},
  \bibinfo{person}{Margaret Martonosi}, \bibinfo{person}{Martin Suchara},
  \bibinfo{person}{Ken Brown}, \bibinfo{person}{Massoud Pedram}, {and}
  \bibinfo{person}{Todd Brun}.} \bibinfo{year}{2012}\natexlab{}.
\newblock \bibinfo{booktitle}{\emph{Scaffold: {Quantum} {Programming}
  {Language}}}.
\newblock \bibinfo{type}{{T}echnical {R}eport}. \bibinfo{institution}{Princeton
  Univ NJ Dept of Computer Science}. \bibinfo{pages}{43} pages.
\newblock
\urldef\tempurl%
\url{https://www.cs.princeton.edu/research/techreps/TR-934-12}
\showURL{%
\tempurl}


\bibitem[Aleksandrowicz et~al\mbox{.}(2019)]%
        {qiskit}
\bibfield{author}{\bibinfo{person}{Gadi Aleksandrowicz},
  \bibinfo{person}{Thomas Alexander}, \bibinfo{person}{Panagiotis Barkoutsos},
  \bibinfo{person}{Luciano Bello}, \bibinfo{person}{Yael Ben-Haim},
  \bibinfo{person}{David Bucher}, \bibinfo{person}{F.~Jose
  Cabrera-Hern{\'a}ndez}, \bibinfo{person}{Jorge Carballo-Franquis},
  \bibinfo{person}{Adrian Chen}, \bibinfo{person}{Chun-Fu Chen},
  {et~al\mbox{.}}} \bibinfo{year}{2019}\natexlab{}.
\newblock \bibinfo{title}{Qiskit: An open-source framework for quantum
  computing}.
\newblock
\newblock
\urldef\tempurl%
\url{https://qiskit.org/}
\showURL{%
\tempurl}


\bibitem[Altenkirch and Grattage(2005)]%
        {altenkirch_qml_2005}
\bibfield{author}{\bibinfo{person}{Thorsten Altenkirch} {and}
  \bibinfo{person}{Jonathan Grattage}.} \bibinfo{year}{2005}\natexlab{}.
\newblock \bibinfo{title}{QML: Quantum data and control}.
\newblock
\newblock
\urldef\tempurl%
\url{https://citeseerx.ist.psu.edu/document?repid=rep1&type=pdf&doi=6db2af1980fc4c54db5e8b974d381c4c583358ee}
\showURL{%
\tempurl}


\bibitem[Altenkirch and Green(2009)]%
        {gay_quantum_2009}
\bibfield{author}{\bibinfo{person}{Thorsten Altenkirch} {and}
  \bibinfo{person}{Alexander~S. Green}.} \bibinfo{year}{2009}\natexlab{}.
\newblock \showarticletitle{The {Quantum} {IO} {Monad}}.
\newblock In \bibinfo{booktitle}{\emph{Semantic {Techniques} in {Quantum}
  {Computation}} (\bibinfo{edition}{1} ed.)},
  \bibfield{editor}{\bibinfo{person}{Simon Gay} {and} \bibinfo{person}{Ian
  Mackie}} (Eds.). \bibinfo{publisher}{Cambridge University Press},
  \bibinfo{address}{Cambridge, UK}, \bibinfo{pages}{173--205}.
\newblock
\showISBNx{978-0-521-51374-6 978-1-139-19331-3}
\urldef\tempurl%
\url{https://doi.org/10.1017/CBO9781139193313.006}
\showDOI{\tempurl}


\bibitem[Amaral et~al\mbox{.}(2020)]%
        {AMARAL2020102584}
\bibfield{author}{\bibinfo{person}{Vasco Amaral}, \bibinfo{person}{Beatriz
  Norberto}, \bibinfo{person}{Miguel Goulão}, \bibinfo{person}{Marco
  Aldinucci}, \bibinfo{person}{Siegfried Benkner}, \bibinfo{person}{Andrea
  Bracciali}, \bibinfo{person}{Paulo Carreira}, \bibinfo{person}{Edgars Celms},
  \bibinfo{person}{Luís Correia}, \bibinfo{person}{Clemens Grelck},
  \bibinfo{person}{Helen Karatza}, \bibinfo{person}{Christoph Kessler},
  \bibinfo{person}{Peter Kilpatrick}, \bibinfo{person}{Hugo Martiniano},
  \bibinfo{person}{Ilias Mavridis}, \bibinfo{person}{Sabri Pllana},
  \bibinfo{person}{Ana Respício}, \bibinfo{person}{José Simão},
  \bibinfo{person}{Luís Veiga}, {and} \bibinfo{person}{Ari Visa}.}
  \bibinfo{year}{2020}\natexlab{}.
\newblock \showarticletitle{Programming languages for data-Intensive HPC
  applications: A systematic mapping study}.
\newblock \bibinfo{journal}{\emph{Parallel Comput.}}  \bibinfo{volume}{91}
  (\bibinfo{year}{2020}), \bibinfo{pages}{102584}.
\newblock
\showISSN{0167-8191}
\urldef\tempurl%
\url{https://doi.org/10.1016/j.parco.2019.102584}
\showDOI{\tempurl}


\bibitem[Amazon Web Services, Inc(2023)]%
        {AmazonBraKet}
Amazon Web Services, Inc \bibinfo{year}{2023}\natexlab{}.
\newblock \bibinfo{booktitle}{\emph{Amazon Braket: Developer Guide}}.
\newblock Amazon Web Services, Inc.
\newblock
\urldef\tempurl%
\url{https://docs.aws.amazon.com/braket/latest/developerguide/braket-get-started.html}
\showURL{%
\tempurl}


\bibitem[Amy and Gheorghiu(2020)]%
        {amy_staq_2020}
\bibfield{author}{\bibinfo{person}{Matthew Amy} {and} \bibinfo{person}{Vlad
  Gheorghiu}.} \bibinfo{year}{2020}\natexlab{}.
\newblock \showarticletitle{staq—A full-stack quantum processing toolkit}.
\newblock \bibinfo{journal}{\emph{Quantum Science and Technology}}
  \bibinfo{volume}{5}, \bibinfo{number}{3} (\bibinfo{year}{2020}),
  \bibinfo{pages}{034016}.
\newblock
\urldef\tempurl%
\url{https://doi.org/10.1088/2058-9565/ab9359}
\showDOI{\tempurl}


\bibitem[Andersen et~al\mbox{.}(2020)]%
        {andersen_repeated_2020}
\bibfield{author}{\bibinfo{person}{Christian~Kraglund Andersen},
  \bibinfo{person}{Ants Remm}, \bibinfo{person}{Stefania Lazar},
  \bibinfo{person}{Sebastian Krinner}, \bibinfo{person}{Nathan Lacroix},
  \bibinfo{person}{Graham~J. Norris}, \bibinfo{person}{Mihai Gabureac},
  \bibinfo{person}{Christopher Eichler}, {and} \bibinfo{person}{Andreas
  Wallraff}.} \bibinfo{year}{2020}\natexlab{}.
\newblock \showarticletitle{Repeated quantum error detection in a surface
  code}.
\newblock \bibinfo{journal}{\emph{Nature Physics}} \bibinfo{volume}{16},
  \bibinfo{number}{8} (\bibinfo{date}{Aug.} \bibinfo{year}{2020}),
  \bibinfo{pages}{875--880}.
\newblock
\showISSN{1745-2473, 1745-2481}
\urldef\tempurl%
\url{https://doi.org/10.1038/s41567-020-0920-y}
\showDOI{\tempurl}


\bibitem[Arute et~al\mbox{.}(2019)]%
        {arute2019quantum}
\bibfield{author}{\bibinfo{person}{Frank Arute}, \bibinfo{person}{Kunal Arya},
  \bibinfo{person}{Ryan Babbush}, \bibinfo{person}{Dave Bacon},
  \bibinfo{person}{Joseph~C. Bardin}, \bibinfo{person}{Rami Barends},
  \bibinfo{person}{Rupak Biswas}, \bibinfo{person}{Sergio Boixo},
  \bibinfo{person}{Fernando G. S.~L. Brandao}, \bibinfo{person}{David~A.
  Buell}, {et~al\mbox{.}}} \bibinfo{year}{2019}\natexlab{}.
\newblock \showarticletitle{Quantum supremacy using a programmable
  superconducting processor}.
\newblock \bibinfo{journal}{\emph{Nature}} \bibinfo{volume}{574},
  \bibinfo{number}{7779} (\bibinfo{year}{2019}), \bibinfo{pages}{505--510}.
\newblock
\urldef\tempurl%
\url{https://doi.org/10.1038/s41586-019-1666-5}
\showDOI{\tempurl}


\bibitem[Arvanitou et~al\mbox{.}(2021)]%
        {Arvanitou2021}
\bibfield{author}{\bibinfo{person}{Elvira-Maria Arvanitou},
  \bibinfo{person}{Apostolos Ampatzoglou}, \bibinfo{person}{Alexander
  Chatzigeorgiou}, {and} \bibinfo{person}{Jeffrey~C. Carver}.}
  \bibinfo{year}{2021}\natexlab{}.
\newblock \showarticletitle{Software engineering practices for scientific
  software development: A systematic mapping study}.
\newblock \bibinfo{journal}{\emph{Journal of Systems and Software}}
  \bibinfo{volume}{172} (\bibinfo{date}{Feb.} \bibinfo{year}{2021}),
  \bibinfo{pages}{110848}.
\newblock
\urldef\tempurl%
\url{https://doi.org/10.1016/j.jss.2020.110848}
\showDOI{\tempurl}


\bibitem[Aspuru-Guzik and Walther(2012)]%
        {AspuruGuzik2012}
\bibfield{author}{\bibinfo{person}{Al{\'{a}}n Aspuru-Guzik} {and}
  \bibinfo{person}{Philip Walther}.} \bibinfo{year}{2012}\natexlab{}.
\newblock \showarticletitle{Photonic quantum simulators}.
\newblock \bibinfo{journal}{\emph{Nature Physics}} \bibinfo{volume}{8},
  \bibinfo{number}{4} (\bibinfo{date}{April} \bibinfo{year}{2012}),
  \bibinfo{pages}{285--291}.
\newblock
\urldef\tempurl%
\url{https://doi.org/10.1038/nphys2253}
\showDOI{\tempurl}


\bibitem[Balensiefer et~al\mbox{.}(2005)]%
        {balensiefer2005qale}
\bibfield{author}{\bibinfo{person}{Steven Balensiefer}, \bibinfo{person}{Lucas
  Kreger-Stickles}, {and} \bibinfo{person}{Mark Oskin}.}
  \bibinfo{year}{2005}\natexlab{}.
\newblock \showarticletitle{QUALE: quantum architecture layout evaluator}. In
  \bibinfo{booktitle}{\emph{Quantum Information and Computation III}},
  \bibfield{editor}{\bibinfo{person}{Eric~J. Donkor},
  \bibinfo{person}{Andrew~R. Pirich}, {and} \bibinfo{person}{Howard~E. Brandt}}
  (Eds.), Vol.~\bibinfo{volume}{5815}. International Society for Optics and
  Photonics, \bibinfo{publisher}{SPIE}, \bibinfo{address}{Washington, USA},
  \bibinfo{pages}{103 -- 114}.
\newblock
\urldef\tempurl%
\url{https://doi.org/10.1117/12.604073}
\showDOI{\tempurl}


\bibitem[Bartsch et~al\mbox{.}(2021)]%
        {bartsch_valeria_2021_5555960}
\bibfield{author}{\bibinfo{person}{Valeria Bartsch}, \bibinfo{person}{Guillaume
  Colin~de Verdière}, \bibinfo{person}{Jean-Philippe Nominé},
  \bibinfo{person}{Daniele Ottaviani}, \bibinfo{person}{Daniele Dragoni},
  \bibinfo{person}{Chayma Bouazza}, \bibinfo{person}{Fabrizio Magugliani},
  \bibinfo{person}{David Bowden}, \bibinfo{person}{Cyril Allouche},
  \bibinfo{person}{Mikael Johansson}, \bibinfo{person}{Olivier Terzo},
  \bibinfo{person}{Andrea Scarabosio}, \bibinfo{person}{Giacomo Vitali},
  \bibinfo{person}{Farida Shagieva}, {and} \bibinfo{person}{Kristel
  Michielsen}.} \bibinfo{year}{2021}\natexlab{}.
\newblock \bibinfo{title}{<QC|HPC>: Quantum for HPC}.
\newblock
\newblock
\urldef\tempurl%
\url{https://doi.org/10.5281/zenodo.5555960}
\showDOI{\tempurl}


\bibitem[Bettelli et~al\mbox{.}(2003)]%
        {bettelli_q_2003}
\bibfield{author}{\bibinfo{person}{Stefano Bettelli}, \bibinfo{person}{Luciano
  Serafini}, {and} \bibinfo{person}{Tommaso Calarco}.}
  \bibinfo{year}{2003}\natexlab{}.
\newblock \showarticletitle{Toward an architecture for quantum programming}.
\newblock \bibinfo{journal}{\emph{The European Physical Journal D - Atomic,
  Molecular and Optical Physics}} \bibinfo{volume}{25}, \bibinfo{number}{2}
  (\bibinfo{date}{Aug.} \bibinfo{year}{2003}), \bibinfo{pages}{181--200}.
\newblock
\showISSN{1434-6060, 1434-6079}
\urldef\tempurl%
\url{https://doi.org/10.1140/epjd/e2003-00242-2}
\showDOI{\tempurl}


\bibitem[Blume et~al\mbox{.}(1994)]%
        {Blume1994}
\bibfield{author}{\bibinfo{person}{William Blume}, \bibinfo{person}{Rudolf
  Eigenmann}, \bibinfo{person}{Jay Hoeflinger}, \bibinfo{person}{David Padua},
  \bibinfo{person}{Paul Petersen}, \bibinfo{person}{Lawrence Rauchwerger},
  {and} \bibinfo{person}{Perg Tu}.} \bibinfo{year}{1994}\natexlab{}.
\newblock \showarticletitle{Automatic Detection of Parallelism: A grand
  challenge for high performance computing}.
\newblock \bibinfo{journal}{\emph{IEEE Parallel \& Distributed Technology:
  Systems \& Applications}} \bibinfo{volume}{2}, \bibinfo{number}{3}
  (\bibinfo{year}{1994}), \bibinfo{pages}{37--}.
\newblock
\urldef\tempurl%
\url{https://doi.org/10.1109/M-PDT.1994.329796}
\showDOI{\tempurl}


\bibitem[Braket(2022)]%
        {aws_qasm3_features}
\bibfield{author}{\bibinfo{person}{Amazon Braket}.}
  \bibinfo{year}{2022}\natexlab{}.
\newblock \bibinfo{booktitle}{\emph{What OpenQASM features does Braket
  support?}}
\newblock Amazon.
\newblock
\urldef\tempurl%
\url{https://docs.aws.amazon.com/braket/latest/developerguide/braket-openqasm-supported-features.html}
\showURL{%
Retrieved February 20, 2023 from \tempurl}


\bibitem[Bravyi et~al\mbox{.}(2022)]%
        {Bravyi2022}
\bibfield{author}{\bibinfo{person}{Sergey Bravyi}, \bibinfo{person}{Oliver
  Dial}, \bibinfo{person}{Jay~M. Gambetta}, \bibinfo{person}{Dar{\'{\i}}o Gil},
  {and} \bibinfo{person}{Zaira Nazario}.} \bibinfo{year}{2022}\natexlab{}.
\newblock \showarticletitle{The future of quantum computing with
  superconducting qubits}.
\newblock \bibinfo{journal}{\emph{Journal of Applied Physics}}
  \bibinfo{volume}{132}, \bibinfo{number}{16} (\bibinfo{date}{Oct.}
  \bibinfo{year}{2022}), \bibinfo{pages}{160902}.
\newblock
\urldef\tempurl%
\url{https://doi.org/10.1063/5.0082975}
\showDOI{\tempurl}


\bibitem[Caleffi et~al\mbox{.}(2022)]%
        {Caleffi2022}
\bibfield{author}{\bibinfo{person}{Marcello Caleffi}, \bibinfo{person}{Michele
  Amoretti}, \bibinfo{person}{Davide Ferrari}, \bibinfo{person}{Daniele Cuomo},
  \bibinfo{person}{Jessica Illiano}, \bibinfo{person}{Antonio Manzalini}, {and}
  \bibinfo{person}{Angela~Sara Cacciapuoti}.} \bibinfo{year}{2022}\natexlab{}.
\newblock \bibinfo{title}{Distributed Quantum Computing: a Survey}.
\newblock
\newblock
\urldef\tempurl%
\url{https://doi.org/10.48550/ARXIV.2212.10609}
\showDOI{\tempurl}


\bibitem[Chhangte and Chakrabarty(2022)]%
        {chhangteMappingQuantumCircuits2022}
\bibfield{author}{\bibinfo{person}{Lalengmawia Chhangte} {and}
  \bibinfo{person}{Alok Chakrabarty}.} \bibinfo{year}{2022}\natexlab{}.
\newblock \showarticletitle{Mapping {{Quantum Circuits}} in {{IBM Q Devices
  Using Progressive Qubit Assignment}} for {{Global Ordering}}}.
\newblock \bibinfo{journal}{\emph{New Gener. Comput.}} \bibinfo{volume}{40},
  \bibinfo{number}{1} (\bibinfo{date}{April} \bibinfo{year}{2022}),
  \bibinfo{pages}{311--338}.
\newblock
\showISSN{0288-3635, 1882-7055}
\urldef\tempurl%
\url{https://doi.org/10.1007/s00354-022-00163-5}
\showDOI{\tempurl}


\bibitem[Chitambar and Gour(2019)]%
        {Chitambar2019}
\bibfield{author}{\bibinfo{person}{Eric Chitambar} {and} \bibinfo{person}{Gilad
  Gour}.} \bibinfo{year}{2019}\natexlab{}.
\newblock \showarticletitle{Quantum resource theories}.
\newblock \bibinfo{journal}{\emph{Rev. Mod. Phys.}}  \bibinfo{volume}{91}
  (\bibinfo{year}{2019}), \bibinfo{pages}{025001}.
\newblock
\urldef\tempurl%
\url{https://doi.org/10.1103/RevModPhys.91.025001}
\showDOI{\tempurl}


\bibitem[Chong et~al\mbox{.}(2017)]%
        {chong_programming_2017}
\bibfield{author}{\bibinfo{person}{Frederic~T. Chong}, \bibinfo{person}{Diana
  Franklin}, {and} \bibinfo{person}{Margaret Martonosi}.}
  \bibinfo{year}{2017}\natexlab{}.
\newblock \showarticletitle{Programming languages and compiler design for
  realistic quantum hardware}.
\newblock \bibinfo{journal}{\emph{Nature}} \bibinfo{volume}{549},
  \bibinfo{number}{7671} (\bibinfo{date}{Sept.} \bibinfo{year}{2017}),
  \bibinfo{pages}{180--187}.
\newblock
\showISSN{1476-4687}
\urldef\tempurl%
\url{https://doi.org/10.1038/nature23459}
\showDOI{\tempurl}


\bibitem[Chuang(2005)]%
        {qasm2circ}
\bibfield{author}{\bibinfo{person}{Isaac Chuang}.}
  \bibinfo{year}{2005}\natexlab{}.
\newblock \bibinfo{booktitle}{\emph{qasm2circ}}.
\newblock MIT.
\newblock
\urldef\tempurl%
\url{https://www.media.mit.edu/quanta/qasm2circ/}
\showURL{%
Retrieved February 13, 2023 from \tempurl}


\bibitem[Ciccozzi et~al\mbox{.}(2022)]%
        {Ciccozzi2022}
\bibfield{author}{\bibinfo{person}{Federico Ciccozzi}, \bibinfo{person}{Lorenzo
  Addazi}, \bibinfo{person}{Sara~Abbaspour Asadollah},
  \bibinfo{person}{Bj\"{o}rn Lisper}, \bibinfo{person}{Abu~Naser Masud}, {and}
  \bibinfo{person}{Saad Mubeen}.} \bibinfo{year}{2022}\natexlab{}.
\newblock \showarticletitle{A Comprehensive Exploration of Languages for
  Parallel Computing}.
\newblock \bibinfo{journal}{\emph{Comput. Surveys}} \bibinfo{volume}{55},
  \bibinfo{number}{2} (\bibinfo{date}{Jan.} \bibinfo{year}{2022}),
  \bibinfo{pages}{1--39}.
\newblock
\urldef\tempurl%
\url{https://doi.org/10.1145/3485008}
\showDOI{\tempurl}


\bibitem[Clairambault and de~Visme(2019)]%
        {clairambault_full_2019}
\bibfield{author}{\bibinfo{person}{Pierre Clairambault} {and}
  \bibinfo{person}{Marc de Visme}.} \bibinfo{year}{2019}\natexlab{}.
\newblock \showarticletitle{Full abstraction for the quantum lambda-calculus}.
\newblock \bibinfo{journal}{\emph{Proceedings of the ACM on Programming
  Languages}} \bibinfo{volume}{4}, \bibinfo{number}{POPL}
  (\bibinfo{year}{2019}), \bibinfo{pages}{63:1--63:28}.
\newblock
\urldef\tempurl%
\url{https://doi.org/10.1145/3371131}
\showDOI{\tempurl}


\bibitem[Cohen and Thompson(2021)]%
        {Cohen2021}
\bibfield{author}{\bibinfo{person}{Sam~R. Cohen} {and} \bibinfo{person}{Jeff~D.
  Thompson}.} \bibinfo{year}{2021}\natexlab{}.
\newblock \showarticletitle{Quantum Computing with Circular Rydberg Atoms}.
\newblock \bibinfo{journal}{\emph{{PRX} Quantum}} \bibinfo{volume}{2},
  \bibinfo{number}{3} (\bibinfo{date}{Aug.} \bibinfo{year}{2021}),
  \bibinfo{numpages}{26}~pages.
\newblock
\urldef\tempurl%
\url{https://doi.org/10.1103/prxquantum.2.030322}
\showDOI{\tempurl}


\bibitem[Cowtan et~al\mbox{.}(2019)]%
        {cowtan2019}
\bibfield{author}{\bibinfo{person}{Alexander Cowtan}, \bibinfo{person}{Silas
  Dilkes}, \bibinfo{person}{Ross Duncan}, \bibinfo{person}{Alexandre
  Krajenbrink}, \bibinfo{person}{Will Simmons}, {and} \bibinfo{person}{Seyon
  Sivarajah}.} \bibinfo{year}{2019}\natexlab{}.
\newblock \showarticletitle{{On the Qubit Routing Problem}}. In
  \bibinfo{booktitle}{\emph{14th Conference on the Theory of Quantum
  Computation, Communication and Cryptography (TQC 2019)}}
  \emph{(\bibinfo{series}{Leibniz International Proceedings in Informatics
  (LIPIcs)}, Vol.~\bibinfo{volume}{135})},
  \bibfield{editor}{\bibinfo{person}{Wim van Dam} {and} \bibinfo{person}{Laura
  Mancinska}} (Eds.). \bibinfo{publisher}{Schloss Dagstuhl--Leibniz-Zentrum
  fuer Informatik}, \bibinfo{address}{Dagstuhl, Germany},
  \bibinfo{pages}{5:1--5:32}.
\newblock
\showISBNx{978-3-95977-112-2}
\showISSN{1868-8969}
\urldef\tempurl%
\url{https://doi.org/10.4230/LIPIcs.TQC.2019.5}
\showDOI{\tempurl}


\bibitem[Cross et~al\mbox{.}(2017)]%
        {cross_open_2017}
\bibfield{author}{\bibinfo{person}{Andrew Cross}, \bibinfo{person}{Lev Bishop},
  \bibinfo{person}{John Smolin}, {and} \bibinfo{person}{Jay Gambetta}.}
  \bibinfo{year}{2017}\natexlab{}.
\newblock \bibinfo{title}{Open {Quantum} {Assembly} {Language}}.
\newblock
\newblock
\urldef\tempurl%
\url{http://arxiv.org/abs/1707.03429}
\showURL{%
\tempurl}


\bibitem[Cross et~al\mbox{.}(2022)]%
        {cross_openqasm_2022}
\bibfield{author}{\bibinfo{person}{Andrew Cross}, \bibinfo{person}{Ali
  Javadi-Abhari}, \bibinfo{person}{Thomas Alexander}, \bibinfo{person}{Niel
  De~Beaudrap}, \bibinfo{person}{Lev~S. Bishop}, \bibinfo{person}{Steven
  Heidel}, \bibinfo{person}{Colm~A. Ryan}, \bibinfo{person}{Prasahnt
  Sivarajah}, \bibinfo{person}{John Smolin}, \bibinfo{person}{Jay~M. Gambetta},
  {and} \bibinfo{person}{Blake~R. Johnson}.} \bibinfo{year}{2022}\natexlab{}.
\newblock \showarticletitle{{OpenQASM} 3: {A} {Broader} and {Deeper} {Quantum}
  {Assembly} {Language}}.
\newblock \bibinfo{journal}{\emph{ACM Transactions on Quantum Computing}}
  \bibinfo{volume}{3}, \bibinfo{number}{3} (\bibinfo{date}{Sept.}
  \bibinfo{year}{2022}), \bibinfo{pages}{1--50}.
\newblock
\showISSN{2643-6809, 2643-6817}
\urldef\tempurl%
\url{https://doi.org/10.1145/3505636}
\showDOI{\tempurl}


\bibitem[Cuomo et~al\mbox{.}(2020)]%
        {Cuomo2020}
\bibfield{author}{\bibinfo{person}{Daniele Cuomo}, \bibinfo{person}{Marcello
  Caleffi}, {and} \bibinfo{person}{Angela~Sara Cacciapuoti}.}
  \bibinfo{year}{2020}\natexlab{}.
\newblock \showarticletitle{Towards a distributed quantum computing ecosystem}.
\newblock \bibinfo{journal}{\emph{{IET} Quantum Communication}}
  \bibinfo{volume}{1}, \bibinfo{number}{1} (\bibinfo{date}{July}
  \bibinfo{year}{2020}), \bibinfo{pages}{3--8}.
\newblock
\urldef\tempurl%
\url{https://doi.org/10.1049/iet-qtc.2020.0002}
\showDOI{\tempurl}


\bibitem[Developers(2022)]%
        {cirq_developers_cirq_2022}
\bibfield{author}{\bibinfo{person}{Cirq Developers}.}
  \bibinfo{year}{2022}\natexlab{}.
\newblock \bibinfo{title}{Cirq}.
\newblock
\newblock
\urldef\tempurl%
\url{https://doi.org/10.5281/zenodo.7465577}
\showDOI{\tempurl}
\newblock
\shownote{. See full list of authors on Github: https://github
  .com/quantumlib/Cirq/graphs/contributors}.


\bibitem[Dong et~al\mbox{.}(2022)]%
        {Dong2022}
\bibfield{author}{\bibinfo{person}{Yulong Dong}, \bibinfo{person}{K.~Birgitta
  Whaley}, {and} \bibinfo{person}{Lin Lin}.} \bibinfo{year}{2022}\natexlab{}.
\newblock \showarticletitle{A quantum hamiltonian simulation benchmark}.
\newblock \bibinfo{journal}{\emph{npj Quantum Information}}
  \bibinfo{volume}{8}, \bibinfo{number}{1} (\bibinfo{date}{Nov.}
  \bibinfo{year}{2022}), \bibinfo{pages}{131}.
\newblock
\urldef\tempurl%
\url{https://doi.org/10.1038/s41534-022-00636-x}
\showDOI{\tempurl}


\bibitem[Dousti et~al\mbox{.}(2015)]%
        {dousti2015squash}
\bibfield{author}{\bibinfo{person}{Mohammad~Javad Dousti},
  \bibinfo{person}{Alireza Shafaei}, {and} \bibinfo{person}{Massoud Pedram}.}
  \bibinfo{year}{2015}\natexlab{}.
\newblock \bibinfo{title}{Squash 2: A Hierarchical Scalable Quantum Mapper
  Considering Ancilla Sharing}.
\newblock
\newblock
\urldef\tempurl%
\url{https://doi.org/10.48550/ARXIV.1512.07402}
\showDOI{\tempurl}


\bibitem[Farhi et~al\mbox{.}(2014)]%
        {farhi2014quantum}
\bibfield{author}{\bibinfo{person}{Edward Farhi}, \bibinfo{person}{Jeffrey
  Goldstone}, {and} \bibinfo{person}{Sam Gutmann}.}
  \bibinfo{year}{2014}\natexlab{}.
\newblock \bibinfo{title}{A Quantum Approximate Optimization Algorithm}.
\newblock
\newblock
\urldef\tempurl%
\url{https://doi.org/10.48550/ARXIV.1411.4028}
\showDOI{\tempurl}


\bibitem[Farhoodi et~al\mbox{.}(2013)]%
        {FARHOODI2013}
\bibfield{author}{\bibinfo{person}{Roshanak Farhoodi}, \bibinfo{person}{Vahid
  Garousi}, \bibinfo{person}{Dietmar Pfahl}, {and} \bibinfo{person}{Jonathan
  Sillito}.} \bibinfo{year}{2013}\natexlab{}.
\newblock \showarticletitle{Development of scientific software: A systematic
  mapping, a bibliometrics study, and a paper repository}.
\newblock \bibinfo{journal}{\emph{International Journal of Software Engineering
  and Knowledge Engineering}} \bibinfo{volume}{23}, \bibinfo{number}{04}
  (\bibinfo{date}{May} \bibinfo{year}{2013}), \bibinfo{pages}{463--506}.
\newblock
\urldef\tempurl%
\url{https://doi.org/10.1142/s0218194013500137}
\showDOI{\tempurl}


\bibitem[Fingerhuth et~al\mbox{.}(2018)]%
        {fingerhuth_open_2018}
\bibfield{author}{\bibinfo{person}{Mark Fingerhuth}, \bibinfo{person}{Tomáš
  Babej}, {and} \bibinfo{person}{Peter Wittek}.}
  \bibinfo{year}{2018}\natexlab{}.
\newblock \showarticletitle{Open source software in quantum computing}.
\newblock \bibinfo{journal}{\emph{PLOS ONE}} \bibinfo{volume}{13},
  \bibinfo{number}{12} (\bibinfo{date}{Dec.} \bibinfo{year}{2018}),
  \bibinfo{pages}{e0208561}.
\newblock
\showISSN{1932-6203}
\urldef\tempurl%
\url{https://doi.org/10.1371/journal.pone.0208561}
\showDOI{\tempurl}


\bibitem[Fu et~al\mbox{.}(2020b)]%
        {fu_tutorial_2020}
\bibfield{author}{\bibinfo{person}{Peng Fu}, \bibinfo{person}{Kohei Kishida},
  \bibinfo{person}{Neil~J. Ross}, {and} \bibinfo{person}{Peter Selinger}.}
  \bibinfo{year}{2020}\natexlab{b}.
\newblock \showarticletitle{A {Tutorial} {Introduction} to {Quantum} {Circuit}
  {Programming} in {Dependently} {Typed} {Proto}-{Quipper}}. In
  \bibinfo{booktitle}{\emph{Reversible {Computation}}}
  \emph{(\bibinfo{series}{Lecture {Notes} in {Computer} {Science}})},
  \bibfield{editor}{\bibinfo{person}{Ivan Lanese} {and}
  \bibinfo{person}{Mariusz Rawski}} (Eds.). \bibinfo{publisher}{Springer
  International Publishing}, \bibinfo{address}{Cham},
  \bibinfo{pages}{153--168}.
\newblock
\showISBNx{978-3-030-52482-1}
\urldef\tempurl%
\url{https://doi.org/10.1007/978-3-030-52482-1_9}
\showDOI{\tempurl}


\bibitem[Fu et~al\mbox{.}(2020a)]%
        {fu_linear_2020}
\bibfield{author}{\bibinfo{person}{Peng Fu}, \bibinfo{person}{Kohei Kishida},
  {and} \bibinfo{person}{Peter Selinger}.} \bibinfo{year}{2020}\natexlab{a}.
\newblock \showarticletitle{Linear {Dependent} {Type} {Theory} for {Quantum}
  {Programming} {Languages}: {Extended} {Abstract}}. In
  \bibinfo{booktitle}{\emph{Proceedings of the 35th {Annual} {ACM}/{IEEE}
  {Symposium} on {Logic} in {Computer} {Science}}}
  \emph{(\bibinfo{series}{{LICS} '20})}. \bibinfo{publisher}{Association for
  Computing Machinery}, \bibinfo{address}{New York, NY, USA},
  \bibinfo{pages}{440--453}.
\newblock
\showISBNx{978-1-4503-7104-9}
\urldef\tempurl%
\url{https://doi.org/10.1145/3373718.3394765}
\showDOI{\tempurl}


\bibitem[Fu et~al\mbox{.}(2019)]%
        {fu_eqasm_2019}
\bibfield{author}{\bibinfo{person}{Xiang Fu}, \bibinfo{person}{Leon Riesebos},
  \bibinfo{person}{Adriaan Rol}, \bibinfo{person}{Jeroen van Straten},
  \bibinfo{person}{Hans van Someren}, \bibinfo{person}{Nader Khammassi},
  \bibinfo{person}{Imran Ashraf}, \bibinfo{person}{Raymond Vermeulen},
  \bibinfo{person}{Vincent Newsum}, \bibinfo{person}{Kelvin Loh},
  \bibinfo{person}{Jacob de Sterke}, \bibinfo{person}{Wouter Vlothuizen},
  \bibinfo{person}{Raymond Schouten}, \bibinfo{person}{Carmen Almudever},
  \bibinfo{person}{Leonardo DiCarlo}, {and} \bibinfo{person}{Koen Bertels}.}
  \bibinfo{year}{2019}\natexlab{}.
\newblock \showarticletitle{eQASM: An Executable Quantum Instruction Set
  Architecture}. In \bibinfo{booktitle}{\emph{2019 IEEE International Symposium
  on High Performance Computer Architecture (HPCA)}}.
  \bibinfo{publisher}{IEEE}, \bibinfo{address}{Washington, DC, USA},
  \bibinfo{pages}{224--237}.
\newblock
\urldef\tempurl%
\url{https://doi.org/10.1109/HPCA.2019.00040}
\showDOI{\tempurl}


\bibitem[Fu et~al\mbox{.}(2017)]%
        {Microarchitecture_Fu}
\bibfield{author}{\bibinfo{person}{Xiang Fu}, \bibinfo{person}{Michiel Rol},
  \bibinfo{person}{Cornelis Bultink}, \bibinfo{person}{Hans van Someren},
  \bibinfo{person}{Nader Khammassi}, \bibinfo{person}{Imran Ashraf},
  \bibinfo{person}{Raymond Vermeulen}, \bibinfo{person}{Jacob de Sterke},
  \bibinfo{person}{Wouter Vlothuizen}, \bibinfo{person}{Raymond Schouten},
  \bibinfo{person}{Carmen Almudever}, \bibinfo{person}{Leonardo DiCarlo}, {and}
  \bibinfo{person}{Koen Bertels}.} \bibinfo{year}{2017}\natexlab{}.
\newblock \showarticletitle{An Experimental Microarchitecture for a
  Superconducting Quantum Processor}. In \bibinfo{booktitle}{\emph{Proceedings
  of the 50th Annual IEEE/ACM International Symposium on Microarchitecture}}
  (Cambridge, Massachusetts) \emph{(\bibinfo{series}{MICRO-50 '17})}.
  \bibinfo{publisher}{Association for Computing Machinery},
  \bibinfo{address}{New York, NY, USA}, \bibinfo{pages}{813–825}.
\newblock
\showISBNx{9781450349529}
\urldef\tempurl%
\url{https://doi.org/10.1145/3123939.3123952}
\showDOI{\tempurl}


\bibitem[Fu et~al\mbox{.}(2021)]%
        {fu2021quingo}
\bibfield{author}{\bibinfo{person}{Xiang Fu}, \bibinfo{person}{Jintao Yu},
  \bibinfo{person}{Xing Su}, \bibinfo{person}{Hanru Jiang},
  \bibinfo{person}{Hua Wu}, \bibinfo{person}{Fucheng Cheng},
  \bibinfo{person}{Xi Deng}, \bibinfo{person}{Jinrong Zhang},
  \bibinfo{person}{Lei Jin}, \bibinfo{person}{Yihang Yang}, \bibinfo{person}{Le
  Xu}, \bibinfo{person}{Chunchao Hu}, \bibinfo{person}{Anqi Huang},
  \bibinfo{person}{Guangyao Huang}, \bibinfo{person}{Xiaogang Qiang},
  \bibinfo{person}{Mingtang Deng}, \bibinfo{person}{Ping Xu},
  \bibinfo{person}{Weixia Xu}, \bibinfo{person}{Wanwei Liu},
  \bibinfo{person}{Yu Zhang}, \bibinfo{person}{Yuxin Deng},
  \bibinfo{person}{Junjie Wu}, {and} \bibinfo{person}{Yuan Feng}.}
  \bibinfo{year}{2021}\natexlab{}.
\newblock \showarticletitle{Quingo: A Programming Framework for Heterogeneous
  Quantum-Classical Computing with NISQ Features}.
\newblock \bibinfo{journal}{\emph{ACM Transactions on Quantum Computing}}
  \bibinfo{volume}{2}, \bibinfo{number}{4}, Article \bibinfo{articleno}{19}
  (\bibinfo{date}{12} \bibinfo{year}{2021}), \bibinfo{numpages}{37}~pages.
\newblock
\showISSN{2643-6809}
\urldef\tempurl%
\url{https://doi.org/10.1145/3483528}
\showDOI{\tempurl}


\bibitem[Garhwal et~al\mbox{.}(2021)]%
        {garhwal_quantum_2021}
\bibfield{author}{\bibinfo{person}{Sunita Garhwal}, \bibinfo{person}{Maryam
  Ghorani}, {and} \bibinfo{person}{Amir Ahmad}.}
  \bibinfo{year}{2021}\natexlab{}.
\newblock \showarticletitle{Quantum {Programming} {Language}: {A} {Systematic}
  {Review} of {Research} {Topic} and {Top} {Cited} {Languages}}.
\newblock \bibinfo{journal}{\emph{Archives of Computational Methods in
  Engineering}} \bibinfo{volume}{28}, \bibinfo{number}{2}
  (\bibinfo{date}{March} \bibinfo{year}{2021}), \bibinfo{pages}{289--310}.
\newblock
\showISSN{1886-1784}
\urldef\tempurl%
\url{https://doi.org/10.1007/s11831-019-09372-6}
\showDOI{\tempurl}


\bibitem[Gay and Nagarajan(2005)]%
        {gay_cqp_2004}
\bibfield{author}{\bibinfo{person}{Simon Gay} {and} \bibinfo{person}{Rajagopal
  Nagarajan}.} \bibinfo{year}{2005}\natexlab{}.
\newblock \showarticletitle{Communicating quantum processes}. In
  \bibinfo{booktitle}{\emph{Proceedings of the 32nd ACM SIGPLAN-SIGACT
  Symposium on Principles of Programming languages}}. \bibinfo{publisher}{ACM},
  \bibinfo{address}{Long Beach, California, USA}, \bibinfo{pages}{145--157}.
\newblock


\bibitem[Gay(2006)]%
        {simon_j_gay_quantum_2006}
\bibfield{author}{\bibinfo{person}{Simon~J. Gay}.}
  \bibinfo{year}{2006}\natexlab{}.
\newblock \showarticletitle{Quantum programming languages: survey and
  bibliography}.
\newblock \bibinfo{journal}{\emph{Mathematical Structures in Computer Science}}
  \bibinfo{volume}{16}, \bibinfo{number}{4} (\bibinfo{year}{2006}),
  \bibinfo{pages}{581–600}.
\newblock
\urldef\tempurl%
\url{https://doi.org/10.1017/S0960129506005378}
\showDOI{\tempurl}


\bibitem[Gill et~al\mbox{.}(2021)]%
        {gill_quantum_2021}
\bibfield{author}{\bibinfo{person}{Sukhpal~Singh Gill}, \bibinfo{person}{Adarsh
  Kumar}, \bibinfo{person}{Harvinder Singh}, \bibinfo{person}{Manmeet Singh},
  \bibinfo{person}{Kamalpreet Kaur}, \bibinfo{person}{Muhammad Usman}, {and}
  \bibinfo{person}{Rajkumar Buyya}.} \bibinfo{year}{2021}\natexlab{}.
\newblock \showarticletitle{Quantum computing: {A} taxonomy, systematic review
  and future directions}.
\newblock \bibinfo{journal}{\emph{Software: Practice and Experience}}
  \bibinfo{volume}{52}, \bibinfo{number}{1} (\bibinfo{date}{July}
  \bibinfo{year}{2021}), \bibinfo{pages}{66--114}.
\newblock
\showISSN{0038-0644, 1097-024X}
\urldef\tempurl%
\url{https://doi.org/10.1002/spe.3039}
\showDOI{\tempurl}


\bibitem[Gonz\'alez-Garc\'{\i}a et~al\mbox{.}(2022)]%
        {gonzales-garcia2022error}
\bibfield{author}{\bibinfo{person}{Guillermo Gonz\'alez-Garc\'{\i}a},
  \bibinfo{person}{Rahul Trivedi}, {and} \bibinfo{person}{J.~Ignacio Cirac}.}
  \bibinfo{year}{2022}\natexlab{}.
\newblock \showarticletitle{Error Propagation in {NISQ} Devices for Solving
  Classical Optimization Problems}.
\newblock \bibinfo{journal}{\emph{PRX Quantum}}  \bibinfo{volume}{3}
  (\bibinfo{date}{Dec} \bibinfo{year}{2022}), \bibinfo{pages}{040326}.
\newblock
Issue 4.
\urldef\tempurl%
\url{https://doi.org/10.1103/PRXQuantum.3.040326}
\showDOI{\tempurl}


\bibitem[Gottesman(1997)]%
        {gottesman1997stabilizer}
\bibfield{author}{\bibinfo{person}{Daniel Gottesman}.}
  \bibinfo{year}{1997}\natexlab{}.
\newblock \bibinfo{booktitle}{\emph{Stabilizer codes and quantum error
  correction}}.
\newblock \bibinfo{publisher}{California Institute of Technology},
  \bibinfo{address}{Pasadena, California}.
\newblock


\bibitem[Green et~al\mbox{.}(2013)]%
        {green_quipper_2013}
\bibfield{author}{\bibinfo{person}{Alexander~S. Green},
  \bibinfo{person}{Peter~LeFanu Lumsdaine}, \bibinfo{person}{Neil~J. Ross},
  \bibinfo{person}{Peter Selinger}, {and} \bibinfo{person}{Benoît Valiron}.}
  \bibinfo{year}{2013}\natexlab{}.
\newblock \showarticletitle{Quipper: a scalable quantum programming language}.
  In \bibinfo{booktitle}{\emph{Proceedings of the 34th {ACM} {SIGPLAN}
  {Conference} on {Programming} {Language} {Design} and {Implementation}}}
  \emph{(\bibinfo{series}{{PLDI} '13})}. \bibinfo{publisher}{Association for
  Computing Machinery}, \bibinfo{address}{New York, NY, USA},
  \bibinfo{pages}{333--342}.
\newblock
\showISBNx{978-1-4503-2014-6}
\urldef\tempurl%
\url{https://doi.org/10.1145/2491956.2462177}
\showDOI{\tempurl}


\bibitem[Gyongyosi and Imre(2019)]%
        {Gyongyosi2019}
\bibfield{author}{\bibinfo{person}{Laszlo Gyongyosi} {and}
  \bibinfo{person}{Sandor Imre}.} \bibinfo{year}{2019}\natexlab{}.
\newblock \showarticletitle{A Survey on quantum computing technology}.
\newblock \bibinfo{journal}{\emph{Computer Science Review}}
  \bibinfo{volume}{31} (\bibinfo{date}{Feb.} \bibinfo{year}{2019}),
  \bibinfo{pages}{51--71}.
\newblock
\urldef\tempurl%
\url{https://doi.org/10.1016/j.cosrev.2018.11.002}
\showDOI{\tempurl}


\bibitem[Haferkamp et~al\mbox{.}(2020)]%
        {Haferkamp2020}
\bibfield{author}{\bibinfo{person}{J. Haferkamp}, \bibinfo{person}{D.
  Hangleiter}, \bibinfo{person}{A. Bouland}, \bibinfo{person}{B. Fefferman},
  \bibinfo{person}{J. Eisert}, {and} \bibinfo{person}{J. Bermejo-Vega}.}
  \bibinfo{year}{2020}\natexlab{}.
\newblock \showarticletitle{Closing gaps of a quantum advantage with short-time
  {H}amiltonian dynamics}.
\newblock \bibinfo{journal}{\emph{Phys. Rev. Lett.}}  \bibinfo{volume}{125}
  (\bibinfo{year}{2020}), \bibinfo{pages}{250501}.
\newblock
\urldef\tempurl%
\url{https://doi.org/10.1103/PhysRevLett.125.250501}
\showDOI{\tempurl}


\bibitem[Hager and Wellein(2010)]%
        {hager2010introduction}
\bibfield{author}{\bibinfo{person}{Georg Hager} {and} \bibinfo{person}{Gerhard
  Wellein}.} \bibinfo{year}{2010}\natexlab{}.
\newblock \bibinfo{booktitle}{\emph{Introduction to High Performance Computing
  for Scientists and Engineers}}.
\newblock \bibinfo{publisher}{CRC Press}, \bibinfo{address}{Broken Sound
  Parkway NW, FL, USA}.
\newblock
\showISBNx{9781439811931}
\showLCCN{2010009624}
\urldef\tempurl%
\url{https://books.google.de/books?id=rkWPojgfeM8C}
\showURL{%
\tempurl}


\bibitem[Hey(1996)]%
        {Hey_HPC}
\bibfield{author}{\bibinfo{person}{Anthony J~G Hey}.}
  \bibinfo{year}{1996}\natexlab{}.
\newblock \bibinfo{title}{{High Performance Computing: Past, Present and
  Future}}.
\newblock
\newblock
\urldef\tempurl%
\url{https://doi.org/10.5170/CERN-1996-008.217}
\showDOI{\tempurl}


\bibitem[Hidary(2021)]%
        {hidaryQuantumComputingApplied2021}
\bibfield{author}{\bibinfo{person}{Jack~D. Hidary}.}
  \bibinfo{year}{2021}\natexlab{}.
\newblock \bibinfo{booktitle}{\emph{Quantum {{Computing}}: {{An Applied
  Approach}}}}.
\newblock \bibinfo{publisher}{{Springer International Publishing}},
  \bibinfo{address}{{Cham}}.
\newblock
\urldef\tempurl%
\url{https://doi.org/10.1007/978-3-030-83274-2}
\showDOI{\tempurl}


\bibitem[Hietala(2016)]%
        {kesha_hietala_quantum_2016}
\bibfield{author}{\bibinfo{person}{Kesha Hietala}.}
  \bibinfo{year}{2016}\natexlab{}.
\newblock \bibinfo{title}{Quantum {Programming} {Languages}}.
\newblock
\newblock
\urldef\tempurl%
\url{https://khieta.github.io/files/drafts/quantum-pl-survey.pdf}
\showURL{%
\tempurl}


\bibitem[Humble et~al\mbox{.}(2021)]%
        {humbleQuantumComputers2021}
\bibfield{author}{\bibinfo{person}{Travis~S. Humble},
  \bibinfo{person}{Alexander McCaskey}, \bibinfo{person}{Dmitry~I. Lyakh},
  \bibinfo{person}{Meenambika Gowrishankar}, \bibinfo{person}{Albert Frisch},
  {and} \bibinfo{person}{Thomas Monz}.} \bibinfo{year}{2021}\natexlab{}.
\newblock \showarticletitle{Quantum Computers for High-Performance Computing}.
\newblock \bibinfo{journal}{\emph{IEEE Micro}} \bibinfo{volume}{41},
  \bibinfo{number}{5} (\bibinfo{date}{sep} \bibinfo{year}{2021}),
  \bibinfo{pages}{15–23}.
\newblock
\showISSN{0272-1732}
\urldef\tempurl%
\url{https://doi.org/10.1109/MM.2021.3099140}
\showDOI{\tempurl}


\bibitem[Ihde et~al\mbox{.}(2022)]%
        {Ihde2022}
\bibfield{author}{\bibinfo{person}{Nina Ihde}, \bibinfo{person}{Paula Marten},
  \bibinfo{person}{Ahmed Eleliemy}, \bibinfo{person}{Gabrielle Poerwawinata},
  \bibinfo{person}{Pedro Silva}, \bibinfo{person}{Ilin Tolovski},
  \bibinfo{person}{Florina~M. Ciorba}, {and} \bibinfo{person}{Tilmann Rabl}.}
  \bibinfo{year}{2022}\natexlab{}.
\newblock \showarticletitle{A Survey of~Big Data, High Performance Computing,
  and~Machine Learning Benchmarks}.
\newblock In \bibinfo{booktitle}{\emph{Lecture Notes in Computer Science}}.
  \bibinfo{publisher}{Springer International Publishing},
  \bibinfo{address}{Springer International Publishing AG, Birkhäuser Verlag
  AG, Springer, Cham Springer International}, \bibinfo{pages}{98--118}.
\newblock
\urldef\tempurl%
\url{https://doi.org/10.1007/978-3-030-94437-7_7}
\showDOI{\tempurl}


\bibitem[Izaac(2023)]%
        {pennylane_catlyst}
\bibfield{author}{\bibinfo{person}{Josh Izaac}.}
  \bibinfo{year}{2023}\natexlab{}.
\newblock \bibinfo{booktitle}{\emph{Pennylane blog - introducing catalyst:
  Quantum just-in-time compilation}}.
\newblock PennyLane.
\newblock
\urldef\tempurl%
\url{https://pennylane.ai/blog/2023/03/introducing-catalyst-quantum-just-in-time-compilation/}
\showURL{%
Retrieved April 14, 2023 from \tempurl}


\bibitem[JavadiAbhari et~al\mbox{.}(2015)]%
        {javadiabhari_scaffcc_2015}
\bibfield{author}{\bibinfo{person}{Ali JavadiAbhari}, \bibinfo{person}{Shruti
  Patil}, \bibinfo{person}{Daniel Kudrow}, \bibinfo{person}{Jeff Heckey},
  \bibinfo{person}{Alexey Lvov}, \bibinfo{person}{Frederic~T. Chong}, {and}
  \bibinfo{person}{Margaret Martonosi}.} \bibinfo{year}{2015}\natexlab{}.
\newblock \showarticletitle{{ScaffCC}: {Scalable} {Compilation} and {Analysis}
  of {Quantum} {Programs}}.
\newblock \bibinfo{journal}{\emph{Parallel Comput.}}  \bibinfo{volume}{45}
  (\bibinfo{date}{June} \bibinfo{year}{2015}), \bibinfo{pages}{2--17}.
\newblock
\showISSN{01678191}
\urldef\tempurl%
\url{https://doi.org/10.1016/j.parco.2014.12.001}
\showDOI{\tempurl}


\bibitem[Johanson and Hasselbring(2018)]%
        {Johanson2018}
\bibfield{author}{\bibinfo{person}{Arne Johanson} {and}
  \bibinfo{person}{Wilhelm Hasselbring}.} \bibinfo{year}{2018}\natexlab{}.
\newblock \showarticletitle{Software Engineering for Computational Science:
  Past, Present, Future}.
\newblock \bibinfo{journal}{\emph{Computing in Science \& Engineering}}
  \bibinfo{volume}{20}, \bibinfo{number}{2} (\bibinfo{year}{2018}),
  \bibinfo{pages}{90--109}.
\newblock
\urldef\tempurl%
\url{https://doi.org/10.1109/MCSE.2018.021651343}
\showDOI{\tempurl}


\bibitem[Jorrand(2007)]%
        {jorrand_programmers_2007}
\bibfield{author}{\bibinfo{person}{Philippe Jorrand}.}
  \bibinfo{year}{2007}\natexlab{}.
\newblock \showarticletitle{A {Programmer}’s {Survey} of the {Quantum}
  {Computing} {Paradigm}}.
\newblock \bibinfo{journal}{\emph{International Journal of Nuclear and Quantum
  Engineering}} \bibinfo{volume}{1}, \bibinfo{number}{8}
  (\bibinfo{year}{2007}), \bibinfo{pages}{409 -- 415}.
\newblock
\showISSN{eISSN: 1307-6892}
\urldef\tempurl%
\url{https://publications.waset.org/vol/8}
\showURL{%
\tempurl}


\bibitem[Jorrand and Lalire(2004)]%
        {jorrand2003}
\bibfield{author}{\bibinfo{person}{Philippe Jorrand} {and}
  \bibinfo{person}{Marie Lalire}.} \bibinfo{year}{2004}\natexlab{}.
\newblock \showarticletitle{Toward a quantum process algebra}. In
  \bibinfo{booktitle}{\emph{Proceedings of the 1st Conference on Computing
  Frontiers}}. \bibinfo{publisher}{ACM}, \bibinfo{address}{New York, NY, USA},
  \bibinfo{pages}{111--119}.
\newblock
\urldef\tempurl%
\url{https://doi.org/10.1145/977091.977108}
\showDOI{\tempurl}


\bibitem[Jozsa(1994)]%
        {Jozsa1994}
\bibfield{author}{\bibinfo{person}{Richard Jozsa}.}
  \bibinfo{year}{1994}\natexlab{}.
\newblock \showarticletitle{Fidelity for Mixed Quantum States}.
\newblock \bibinfo{journal}{\emph{Journal of Modern Optics}}
  \bibinfo{volume}{41}, \bibinfo{number}{12} (\bibinfo{year}{1994}),
  \bibinfo{pages}{2315--2323}.
\newblock
\urldef\tempurl%
\url{https://doi.org/10.1080/09500349414552171}
\showDOI{\tempurl}


\bibitem[Khammassi et~al\mbox{.}(2022)]%
        {khammassi_openql_2022}
\bibfield{author}{\bibinfo{person}{Nader Khammassi}, \bibinfo{person}{Imran
  Ashraf}, \bibinfo{person}{Hans van Someren}, \bibinfo{person}{Razvan Nane},
  \bibinfo{person}{Anna Krol}, \bibinfo{person}{Adriaan Rol},
  \bibinfo{person}{Lingling Lao}, \bibinfo{person}{Koen Bertels}, {and}
  \bibinfo{person}{Carmen Almudever}.} \bibinfo{year}{2022}\natexlab{}.
\newblock \showarticletitle{{OpenQL}: {A} {Portable} {Quantum} {Programming}
  {Framework} for {Quantum} {Accelerators}}.
\newblock \bibinfo{journal}{\emph{ACM Journal on Emerging Technologies in
  Computing Systems}} \bibinfo{volume}{18}, \bibinfo{number}{1}
  (\bibinfo{date}{Jan.} \bibinfo{year}{2022}), \bibinfo{pages}{1--24}.
\newblock
\showISSN{1550-4832, 1550-4840}
\urldef\tempurl%
\url{https://doi.org/10.1145/3474222}
\showDOI{\tempurl}


\bibitem[Khammassi et~al\mbox{.}(2018)]%
        {khammassi_cqasm_2018}
\bibfield{author}{\bibinfo{person}{Nader Khammassi}, \bibinfo{person}{Gian~G.
  Guerreschi}, \bibinfo{person}{Imran. Ashraf}, \bibinfo{person}{Justin~W.
  Hogaboam}, \bibinfo{person}{Carmen~G. Almudever}, {and} \bibinfo{person}{Koen
  Bertels}.} \bibinfo{year}{2018}\natexlab{}.
\newblock \bibinfo{title}{{cQASM} v1.0: {Towards} a {Common} {Quantum}
  {Assembly} {Language}}.
\newblock
\newblock
\urldef\tempurl%
\url{http://arxiv.org/abs/1805.09607}
\showURL{%
\tempurl}


\bibitem[Killoran et~al\mbox{.}(2019)]%
        {killoran_strawberry_2019}
\bibfield{author}{\bibinfo{person}{Nathan Killoran}, \bibinfo{person}{Josh
  Izaac}, \bibinfo{person}{Nicolás Quesada}, \bibinfo{person}{Ville Bergholm},
  \bibinfo{person}{Matthew Amy}, {and} \bibinfo{person}{Christian Weedbrook}.}
  \bibinfo{year}{2019}\natexlab{}.
\newblock \showarticletitle{Strawberry {Fields}: {A} {Software} {Platform} for
  {Photonic} {Quantum} {Computing}}.
\newblock \bibinfo{journal}{\emph{Quantum}}  \bibinfo{volume}{3}
  (\bibinfo{date}{March} \bibinfo{year}{2019}), \bibinfo{pages}{129}.
\newblock
\showISSN{2521-327X}
\urldef\tempurl%
\url{https://doi.org/10.22331/q-2019-03-11-129}
\showDOI{\tempurl}


\bibitem[Knill(1996)]%
        {Knill_1996}
\bibfield{author}{\bibinfo{person}{Emmanuel Knill}.}
  \bibinfo{year}{1996}\natexlab{}.
\newblock \bibinfo{booktitle}{\emph{Conventions for quantum pseudocode}}.
\newblock \bibinfo{type}{{T}echnical {R}eport} LA-UR-96-2724.
  \bibinfo{institution}{Los Alamos National Lab.}
\newblock
\urldef\tempurl%
\url{https://doi.org/10.2172/366453}
\showDOI{\tempurl}


\bibitem[Knizia and Chan(2012)]%
        {Knizia2012}
\bibfield{author}{\bibinfo{person}{Gerald Knizia} {and} \bibinfo{person}{Garnet
  Kin-Lic Chan}.} \bibinfo{year}{2012}\natexlab{}.
\newblock \showarticletitle{Density Matrix Embedding: A Simple Alternative to
  Dynamical Mean-Field Theory}.
\newblock \bibinfo{journal}{\emph{Phys. Rev. Lett.}}  \bibinfo{volume}{109}
  (\bibinfo{date}{Nov} \bibinfo{year}{2012}), \bibinfo{pages}{186404}.
\newblock
Issue 18.
\urldef\tempurl%
\url{https://doi.org/10.1103/PhysRevLett.109.186404}
\showDOI{\tempurl}


\bibitem[Koczor(2021)]%
        {koczor2021exponential}
\bibfield{author}{\bibinfo{person}{B\'alint Koczor}.}
  \bibinfo{year}{2021}\natexlab{}.
\newblock \showarticletitle{Exponential Error Suppression for Near-Term Quantum
  Devices}.
\newblock \bibinfo{journal}{\emph{Phys. Rev. X}}  \bibinfo{volume}{11}
  (\bibinfo{date}{Sep} \bibinfo{year}{2021}), \bibinfo{pages}{031057}.
\newblock
Issue 3.
\urldef\tempurl%
\url{https://doi.org/10.1103/PhysRevX.11.031057}
\showDOI{\tempurl}


\bibitem[Krantz et~al\mbox{.}(2019)]%
        {Krantz2019}
\bibfield{author}{\bibinfo{person}{Philip Krantz}, \bibinfo{person}{Morten
  Kjaergaard}, \bibinfo{person}{Fei Yan}, \bibinfo{person}{Terry~P Orlando},
  \bibinfo{person}{Simon Gustavsson}, {and} \bibinfo{person}{William~D
  Oliver}.} \bibinfo{year}{2019}\natexlab{}.
\newblock \showarticletitle{A quantum engineer{\textquotesingle}s guide to
  superconducting qubits}.
\newblock \bibinfo{journal}{\emph{Applied Physics Reviews}}
  \bibinfo{volume}{6}, \bibinfo{number}{2} (\bibinfo{date}{June}
  \bibinfo{year}{2019}), \bibinfo{pages}{021318}.
\newblock
\urldef\tempurl%
\url{https://doi.org/10.1063/1.5089550}
\showDOI{\tempurl}


\bibitem[Krämer et~al\mbox{.}(2018)]%
        {kramer_quantumopticsjl_2018}
\bibfield{author}{\bibinfo{person}{Sebastian Krämer}, \bibinfo{person}{David
  Plankensteiner}, \bibinfo{person}{Laurin Ostermann}, {and}
  \bibinfo{person}{Helmut Ritsch}.} \bibinfo{year}{2018}\natexlab{}.
\newblock \showarticletitle{{QuantumOptics}.jl: {A} {Julia} framework for
  simulating open quantum systems}.
\newblock \bibinfo{journal}{\emph{Computer Physics Communications}}
  \bibinfo{volume}{227} (\bibinfo{date}{June} \bibinfo{year}{2018}),
  \bibinfo{pages}{109--116}.
\newblock
\showISSN{00104655}
\urldef\tempurl%
\url{https://doi.org/10.1016/j.cpc.2018.02.004}
\showDOI{\tempurl}


\bibitem[Ladd et~al\mbox{.}(2010)]%
        {Ladd2010}
\bibfield{author}{\bibinfo{person}{Thaddeus~D Ladd}, \bibinfo{person}{Fedor
  Jelezko}, \bibinfo{person}{Raymond Laflamme}, \bibinfo{person}{Yasunobu
  Nakamura}, \bibinfo{person}{Christopher Monroe}, {and}
  \bibinfo{person}{Jeremy~Lloyd O’Brien}.} \bibinfo{year}{2010}\natexlab{}.
\newblock \showarticletitle{Quantum computers}.
\newblock \bibinfo{journal}{\emph{Nature}} \bibinfo{volume}{464},
  \bibinfo{number}{7285} (\bibinfo{date}{March} \bibinfo{year}{2010}),
  \bibinfo{pages}{45--53}.
\newblock
\urldef\tempurl%
\url{https://doi.org/10.1038/nature08812}
\showDOI{\tempurl}


\bibitem[Laguna et~al\mbox{.}(2019)]%
        {Laguna2019}
\bibfield{author}{\bibinfo{person}{Ignacio Laguna}, \bibinfo{person}{Ryan
  Marshall}, \bibinfo{person}{Kathryn Mohror}, \bibinfo{person}{Martin
  Ruefenacht}, \bibinfo{person}{Anthony Skjellum}, {and}
  \bibinfo{person}{Nawrin Sultana}.} \bibinfo{year}{2019}\natexlab{}.
\newblock \showarticletitle{A large-scale study of {MPI} usage in open-source
  {HPC} applications}. In \bibinfo{booktitle}{\emph{Proceedings of the
  International Conference for High Performance Computing, Networking, Storage
  and Analysis}} (Denver, Colorado). \bibinfo{publisher}{{ACM}},
  \bibinfo{address}{New York, NY, USA}, Article \bibinfo{articleno}{31},
  \bibinfo{numpages}{14}~pages.
\newblock
\urldef\tempurl%
\url{https://doi.org/10.1145/3295500.3356176}
\showDOI{\tempurl}


\bibitem[Lalire and Jorrand(2004)]%
        {lalire_qpalg_2004}
\bibfield{author}{\bibinfo{person}{Marie Lalire} {and}
  \bibinfo{person}{Philippe Jorrand}.} \bibinfo{year}{2004}\natexlab{}.
\newblock \bibinfo{title}{{A} {Process} {Algebraic} {Approach} to {Concurrent}
  and {Distributed} {Quantum} {Computation}: {Operational} {Semantics}}.
\newblock
\newblock
\urldef\tempurl%
\url{http://arxiv.org/abs/quant-ph/0407005}
\showURL{%
\tempurl}


\bibitem[Lapets et~al\mbox{.}(2013)]%
        {lapets_quafl_2013}
\bibfield{author}{\bibinfo{person}{Andrei Lapets}, \bibinfo{person}{Marcus~P.
  da Silva}, \bibinfo{person}{Mike Thome}, \bibinfo{person}{Aaron Adler},
  \bibinfo{person}{Jacob Beal}, {and} \bibinfo{person}{Martin Roetteler}.}
  \bibinfo{year}{2013}\natexlab{}.
\newblock \showarticletitle{{QuaFL}: a typed {DSL} for quantum programming}. In
  \bibinfo{booktitle}{\emph{Proceedings of the 1st annual workshop on
  {Functional} programming concepts in domain-specific languages - {FPCDSL}
  '13}}. \bibinfo{publisher}{ACM Press}, \bibinfo{address}{Boston,
  Massachusetts, USA}, \bibinfo{pages}{19}.
\newblock
\showISBNx{978-1-4503-2380-2}
\urldef\tempurl%
\url{https://doi.org/10.1145/2505351.2505357}
\showDOI{\tempurl}


\bibitem[Liu et~al\mbox{.}(2022)]%
        {diamond_qubit}
\bibfield{author}{\bibinfo{person}{Ya-Chi Liu}, \bibinfo{person}{Yi-Chung
  Dzeng}, {and} \bibinfo{person}{Chao-Cheng Ting}.}
  \bibinfo{year}{2022}\natexlab{}.
\newblock \showarticletitle{Nitrogen Vacancy-Centered Diamond Qubit: The
  fabrication, design, and application in quantum computing}.
\newblock \bibinfo{journal}{\emph{IEEE Nanotechnology Magazine}}
  \bibinfo{volume}{16}, \bibinfo{number}{4} (\bibinfo{year}{2022}),
  \bibinfo{pages}{37--43}.
\newblock
\urldef\tempurl%
\url{https://doi.org/10.1109/MNANO.2022.3175405}
\showDOI{\tempurl}


\bibitem[Lloyd et~al\mbox{.}(2014)]%
        {lloyd2014quantum}
\bibfield{author}{\bibinfo{person}{Seth Lloyd}, \bibinfo{person}{Masoud
  Mohseni}, {and} \bibinfo{person}{Patrick Rebentrost}.}
  \bibinfo{year}{2014}\natexlab{}.
\newblock \showarticletitle{Quantum principal component analysis}.
\newblock \bibinfo{journal}{\emph{Nature Physics}} \bibinfo{volume}{10},
  \bibinfo{number}{9} (\bibinfo{year}{2014}), \bibinfo{pages}{631--633}.
\newblock
\urldef\tempurl%
\url{https://doi.org/10.1038/nphys3029}
\showDOI{\tempurl}


\bibitem[Mauerer(2005)]%
        {mauerer_cqpl_2005}
\bibfield{author}{\bibinfo{person}{Wolfgang Mauerer}.}
  \bibinfo{year}{2005}\natexlab{}.
\newblock \bibinfo{title}{Semantics and simulation of communication in quantum
  programming}.
\newblock
\newblock
\urldef\tempurl%
\url{http://arxiv.org/abs/quant-ph/0511145}
\showURL{%
\tempurl}


\bibitem[Maymin(1997)]%
        {maymin_extending_1997}
\bibfield{author}{\bibinfo{person}{Philip Maymin}.}
  \bibinfo{year}{1997}\natexlab{}.
\newblock \bibinfo{title}{Extending the {Lambda} {Calculus} to {Express}
  {Randomized} and {Quantumized} {Algorithms}}.
\newblock
\newblock
\urldef\tempurl%
\url{http://arxiv.org/abs/quant-ph/9612052}
\showURL{%
\tempurl}


\bibitem[Mccaskey et~al\mbox{.}(2021)]%
        {mccaskey_extending_2021}
\bibfield{author}{\bibinfo{person}{Alexander Mccaskey}, \bibinfo{person}{Thien
  Nguyen}, \bibinfo{person}{Anthony Santana}, \bibinfo{person}{Daniel
  Claudino}, \bibinfo{person}{Tyler Kharazi}, {and} \bibinfo{person}{Hal
  Finkel}.} \bibinfo{year}{2021}\natexlab{}.
\newblock \showarticletitle{Extending C++ for Heterogeneous Quantum-Classical
  Computing}.
\newblock \bibinfo{journal}{\emph{ACM Transactions on Quantum Computing}}
  \bibinfo{volume}{2}, \bibinfo{number}{2}, Article \bibinfo{articleno}{6}
  (\bibinfo{date}{jul} \bibinfo{year}{2021}), \bibinfo{numpages}{36}~pages.
\newblock
\showISSN{2643-6809}
\urldef\tempurl%
\url{https://doi.org/10.1145/3462670}
\showDOI{\tempurl}


\bibitem[McCaskey et~al\mbox{.}(2020)]%
        {mccaskey_xacc_2020}
\bibfield{author}{\bibinfo{person}{Alexander~J. McCaskey},
  \bibinfo{person}{Dmitry~I. Lyakh}, \bibinfo{person}{Eugene~F. Dumitrescu},
  \bibinfo{person}{Sarah~S. Powers}, {and} \bibinfo{person}{Travis~S. Humble}.}
  \bibinfo{year}{2020}\natexlab{}.
\newblock \showarticletitle{{XACC}: a system-level software infrastructure for
  heterogeneous quantum–classical computing}.
\newblock \bibinfo{journal}{\emph{Quantum Science and Technology}}
  \bibinfo{volume}{5} (\bibinfo{year}{2020}), \bibinfo{pages}{024002}.
\newblock
\showISSN{2058-9565}
\urldef\tempurl%
\url{https://doi.org/10.1088/2058-9565/ab6bf6}
\showDOI{\tempurl}


\bibitem[Mermin(2007)]%
        {merminQuantumComputerScience2007}
\bibfield{author}{\bibinfo{person}{Nathaniel~David Mermin}.}
  \bibinfo{year}{2007}\natexlab{}.
\newblock \bibinfo{booktitle}{\emph{Quantum {{Computer Science}}: {{An
  Introduction}}} (\bibinfo{edition}{1} ed.)}.
\newblock \bibinfo{publisher}{{Cambridge University Press}},
  \bibinfo{address}{Cambridge}.
\newblock
\urldef\tempurl%
\url{https://doi.org/10.1017/CBO9780511813870}
\showDOI{\tempurl}


\bibitem[{Message Passing Interface Forum}(2021)]%
        {mpi40}
\bibfield{author}{\bibinfo{person}{{Message Passing Interface Forum}}.}
  \bibinfo{year}{2021}\natexlab{}.
\newblock \bibinfo{booktitle}{\emph{{MPI}: A Message-Passing Interface Standard
  Version 4.0}}.
\newblock Message Passing Interface Forum.
\newblock
\urldef\tempurl%
\url{https://www.mpi-forum.org/docs/mpi-4.0/mpi40-report.pdf}
\showURL{%
\tempurl}


\bibitem[Mintz et~al\mbox{.}(2020)]%
        {mintz_qcor_2019}
\bibfield{author}{\bibinfo{person}{Tiffany~M Mintz},
  \bibinfo{person}{Alexander~J Mccaskey}, \bibinfo{person}{Eugene~F
  Dumitrescu}, \bibinfo{person}{Shirley~V Moore}, \bibinfo{person}{Sarah
  Powers}, {and} \bibinfo{person}{Pavel Lougovski}.}
  \bibinfo{year}{2020}\natexlab{}.
\newblock \showarticletitle{Qcor: A language extension specification for the
  heterogeneous quantum-classical model of computation}.
\newblock \bibinfo{journal}{\emph{ACM Journal on Emerging Technologies in
  Computing Systems (JETC)}} \bibinfo{volume}{16}, \bibinfo{number}{2}
  (\bibinfo{year}{2020}), \bibinfo{pages}{1--17}.
\newblock
\urldef\tempurl%
\url{https://doi.org/10.1145/3380964}
\showDOI{\tempurl}


\bibitem[Miszczak(2011)]%
        {miszczak_models_2011}
\bibfield{author}{\bibinfo{person}{Jaroslaw~Adam Miszczak}.}
  \bibinfo{year}{2011}\natexlab{}.
\newblock \showarticletitle{Models of quantum computation and quantum
  programming languages}.
\newblock \bibinfo{journal}{\emph{Bulletin of the Polish Academy of Sciences:
  Technical Sciences}} \bibinfo{volume}{59}, \bibinfo{number}{3}
  (\bibinfo{year}{2011}), \bibinfo{pages}{305--324}.
\newblock
\showISSN{0239-7528}
\urldef\tempurl%
\url{https://doi.org/10.2478/v10175-011-0039-5}
\showDOI{\tempurl}


\bibitem[Mlnar{\i}k(2006)]%
        {mlnarik2006introduction}
\bibfield{author}{\bibinfo{person}{Hynek Mlnar{\i}k}.}
  \bibinfo{year}{2006}\natexlab{}.
\newblock \bibinfo{title}{Introduction to LanQ--an Imperative Quantum
  Programming Language}.
\newblock
\newblock
\urldef\tempurl%
\url{https://lanq.sourceforge.net/doc/introToLanQ.pdf}
\showURL{%
\tempurl}


\bibitem[Nam et~al\mbox{.}(2018)]%
        {nam_automated_2018}
\bibfield{author}{\bibinfo{person}{Yunseong Nam}, \bibinfo{person}{Neil~J.
  Ross}, \bibinfo{person}{Yuan Su}, \bibinfo{person}{Andrew~M. Childs}, {and}
  \bibinfo{person}{Dmitri Maslov}.} \bibinfo{year}{2018}\natexlab{}.
\newblock \showarticletitle{Automated optimization of large quantum circuits
  with continuous parameters}.
\newblock \bibinfo{journal}{\emph{npj Quantum Information}}
  \bibinfo{volume}{4}, \bibinfo{number}{1} (\bibinfo{date}{May}
  \bibinfo{year}{2018}), \bibinfo{pages}{1--12}.
\newblock
\showISSN{2056-6387}
\urldef\tempurl%
\url{https://doi.org/10.1038/s41534-018-0072-4}
\showDOI{\tempurl}


\bibitem[Nguyen and McCaskey(2022)]%
        {nguyen2022extending}
\bibfield{author}{\bibinfo{person}{Thien Nguyen} {and}
  \bibinfo{person}{Alexander~J. McCaskey}.} \bibinfo{year}{2022}\natexlab{}.
\newblock \showarticletitle{Extending Python for Quantum-Classical Computing
  via Quantum Just-in-Time Compilation}.
\newblock \bibinfo{journal}{\emph{ACM Transactions on Quantum Computing}}
  \bibinfo{volume}{3}, \bibinfo{number}{4}, Article \bibinfo{articleno}{24}
  (\bibinfo{date}{jul} \bibinfo{year}{2022}), \bibinfo{numpages}{25}~pages.
\newblock
\showISSN{2643-6809}
\urldef\tempurl%
\url{https://doi.org/10.1145/3544496}
\showDOI{\tempurl}


\bibitem[Nielsen and Chuang(2012)]%
        {nielsenQuantumComputationQuantum2012}
\bibfield{author}{\bibinfo{person}{Michael~A. Nielsen} {and}
  \bibinfo{person}{Isaac~L. Chuang}.} \bibinfo{year}{2012}\natexlab{}.
\newblock \bibinfo{booktitle}{\emph{Quantum {{Computation}} and {{Quantum
  Information}}: 10th {{Anniversary Edition}}} (\bibinfo{edition}{1} ed.)}.
\newblock \bibinfo{publisher}{{Cambridge University Press}},
  \bibinfo{address}{Cambridge}.
\newblock
\urldef\tempurl%
\url{https://doi.org/10.1017/CBO9780511976667}
\showDOI{\tempurl}


\bibitem[NVIDIA(2023)]%
        {cuda_quantum}
\bibfield{author}{\bibinfo{person}{NVIDIA}.} \bibinfo{year}{2023}\natexlab{}.
\newblock \bibinfo{booktitle}{\emph{CUDA Quantum for Hybrid Quantum-Classical
  Computing | NVIDIA Developer}}.
\newblock NVIDIA.
\newblock
\urldef\tempurl%
\url{https://developer.nvidia.com/cuda-quantum}
\showURL{%
Retrieved April 14, 2023 from \tempurl}


\bibitem[O'Hearn and Tennent(1997)]%
        {Reynolds1997}
\bibfield{author}{\bibinfo{person}{Peter~W. O'Hearn} {and}
  \bibinfo{person}{Robert~D. Tennent}.} \bibinfo{year}{1997}\natexlab{}.
\newblock \bibinfo{booktitle}{\emph{The Essence of Algol}}.
\newblock \bibinfo{publisher}{Birkh{\"a}user Boston}, \bibinfo{address}{Boston,
  MA}. 67--88 pages.
\newblock
\showISBNx{978-1-4612-4118-8}
\urldef\tempurl%
\url{https://doi.org/10.1007/978-1-4612-4118-8_4}
\showDOI{\tempurl}


\bibitem[\"{O}mer(1998)]%
        {bernhard_omer_qcl_1998}
\bibfield{author}{\bibinfo{person}{Bernhard \"{O}mer}.}
  \bibinfo{year}{1998}\natexlab{}.
\newblock \emph{\bibinfo{title}{{A} {Procedural} {Formalism} for {Quantum}
  {Computing}}}.
\newblock \bibinfo{thesistype}{Master's\ thesis}. \bibinfo{school}{Technical
  University of Vienna}.
\newblock
\urldef\tempurl%
\url{http://tph.tuwien.ac.at/~oemer/doc/qcldoc.pdf}
\showURL{%
\tempurl}


\bibitem[\"{O}mer(2000)]%
        {bernhard_omer_qcl_2000}
\bibfield{author}{\bibinfo{person}{Bernhard \"{O}mer}.}
  \bibinfo{year}{2000}\natexlab{}.
\newblock \emph{\bibinfo{title}{{Quantum} {Programming} in {QCL}}}.
\newblock \bibinfo{thesistype}{Master's\ thesis}. \bibinfo{school}{Technical
  University of Vienna}.
\newblock
\urldef\tempurl%
\url{http://tph.tuwien.ac.at/~oemer/doc/quprog.pdf}
\showURL{%
\tempurl}


\bibitem[\"{O}mer(2003)]%
        {bernhard_omer_qcl_2003}
\bibfield{author}{\bibinfo{person}{Bernhard \"{O}mer}.}
  \bibinfo{year}{2003}\natexlab{}.
\newblock \emph{\bibinfo{title}{{Structured} {Quantum} {Programming}}}.
\newblock \bibinfo{thesistype}{Ph.\,D. Dissertation}.
  \bibinfo{school}{Technical University of Vienna}.
\newblock
\urldef\tempurl%
\url{http://tph.tuwien.ac.at/~oemer/doc/structquprog.pdf}
\showURL{%
\tempurl}


\bibitem[{OpenMP Architecture Review Board}(2008)]%
        {openmp08}
\bibfield{author}{\bibinfo{person}{{OpenMP Architecture Review Board}}.}
  \bibinfo{year}{2008}\natexlab{}.
\newblock \bibinfo{title}{{OpenMP} Application Program Interface Version 3.0}.
\newblock
\newblock
\urldef\tempurl%
\url{http://www.openmp.org/mp-documents/spec30.pdf}
\showURL{%
\tempurl}


\bibitem[Paolini et~al\mbox{.}(2019)]%
        {paolini_quantum_2019}
\bibfield{author}{\bibinfo{person}{Luca Paolini}, \bibinfo{person}{Luca
  Roversi}, {and} \bibinfo{person}{Margherita Zorzi}.}
  \bibinfo{year}{2019}\natexlab{}.
\newblock \showarticletitle{Quantum {Programming} {Made} {Easy}}.
\newblock \bibinfo{journal}{\emph{Electronic Proceedings in Theoretical
  Computer Science}}  \bibinfo{volume}{292} (\bibinfo{date}{April}
  \bibinfo{year}{2019}), \bibinfo{pages}{133--147}.
\newblock
\showISSN{2075-2180}
\urldef\tempurl%
\url{https://doi.org/10.4204/EPTCS.292.8}
\showDOI{\tempurl}


\bibitem[Paolini and Zorzi(2017)]%
        {paolini_qpcf_2017}
\bibfield{author}{\bibinfo{person}{Luca Paolini} {and}
  \bibinfo{person}{Margherita Zorzi}.} \bibinfo{year}{2017}\natexlab{}.
\newblock \showarticletitle{{qPCF}: A Language for Quantum Circuit
  Computations}. In \bibinfo{booktitle}{\emph{Theory and Applications of Models
  of Computation: 14th Annual Conference, TAMC 2017}}.
  \bibinfo{publisher}{Springer}, \bibinfo{address}{Bern, Switzerland},
  \bibinfo{pages}{455--469}.
\newblock
\urldef\tempurl%
\url{https://doi.org/10.1007/978-3-319-55911-7_33}
\showDOI{\tempurl}


\bibitem[Paykin et~al\mbox{.}(2017)]%
        {paykin_qwire_2017}
\bibfield{author}{\bibinfo{person}{Jennifer Paykin}, \bibinfo{person}{Robert
  Rand}, {and} \bibinfo{person}{Steve Zdancewic}.}
  \bibinfo{year}{2017}\natexlab{}.
\newblock \showarticletitle{{QWIRE}: a core language for quantum circuits}. In
  \bibinfo{booktitle}{\emph{Proceedings of the 44th {ACM} {SIGPLAN} {Symposium}
  on {Principles} of {Programming} {Languages}}}. \bibinfo{publisher}{ACM},
  \bibinfo{address}{Paris France}, \bibinfo{pages}{846--858}.
\newblock
\showISBNx{978-1-4503-4660-3}
\urldef\tempurl%
\url{https://doi.org/10.1145/3009837.3009894}
\showDOI{\tempurl}


\bibitem[Pereira et~al\mbox{.}(2021)]%
        {pereira2021ranking}
\bibfield{author}{\bibinfo{person}{Rui Pereira}, \bibinfo{person}{Marco Couto},
  \bibinfo{person}{Francisco Ribeiro}, \bibinfo{person}{Rui Rua},
  \bibinfo{person}{J{\'a}come Cunha}, \bibinfo{person}{Jo{\~a}o~Paulo
  Fernandes}, {and} \bibinfo{person}{Jo{\~a}o Saraiva}.}
  \bibinfo{year}{2021}\natexlab{}.
\newblock \showarticletitle{Ranking programming languages by energy
  efficiency}.
\newblock \bibinfo{journal}{\emph{Science of Computer Programming}}
  \bibinfo{volume}{205} (\bibinfo{year}{2021}), \bibinfo{pages}{102609}.
\newblock
\urldef\tempurl%
\url{https://doi.org/10.1016/j.scico.2021.102609}
\showDOI{\tempurl}


\bibitem[Polychronopoulos et~al\mbox{.}(1989)]%
        {POLYCHRONOPOULOS1989}
\bibfield{author}{\bibinfo{person}{Constantine~D. Polychronopoulos},
  \bibinfo{person}{Milind Girkar}, \bibinfo{person}{Mohammad~Reza HAGHIGHAT},
  \bibinfo{person}{Chia~Ling Lee}, \bibinfo{person}{Bruce Leung}, {and}
  \bibinfo{person}{Dale Schouten}.} \bibinfo{year}{1989}\natexlab{}.
\newblock \showarticletitle{{PARAFRASE}-2: An environment for Parallelizing,
  Partitioning, Synchronizing, and Scheduling Programs on Multiprocessors}.
\newblock \bibinfo{journal}{\emph{International Journal of High Speed
  Computing}} \bibinfo{volume}{01}, \bibinfo{number}{01} (\bibinfo{date}{May}
  \bibinfo{year}{1989}), \bibinfo{pages}{45--72}.
\newblock
\urldef\tempurl%
\url{https://doi.org/10.1142/s0129053389000044}
\showDOI{\tempurl}


\bibitem[Purkeypile(2009)]%
        {purkeypile_cove_2009}
\bibfield{author}{\bibinfo{person}{Matt Purkeypile}.}
  \bibinfo{year}{2009}\natexlab{}.
\newblock \bibinfo{title}{Cove: {A} {Practical} {Quantum} {Computer}
  {Programming} {Framework}}.
\newblock
\newblock
\urldef\tempurl%
\url{http://arxiv.org/abs/0911.2423}
\showURL{%
\tempurl}


\bibitem[{QIR Alliance}(2021)]%
        {QIRSpec2021}
\bibfield{author}{\bibinfo{person}{{QIR Alliance}}.}
  \bibinfo{year}{2021}\natexlab{}.
\newblock \bibinfo{booktitle}{\emph{{QIR Specification}}}.
\newblock QIR Alliance.
\newblock
\urldef\tempurl%
\url{https://github.com/qir-alliance/qir-spec}
\showURL{%
\tempurl}
\newblock
\shownote{Also see \url{https://qir-alliance.org}}.


\bibitem[Quantum(2021)]%
        {ibm_qasm3_features}
\bibfield{author}{\bibinfo{person}{IBM Quantum}.}
  \bibinfo{year}{2021}\natexlab{}.
\newblock \bibinfo{booktitle}{\emph{OpenQASM 3 language features}}.
\newblock IBM.
\newblock
\urldef\tempurl%
\url{https://quantum-computing.ibm.com/services/resources/docs/resources/manage/systems/dynamic-circuits/feature-table}
\showURL{%
Retrieved February 20, 2023 from \tempurl}


\bibitem[Rand et~al\mbox{.}(2019)]%
        {rand_reqwire_2019}
\bibfield{author}{\bibinfo{person}{Robert Rand}, \bibinfo{person}{Jennifer
  Paykin}, \bibinfo{person}{Dong-Ho Lee}, {and} \bibinfo{person}{Steve
  Zdancewic}.} \bibinfo{year}{2019}\natexlab{}.
\newblock \showarticletitle{{ReQWIRE}: {Reasoning} about {Reversible} {Quantum}
  {Circuits}}.
\newblock \bibinfo{journal}{\emph{Electronic Proceedings in Theoretical
  Computer Science}}  \bibinfo{volume}{287} (\bibinfo{date}{Jan.}
  \bibinfo{year}{2019}), \bibinfo{pages}{299--312}.
\newblock
\showISSN{2075-2180}
\urldef\tempurl%
\url{https://doi.org/10.4204/EPTCS.287.17}
\showDOI{\tempurl}


\bibitem[Ravi et~al\mbox{.}(2021)]%
        {ravi2021adaptive}
\bibfield{author}{\bibinfo{person}{Gokul~Subramanian Ravi},
  \bibinfo{person}{Kaitlin~N. Smith}, \bibinfo{person}{Prakash Murali}, {and}
  \bibinfo{person}{Frederic~T. Chong}.} \bibinfo{year}{2021}\natexlab{}.
\newblock \showarticletitle{Adaptive job and resource management for the
  growing quantum cloud}. In \bibinfo{booktitle}{\emph{2021 IEEE International
  Conference on Quantum Computing and Engineering (QCE)}}.
  \bibinfo{publisher}{IEE}, \bibinfo{address}{Broomfield, CO, USA},
  \bibinfo{pages}{301--312}.
\newblock
\urldef\tempurl%
\url{https://doi.org/10.1109/QCE52317.2021.00047}
\showDOI{\tempurl}


\bibitem[Riel(2021)]%
        {ibm_roadmap}
\bibfield{author}{\bibinfo{person}{Heike Riel}.}
  \bibinfo{year}{2021}\natexlab{}.
\newblock \showarticletitle{Quantum Computing Technology}. In
  \bibinfo{booktitle}{\emph{2021 IEEE International Electron Devices Meeting
  (IEDM)}}. \bibinfo{publisher}{IEEE}, \bibinfo{address}{San Francisco, CA,
  USA}, \bibinfo{pages}{1.3.1--1.3.7}.
\newblock
\urldef\tempurl%
\url{https://doi.org/10.1109/IEDM19574.2021.9720538}
\showDOI{\tempurl}


\bibitem[Rios and Selinger(2018)]%
        {rios_categorical_2018}
\bibfield{author}{\bibinfo{person}{Francisco Rios} {and} \bibinfo{person}{Peter
  Selinger}.} \bibinfo{year}{2018}\natexlab{}.
\newblock \showarticletitle{A {Categorical} {Model} for a {Quantum} {Circuit}
  {Description} {Language} ({Extended} {Abstract})}.
\newblock \bibinfo{journal}{\emph{Electronic Proceedings in Theoretical
  Computer Science}}  \bibinfo{volume}{266} (\bibinfo{date}{Feb.}
  \bibinfo{year}{2018}), \bibinfo{pages}{164--178}.
\newblock
\showISSN{2075-2180}
\urldef\tempurl%
\url{https://doi.org/10.4204/EPTCS.266.11}
\showDOI{\tempurl}


\bibitem[Romero and Aspuru-Guzik(2021)]%
        {romero2021variational}
\bibfield{author}{\bibinfo{person}{Jonathan Romero} {and}
  \bibinfo{person}{Alán Aspuru-Guzik}.} \bibinfo{year}{2021}\natexlab{}.
\newblock \showarticletitle{Variational Quantum Generators: Generative
  Adversarial Quantum Machine Learning for Continuous Distributions}.
\newblock \bibinfo{journal}{\emph{Advanced Quantum Technologies}}
  \bibinfo{volume}{4}, \bibinfo{number}{1} (\bibinfo{year}{2021}),
  \bibinfo{pages}{2000003}.
\newblock
\urldef\tempurl%
\url{https://doi.org/10.1002/qute.202000003}
\showDOI{\tempurl}


\bibitem[Romero et~al\mbox{.}(2017)]%
        {Romero_2017}
\bibfield{author}{\bibinfo{person}{Jonathan Romero},
  \bibinfo{person}{Jonathan~P Olson}, {and} \bibinfo{person}{Alan
  Aspuru-Guzik}.} \bibinfo{year}{2017}\natexlab{}.
\newblock \showarticletitle{Quantum autoencoders for efficient compression of
  quantum data}.
\newblock \bibinfo{journal}{\emph{Quantum Science and Technology}}
  \bibinfo{volume}{2}, \bibinfo{number}{4} (\bibinfo{date}{aug}
  \bibinfo{year}{2017}), \bibinfo{pages}{045001}.
\newblock
\urldef\tempurl%
\url{https://doi.org/10.1088/2058-9565/aa8072}
\showDOI{\tempurl}


\bibitem[Ross(2017)]%
        {ross_algebraic_2017}
\bibfield{author}{\bibinfo{person}{Neil~J. Ross}.}
  \bibinfo{year}{2017}\natexlab{}.
\newblock \bibinfo{title}{Algebraic and {Logical} {Methods} in {Quantum}
  {Computation}}.
\newblock
\newblock
\urldef\tempurl%
\url{http://arxiv.org/abs/1510.02198}
\showURL{%
\tempurl}


\bibitem[R{\"u}diger(2007)]%
        {rudiger_quantum_2006}
\bibfield{author}{\bibinfo{person}{Roland R{\"u}diger}.}
  \bibinfo{year}{2007}\natexlab{}.
\newblock \showarticletitle{Quantum programming languages: An introductory
  overview}.
\newblock \bibinfo{journal}{\emph{Comput. J.}} \bibinfo{volume}{50},
  \bibinfo{number}{2} (\bibinfo{year}{2007}), \bibinfo{pages}{134--150}.
\newblock


\bibitem[Ruefenacht et~al\mbox{.}(2022)]%
        {ruefenacht2022ea}
\bibfield{author}{\bibinfo{person}{Martin Ruefenacht},
  \bibinfo{person}{Bruno~G. Taketani}, \bibinfo{person}{Pasi Lähteenmäki},
  \bibinfo{person}{Ville Bergholm}, \bibinfo{person}{Dieter Kranzlmüller},
  \bibinfo{person}{Laura Schulz}, {and} \bibinfo{person}{Martin Schulz}.}
  \bibinfo{year}{2022}\natexlab{}.
\newblock \bibinfo{title}{Bringing quantum acceleration to supercomputers}.
\newblock
\newblock
\urldef\tempurl%
\url{https://meetiqm.com/uploads/documents/IQM_HPC-QC-Integration-Whitepaper.pdf}
\showURL{%
\tempurl}


\bibitem[Selinger(2004a)]%
        {goos_brief_2004}
\bibfield{author}{\bibinfo{person}{Peter Selinger}.}
  \bibinfo{year}{2004}\natexlab{a}.
\newblock \showarticletitle{A {Brief} {Survey} of {Quantum} {Programming}
  {Languages}}.
\newblock In \bibinfo{booktitle}{\emph{Functional and {Logic} {Programming}}},
  \bibfield{editor}{\bibinfo{person}{Gerhard Goos}, \bibinfo{person}{Juris
  Hartmanis}, \bibinfo{person}{Jan van Leeuwen}, \bibinfo{person}{Yukiyoshi
  Kameyama}, {and} \bibinfo{person}{Peter~J. Stuckey}} (Eds.).
  Vol.~\bibinfo{volume}{2998}. \bibinfo{publisher}{Springer Berlin Heidelberg},
  \bibinfo{address}{Berlin, Heidelberg}, \bibinfo{pages}{1--6}.
\newblock
\showISBNx{978-3-540-21402-1 978-3-540-24754-8}
\urldef\tempurl%
\url{https://doi.org/10.1007/978-3-540-24754-8_1}
\showDOI{\tempurl}


\bibitem[Selinger(2004b)]%
        {selinger_qfc_2004}
\bibfield{author}{\bibinfo{person}{Peter Selinger}.}
  \bibinfo{year}{2004}\natexlab{b}.
\newblock \showarticletitle{Towards a quantum programming language}.
\newblock \bibinfo{journal}{\emph{Mathematical Structures in Computer Science}}
  \bibinfo{volume}{14}, \bibinfo{number}{4} (\bibinfo{date}{Aug.}
  \bibinfo{year}{2004}), \bibinfo{pages}{527--586}.
\newblock
\showISSN{0960-1295, 1469-8072}
\urldef\tempurl%
\url{https://doi.org/10.1017/S0960129504004256}
\showDOI{\tempurl}


\bibitem[Selinger and Valiron(2006)]%
        {selinger_lambda_2006}
\bibfield{author}{\bibinfo{person}{Peter Selinger} {and}
  \bibinfo{person}{Benoit Valiron}.} \bibinfo{year}{2006}\natexlab{}.
\newblock \showarticletitle{A lambda calculus for quantum computation with
  classical control}.
\newblock \bibinfo{journal}{\emph{Mathematical Structures in Computer Science}}
  \bibinfo{volume}{16}, \bibinfo{number}{3} (\bibinfo{date}{June}
  \bibinfo{year}{2006}), \bibinfo{pages}{527--552}.
\newblock
\showISSN{0960-1295, 1469-8072}
\urldef\tempurl%
\url{https://doi.org/10.1017/S0960129506005238}
\showDOI{\tempurl}


\bibitem[Shor(1994)]%
        {shor1994algorithms}
\bibfield{author}{\bibinfo{person}{Peter~W. Shor}.}
  \bibinfo{year}{1994}\natexlab{}.
\newblock \showarticletitle{Algorithms for quantum computation: discrete
  logarithms and factoring}. In \bibinfo{booktitle}{\emph{Proceedings 35th
  annual symposium on foundations of computer science}}.
  \bibinfo{publisher}{IEEE}, \bibinfo{address}{Santa Fe, NM, USA},
  \bibinfo{pages}{124--134}.
\newblock
\urldef\tempurl%
\url{https://doi.org/10.1109/SFCS.1994.365700}
\showDOI{\tempurl}


\bibitem[Shor(1995)]%
        {shor1995scheme}
\bibfield{author}{\bibinfo{person}{Peter~W Shor}.}
  \bibinfo{year}{1995}\natexlab{}.
\newblock \showarticletitle{Scheme for reducing decoherence in quantum computer
  memory}.
\newblock \bibinfo{journal}{\emph{Physical review A}} \bibinfo{volume}{52},
  \bibinfo{number}{4} (\bibinfo{year}{1995}), \bibinfo{pages}{R2493}.
\newblock
\urldef\tempurl%
\url{https://doi.org/10.1103/PhysRevA.52.R2493}
\showDOI{\tempurl}


\bibitem[Sivarajah et~al\mbox{.}(2020)]%
        {sivarajah_tket_2020}
\bibfield{author}{\bibinfo{person}{Seyon Sivarajah}, \bibinfo{person}{Silas
  Dilkes}, \bibinfo{person}{Alexander Cowtan}, \bibinfo{person}{Will Simmons},
  \bibinfo{person}{Alec Edgington}, {and} \bibinfo{person}{Ross Duncan}.}
  \bibinfo{year}{2020}\natexlab{}.
\newblock \showarticletitle{t{\textbar}ket⟩ : {A} {Retargetable} {Compiler}
  for {NISQ} {Devices}}.
\newblock \bibinfo{journal}{\emph{Quantum Science and Technology}}
  \bibinfo{volume}{6}, \bibinfo{number}{1} (\bibinfo{year}{2020}),
  \bibinfo{pages}{014003}.
\newblock
\showISSN{2058-9565}
\urldef\tempurl%
\url{https://doi.org/10.1088/2058-9565/ab8e92}
\showDOI{\tempurl}


\bibitem[Smith et~al\mbox{.}(2017)]%
        {smith_practical_2017}
\bibfield{author}{\bibinfo{person}{Robert~S. Smith},
  \bibinfo{person}{Michael~J. Curtis}, {and} \bibinfo{person}{William~J.
  Zeng}.} \bibinfo{year}{2017}\natexlab{}.
\newblock \bibinfo{title}{A {Practical} {Quantum} {Instruction} {Set}
  {Architecture}}.
\newblock
\newblock
\urldef\tempurl%
\url{http://arxiv.org/abs/1608.03355}
\showURL{%
\tempurl}


\bibitem[Sofge(2008)]%
        {sofge_survey_2008}
\bibfield{author}{\bibinfo{person}{Donald~A Sofge}.}
  \bibinfo{year}{2008}\natexlab{}.
\newblock \showarticletitle{A survey of quantum programming languages: History,
  methods, and tools}. In \bibinfo{booktitle}{\emph{Second International
  Conference on Quantum, Nano and Micro Technologies (ICQNM 2008)}}.
  \bibinfo{publisher}{IEEE}, \bibinfo{address}{Sainte Luce, Martinique,
  France}, \bibinfo{pages}{66--71}.
\newblock
\urldef\tempurl%
\url{https://doi.org/10.1109/ICQNM.2008.15}
\showDOI{\tempurl}


\bibitem[Steiger et~al\mbox{.}(2018)]%
        {steiger_projectq_2018}
\bibfield{author}{\bibinfo{person}{Damian~S. Steiger}, \bibinfo{person}{Thomas
  Häner}, {and} \bibinfo{person}{Matthias Troyer}.}
  \bibinfo{year}{2018}\natexlab{}.
\newblock \showarticletitle{{ProjectQ}: {An} {Open} {Source} {Software}
  {Framework} for {Quantum} {Computing}}.
\newblock \bibinfo{journal}{\emph{Quantum}}  \bibinfo{volume}{2}
  (\bibinfo{date}{Jan.} \bibinfo{year}{2018}), \bibinfo{pages}{49}.
\newblock
\showISSN{2521-327X}
\urldef\tempurl%
\url{https://doi.org/10.22331/q-2018-01-31-49}
\showDOI{\tempurl}


\bibitem[Svore et~al\mbox{.}(2006)]%
        {svore_layered_2006}
\bibfield{author}{\bibinfo{person}{Krysta~M. Svore}, \bibinfo{person}{Alfred~V.
  Aho}, \bibinfo{person}{Andrew~W. Cross}, \bibinfo{person}{Isaac Chuang},
  {and} \bibinfo{person}{Igor~L. Markov}.} \bibinfo{year}{2006}\natexlab{}.
\newblock \showarticletitle{A layered software architecture for quantum
  computing design tools}.
\newblock \bibinfo{journal}{\emph{Computer}} \bibinfo{volume}{39},
  \bibinfo{number}{1} (\bibinfo{date}{Jan.} \bibinfo{year}{2006}),
  \bibinfo{pages}{74--83}.
\newblock
\showISSN{0018-9162}
\urldef\tempurl%
\url{https://doi.org/10.1109/MC.2006.4}
\showDOI{\tempurl}


\bibitem[Svore et~al\mbox{.}(2018)]%
        {svore2018q}
\bibfield{author}{\bibinfo{person}{Krysta~M. Svore}, \bibinfo{person}{Alan
  Geller}, \bibinfo{person}{Matthias Troyer}, \bibinfo{person}{John Azariah},
  \bibinfo{person}{Christopher Granade}, \bibinfo{person}{Bettina Heim},
  \bibinfo{person}{Vadym Kliuchnikov}, \bibinfo{person}{Mariia Mykhailova},
  \bibinfo{person}{Andres Paz}, {and} \bibinfo{person}{Martin Roetteler}.}
  \bibinfo{year}{2018}\natexlab{}.
\newblock \showarticletitle{Q\#: {Enabling} scalable quantum computing and
  development with a high-level domain-specific language}. In
  \bibinfo{booktitle}{\emph{Proceedings of the {Real} {World} {Domain}
  {Specific} {Languages}}}. \bibinfo{publisher}{Association for Computing
  Machinery}, \bibinfo{address}{Vienna Austria}, \bibinfo{pages}{1--10}.
\newblock
\urldef\tempurl%
\url{https://doi.org/10.1145/3183895.3183901}
\showDOI{\tempurl}


\bibitem[Team(2022)]%
        {coq_2022}
\bibfield{author}{\bibinfo{person}{The Coq~Development Team}.}
  \bibinfo{year}{2022}\natexlab{}.
\newblock \bibinfo{booktitle}{\emph{The Coq Proof Assistant}}.
\newblock The Coq Development Team.
\newblock
\urldef\tempurl%
\url{https://doi.org/10.5281/zenodo.7313584}
\showDOI{\tempurl}


\bibitem[Temme et~al\mbox{.}(2017)]%
        {temme2017error}
\bibfield{author}{\bibinfo{person}{Kristan Temme}, \bibinfo{person}{Sergey
  Bravyi}, {and} \bibinfo{person}{Jay~M. Gambetta}.}
  \bibinfo{year}{2017}\natexlab{}.
\newblock \showarticletitle{Error Mitigation for Short-Depth Quantum Circuits}.
\newblock \bibinfo{journal}{\emph{Phys. Rev. Lett.}}  \bibinfo{volume}{119}
  (\bibinfo{date}{Nov} \bibinfo{year}{2017}), \bibinfo{pages}{180509}.
\newblock
Issue 18.
\urldef\tempurl%
\url{https://doi.org/10.1103/PhysRevLett.119.180509}
\showDOI{\tempurl}


\bibitem[Unruh(2006)]%
        {unruh_quantum_2006}
\bibfield{author}{\bibinfo{person}{Dominique Unruh}.}
  \bibinfo{year}{2006}\natexlab{}.
\newblock \showarticletitle{Quantum {Programming} {Languages}}.
\newblock \bibinfo{journal}{\emph{Informatik-Forschung und Entwicklung}}
  \bibinfo{volume}{21}, \bibinfo{number}{1-2} (\bibinfo{year}{2006}),
  \bibinfo{pages}{55--63}.
\newblock
\urldef\tempurl%
\url{https://doi.org/10.1007/s00450-006-0012-y}
\showDOI{\tempurl}


\bibitem[Valiron(2013)]%
        {Valiron_2012}
\bibfield{author}{\bibinfo{person}{Beno{\^\i}t Valiron}.}
  \bibinfo{year}{2013}\natexlab{}.
\newblock \showarticletitle{Quantum computation: From a programmer’s
  perspective}.
\newblock \bibinfo{journal}{\emph{New Generation Computing}}
  \bibinfo{volume}{31}, \bibinfo{number}{1} (\bibinfo{year}{2013}),
  \bibinfo{pages}{1--26}.
\newblock


\bibitem[van Tonder(2004)]%
        {van_tonder_lambda_2004}
\bibfield{author}{\bibinfo{person}{André van Tonder}.}
  \bibinfo{year}{2004}\natexlab{}.
\newblock \showarticletitle{A {Lambda} {Calculus} for {Quantum} {Computation}}.
\newblock \bibinfo{journal}{\emph{SIAM J. Comput.}} \bibinfo{volume}{33},
  \bibinfo{number}{5} (\bibinfo{date}{Jan.} \bibinfo{year}{2004}),
  \bibinfo{pages}{1109--1135}.
\newblock
\showISSN{0097-5397, 1095-7111}
\urldef\tempurl%
\url{https://doi.org/10.1137/S0097539703432165}
\showDOI{\tempurl}


\bibitem[van Tonder and Dorca(2003)]%
        {tonder_lambda_2003}
\bibfield{author}{\bibinfo{person}{André van Tonder} {and}
  \bibinfo{person}{Miquel Dorca}.} \bibinfo{year}{2003}\natexlab{}.
\newblock \bibinfo{title}{Quantum Computation, Categorical Semantics and Linear
  Logic}.
\newblock
\newblock
\urldef\tempurl%
\url{https://doi.org/10.48550/ARXIV.QUANT-PH/0312174}
\showDOI{\tempurl}


\bibitem[Voichick et~al\mbox{.}(2023)]%
        {voichick_qunity_2022}
\bibfield{author}{\bibinfo{person}{Finn Voichick}, \bibinfo{person}{Liyi Li},
  \bibinfo{person}{Robert Rand}, {and} \bibinfo{person}{Michael Hicks}.}
  \bibinfo{year}{2023}\natexlab{}.
\newblock \showarticletitle{Qunity: A Unified Language for Quantum and
  Classical Computing}.
\newblock \bibinfo{journal}{\emph{Proceedings of the ACM on Programming
  Languages}} \bibinfo{volume}{7}, \bibinfo{number}{POPL}
  (\bibinfo{year}{2023}), \bibinfo{pages}{921--951}.
\newblock
\urldef\tempurl%
\url{https://doi.org/10.1145/3571225}
\showDOI{\tempurl}


\bibitem[Wang et~al\mbox{.}(2020)]%
        {chemical2020}
\bibfield{author}{\bibinfo{person}{Christopher~S. Wang},
  \bibinfo{person}{Jacob~C. Curtis}, \bibinfo{person}{Brian~J. Lester},
  \bibinfo{person}{Yaxing Zhang}, \bibinfo{person}{Yvonne~Y. Gao},
  \bibinfo{person}{Jessica Freeze}, \bibinfo{person}{Victor~S. Batista},
  \bibinfo{person}{Patrick~H. Vaccaro}, \bibinfo{person}{Isaac~L. Chuang},
  \bibinfo{person}{Luigi Frunzio}, \bibinfo{person}{Liang Jiang},
  \bibinfo{person}{Steven~M. Girvin}, {and} \bibinfo{person}{Robert~J.
  Schoelkopf}.} \bibinfo{year}{2020}\natexlab{}.
\newblock \showarticletitle{Efficient Multiphoton Sampling of Molecular
  Vibronic Spectra on a Superconducting Bosonic Processor}.
\newblock \bibinfo{journal}{\emph{Phys. Rev. X}}  \bibinfo{volume}{10}
  (\bibinfo{date}{06} \bibinfo{year}{2020}), \bibinfo{pages}{021060}.
\newblock
Issue 2.
\urldef\tempurl%
\url{https://doi.org/10.1103/PhysRevX.10.021060}
\showDOI{\tempurl}


\bibitem[Wang et~al\mbox{.}(2014)]%
        {wang2014exploitation}
\bibfield{author}{\bibinfo{person}{Zheng Wang}, \bibinfo{person}{Daniel
  Powell}, \bibinfo{person}{Bj{\"o}rn Franke}, {and} \bibinfo{person}{Michael
  O'Boyle}.} \bibinfo{year}{2014}\natexlab{}.
\newblock \showarticletitle{Exploitation of GPUs for the Parallelisation of
  Probably Parallel Legacy Code}. In \bibinfo{booktitle}{\emph{Compiler
  Construction}}, \bibfield{editor}{\bibinfo{person}{Albert Cohen}} (Ed.).
  \bibinfo{publisher}{Springer Berlin Heidelberg}, \bibinfo{address}{Berlin,
  Heidelberg}, \bibinfo{pages}{154--173}.
\newblock
\showISBNx{978-3-642-54807-9}
\urldef\tempurl%
\url{https://doi.org/10.1007/978-3-642-54807-9_9}
\showDOI{\tempurl}


\bibitem[Weder et~al\mbox{.}(2023)]%
        {wederProvenancePreservingAnalysisRewrite2023}
\bibfield{author}{\bibinfo{person}{Benjamin Weder}, \bibinfo{person}{Johanna
  Barzen}, \bibinfo{person}{Martin Beisel}, {and} \bibinfo{person}{Frank
  Leymann}.} \bibinfo{year}{2023}\natexlab{}.
\newblock \showarticletitle{Provenance-{{Preserving Analysis}} and {{Rewrite}}
  of {{Quantum Workflows}} for {{Hybrid Quantum Algorithms}}}.
\newblock \bibinfo{journal}{\emph{SN COMPUT. SCI.}} \bibinfo{volume}{4},
  \bibinfo{number}{3} (\bibinfo{date}{Feb.} \bibinfo{year}{2023}),
  \bibinfo{pages}{233}.
\newblock
\urldef\tempurl%
\url{https://doi.org/10.1007/s42979-022-01625-9}
\showDOI{\tempurl}


\bibitem[Wille et~al\mbox{.}(2019)]%
        {willeMappingQuantumCircuits2019}
\bibfield{author}{\bibinfo{person}{Robert Wille}, \bibinfo{person}{Lukas
  Burgholzer}, {and} \bibinfo{person}{Alwin Zulehner}.}
  \bibinfo{year}{2019}\natexlab{}.
\newblock \showarticletitle{Mapping Quantum Circuits to IBM QX Architectures
  Using the Minimal Number of SWAP and H Operations}. In
  \bibinfo{booktitle}{\emph{Proceedings of the 56th Annual Design Automation
  Conference 2019}} (Las Vegas, NV, USA) \emph{(\bibinfo{series}{DAC '19})}.
  \bibinfo{publisher}{Association for Computing Machinery},
  \bibinfo{address}{New York, NY, USA}, Article \bibinfo{articleno}{142},
  \bibinfo{numpages}{6}~pages.
\newblock
\showISBNx{9781450367257}
\urldef\tempurl%
\url{https://doi.org/10.1145/3316781.3317859}
\showDOI{\tempurl}
\newblock
\shownote{ISSN: 0738-100X}.


\bibitem[Wootters and Zurek(1982)]%
        {Wootters1982}
\bibfield{author}{\bibinfo{person}{William~K. Wootters} {and}
  \bibinfo{person}{Wojciech~H. Zurek}.} \bibinfo{year}{1982}\natexlab{}.
\newblock \showarticletitle{A single quantum cannot be cloned}.
\newblock \bibinfo{journal}{\emph{Nature}} \bibinfo{volume}{299},
  \bibinfo{number}{5886} (\bibinfo{year}{1982}), \bibinfo{pages}{802--803}.
\newblock
\showISSN{1476-4687}
\urldef\tempurl%
\url{https://doi.org/10.1038/299802a0}
\showDOI{\tempurl}


\bibitem[Xu and Song(2008)]%
        {xu_ndqjava_2008}
\bibfield{author}{\bibinfo{person}{Jiafu Xu} {and} \bibinfo{person}{Fanming
  Song}.} \bibinfo{year}{2008}\natexlab{}.
\newblock \showarticletitle{Quantum programming languages}.
\newblock \bibinfo{journal}{\emph{Frontiers of Computer Science in China}}
  \bibinfo{volume}{2}, \bibinfo{number}{2} (\bibinfo{date}{June}
  \bibinfo{year}{2008}), \bibinfo{pages}{161--166}.
\newblock
\showISSN{1673-7350, 1673-7466}
\urldef\tempurl%
\url{https://doi.org/10.1007/s11704-008-0013-z}
\showDOI{\tempurl}


\bibitem[Yin et~al\mbox{.}(2020)]%
        {securecom2020}
\bibfield{author}{\bibinfo{person}{Juan Yin}, \bibinfo{person}{Yu-Huai Li},
  \bibinfo{person}{Sheng-Kai Liao}, \bibinfo{person}{Meng Yang},
  \bibinfo{person}{Yuan Cao}, \bibinfo{person}{Liang Zhang},
  \bibinfo{person}{Ji-Gang Ren}, \bibinfo{person}{Wen-Qi Cai},
  \bibinfo{person}{Wei-Yue Liu}, \bibinfo{person}{Shuang-Lin Li},
  {et~al\mbox{.}}} \bibinfo{year}{2020}\natexlab{}.
\newblock \showarticletitle{Entanglement-based secure quantum cryptography over
  1,120 kilometres}.
\newblock \bibinfo{journal}{\emph{Nature}} \bibinfo{volume}{582},
  \bibinfo{number}{7813} (\bibinfo{year}{2020}), \bibinfo{pages}{501--505}.
\newblock
\urldef\tempurl%
\url{https://doi.org/10.1038/s41586-020-2401-y}
\showDOI{\tempurl}


\bibitem[Ying et~al\mbox{.}(2012)]%
        {ying_quantum_2012}
\bibfield{author}{\bibinfo{person}{MingSheng Ying}, \bibinfo{person}{Yuan
  Feng}, \bibinfo{person}{RunYao Duan}, \bibinfo{person}{YangJia Li}, {and}
  \bibinfo{person}{NengKun Yu}.} \bibinfo{year}{2012}\natexlab{}.
\newblock \showarticletitle{Quantum programming: {From} theories to
  implementations}.
\newblock \bibinfo{journal}{\emph{Chinese Science Bulletin}}
  \bibinfo{volume}{57}, \bibinfo{number}{16} (\bibinfo{date}{June}
  \bibinfo{year}{2012}), \bibinfo{pages}{1903--1909}.
\newblock
\showISSN{1001-6538, 1861-9541}
\urldef\tempurl%
\url{https://doi.org/10.1007/s11434-012-5147-6}
\showDOI{\tempurl}


\bibitem[Yoo et~al\mbox{.}(2003)]%
        {Yoo2003SLURMSL}
\bibfield{author}{\bibinfo{person}{Andy~B. Yoo}, \bibinfo{person}{Morris~A.
  Jette}, {and} \bibinfo{person}{Mark Grondona}.}
  \bibinfo{year}{2003}\natexlab{}.
\newblock \showarticletitle{SLURM: Simple Linux Utility for Resource
  Management}. In \bibinfo{booktitle}{\emph{Job Scheduling Strategies for
  Parallel Processing}}, \bibfield{editor}{\bibinfo{person}{Dror Feitelson},
  \bibinfo{person}{Larry Rudolph}, {and} \bibinfo{person}{Uwe Schwiegelshohn}}
  (Eds.). \bibinfo{publisher}{Springer Berlin Heidelberg},
  \bibinfo{address}{Berlin, Heidelberg}, \bibinfo{pages}{44--60}.
\newblock
\showISBNx{978-3-540-39727-4}
\urldef\tempurl%
\url{https://doi.org/10.1007/10968987_3}
\showDOI{\tempurl}


\bibitem[Yuan et~al\mbox{.}(2022)]%
        {yuan_twist_2022}
\bibfield{author}{\bibinfo{person}{Charles Yuan}, \bibinfo{person}{Christopher
  McNally}, {and} \bibinfo{person}{Michael Carbin}.}
  \bibinfo{year}{2022}\natexlab{}.
\newblock \showarticletitle{Twist: sound reasoning for purity and entanglement
  in {Quantum} programs}.
\newblock \bibinfo{journal}{\emph{Proceedings of the ACM on Programming
  Languages}} \bibinfo{volume}{6}, \bibinfo{number}{POPL} (\bibinfo{date}{Jan.}
  \bibinfo{year}{2022}), \bibinfo{pages}{1--32}.
\newblock
\showISSN{2475-1421}
\urldef\tempurl%
\url{https://doi.org/10.1145/3498691}
\showDOI{\tempurl}


\bibitem[Zhao(2021)]%
        {zhao_quantum_2021}
\bibfield{author}{\bibinfo{person}{Jianjun Zhao}.}
  \bibinfo{year}{2021}\natexlab{}.
\newblock \bibinfo{title}{Quantum {Software} {Engineering}: {Landscapes} and
  {Horizons}}.
\newblock
\newblock
\urldef\tempurl%
\url{http://arxiv.org/abs/2007.07047}
\showURL{%
\tempurl}


\bibitem[Zorzi(2019)]%
        {zorzi_quantum_2019}
\bibfield{author}{\bibinfo{person}{Margherita Zorzi}.}
  \bibinfo{year}{2019}\natexlab{}.
\newblock \showarticletitle{Quantum {Calculi}—{From} {Theory} to {Language}
  {Design}}.
\newblock \bibinfo{journal}{\emph{Applied Sciences}} \bibinfo{volume}{9},
  \bibinfo{number}{24} (\bibinfo{date}{Jan.} \bibinfo{year}{2019}),
  \bibinfo{pages}{5472}.
\newblock
\showISSN{2076-3417}
\urldef\tempurl%
\url{https://doi.org/10.3390/app9245472}
\showDOI{\tempurl}


\end{thebibliography}

\end{document}